\documentclass[11pt]{article}
\usepackage[a4paper,bottom=4.2cm,top=1cm,head=3cm,width=18cm,dvipdfm]{geometry}
\usepackage[ps2pdf,bookmarks=true,bookmarksnumbered=true,breaklinks=true,pdfpagemode=Fullscreen,pdfstartview=FitBH]{hyperref}
\evensidemargin=\oddsidemargin

\usepackage[dotinlabels]{titletoc}
\usepackage{titlesec}

\usepackage{graphicx}% Include figure files
\usepackage{dcolumn}% Align table columns on decimal point
\usepackage{bm}% bold math
\usepackage{amssymb}
\usepackage{mathrsfs}
\usepackage{amsmath}
\usepackage{epsfig}
\usepackage{here}
\usepackage[dvips]{color}
\usepackage{hhline}
\usepackage{srcltx}

\allowdisplaybreaks[4]

\renewcommand{\thefootnote}{\fnsymbol{footnote}}
\newcommand{\beq}{\begin{equation}}
\newcommand{\eeq}{\end{equation}}
\newcommand{\bq}{\begin{equation}}
\newcommand{\eq}{\end{equation}}
\newcommand{\ba}{\begin{array}}
\newcommand{\ea}{\end{array}}
\newcommand{\beqa}{\begin{eqnarray}}
\newcommand{\eeqa}{\end{eqnarray}}
\newcommand{\beqs}{\begin{subequations}}
\newcommand{\eeqs}{\end{subequations}}

\def\nn{\nonumber}
\def\non{\nonumber}
\newcommand{\bd}{\boldsymbol}

\newcommand{\ZZ}{\mathbb{Z}}

\def\bd{\boldsymbol}
\def\hf{\frac{1}{2}}

\def\om{\omega}

\def\O{{\cal O}}

\def\NN{{\mathcal N}}
\def\RR{{\mathcal R}}

\def\Phit{\widetilde{\Phi}}

\def\mt{\widetilde{m}}
\def\tm{{\widetilde{m}}}

\def\End{\end{document}}

\def\to{\rightarrow}

\def\dis{\displaystyle}
\def\f{\frac}
\def\ov{\overline}

\def\[{\left[}
\def\]{\right]}
\def\({\left(}
\def\){\right)}

\def\a{\alpha}

\def\ab{\bar{\alpha}}

\def\U1EM{U(1)_{\rm em}}
\def\O{\mathcal O}

\def\leqq{\leqslant}
\def\geqq{\geqslant}

\def\N{{\cal N}}
\def\O{\mathcal{O}}
\def\mD{m_D^{}}
\def\mDT{m_D^T}

\def\DetMR{\Vert M_R\Vert}

\def\sR{s_R^{}}
\def\cR{c_R^{}}

\def\[{\left[}
\def\]{\right]}
\def\dis{\displaystyle}

\def\N{{\cal N}}
\def\d{\delta}
\def\da{\delta_a}
\def\dx{\delta_x}

\def\mh{\widehat{m}}

\def\pR{p_R^{}}
\def\pb{\bar{p}}
\def\pr{p_r^{}}
\def\a{\alpha}

\def\ep{\epsilon}
\def\deg{\circ}

\def\d{\delta}

 \def\tbc{\theta_{23}^{~}}
 \def\tac{\theta_{13}^{~}}
 \def\ts{\theta_{s}}
 \def\ta{\theta_{a}}
 \def\tx{\theta_{x}}
 \def\ma{m_1^{}}
 \def\mb{m_2^{}}
 \def\mc{m_3^{}}
 \def\mutau{\mu\!-\!\tau}
 \def\Zmutau{\mathbb{Z}_2^{\mu\tau}}
 \def\nuL{\nu_L^{}}

 \def\ahat{\widehat{\alpha}}

 \def\thisday{December, 2009}

\begin{document}
 \thispagestyle{empty}
 \setcounter{footnote}{0}
 \titlelabel{\thetitle.\quad \hspace{-0.8em}}
\titlecontents{section}
              [1.5em]
              {\vspace{4mm} \large \bf}
              {\contentslabel{1em}}
              {\hspace*{-1em}}
              {\titlerule*[.5pc]{.}\contentspage}
\titlecontents{subsection}
              [3.5em]
              {\vspace{2mm}}
              {\contentslabel{1.8em}}
              {\hspace*{.3em}}
              {\titlerule*[.5pc]{.}\contentspage}
\titlecontents{subsubsection}
              [5.5em]
              {\vspace{2mm}}
              {\contentslabel{2.5em}}
              {\hspace*{.3em}}
              {\titlerule*[.5pc]{.}\contentspage}

 \begin{flushright}
 \thisday  \hfill arXiv:\,1001.0940
 \end{flushright}
 \vspace*{10mm}

 \begin{center}
 {\bf {\Large
  Common Origin of Soft $\boldsymbol{\mu\!-\!\tau}$ and CP Breaking \\[2mm]
  in Neutrino Seesaw and the Origin of Matter}}

 \vspace*{8mm}

 {\sc Shao-Feng Ge}\,\footnote{gesf02@mails.tsinghua.edu.cn},~~~
 {\sc Hong-Jian He}\,\footnote{hjhe@tsinghua.edu.cn},~~~
 {\sc Fu-Rong Yin}\,\footnote{yfr@tsinghua.edu.cn}
%\end{flushright}

\vspace*{3mm}

Center for High Energy Physics and Institute of Modern Physics,
\\
Tsinghua University, Beijing 100084, China
\\
and
\\
Kavli Institute for Theoretical Physics China, \\
Chinese Academy of Sciences, Beijing 100190, China
\end{center}

 \vspace*{3mm}
 \begin{abstract}
 \baselineskip 17pt
 %\hspace*{-0.35cm}
 \noindent
 Neutrino oscillation data strongly support $\mutau$ symmetry as a good approximate flavor symmetry
 of the neutrino sector, which has to appear in any viable theory for neutrino mass-generation.
 The $\mutau$ breaking is not only small, but also the source of Dirac CP-violation.
 We conjecture that both discrete $\mutau$ and CP symmetries are fundamental symmetries of
 the seesaw Lagrangian (respected by interaction terms), and they are {\it only softly broken,
 arising from a common origin via a unique dimension-3 Majorana mass-term of the heavy right-handed
 neutrinos.}  From this conceptually attractive and simple construction, we can predict the soft
 $\mutau$ breaking at low energies, leading to quantitative {\it correlations} between the apparently
 {\it two small deviations} $\,\theta_{23}^{} \!-45^\circ\,$ and $\,\theta_{13}^{}\!-0^\circ\,$.\,
 This nontrivially connects the on-going measurements of mixing angle $\,\theta_{23}\,$ with the
 upcoming experimental probes of $\,\theta_{13}$\,.\,  We find that any deviation of
 $\,\theta_{23}^{} \!-45^\circ\,$ must put a {\it lower limit} on $\,\theta_{13}\,$.\,
 Furthermore, we deduce the low energy Dirac and Majorana CP violations from a common soft-breaking
 phase associated with $\mutau$ breaking in the neutrino seesaw.  Finally, from the soft CP breaking
 in neutrino seesaw we derive the cosmological CP violation for the baryon asymmetry via leptogenesis.
 We fully reconstruct the leptogenesis CP-asymmetry from the low energy Dirac CP phase and establish
 a direct link between the cosmological CP-violation and the low energy Jarlskog invariant.
 We predict new lower and upper bounds on the $\,\theta_{13}$\, mixing angle,
 $\,1^\deg \lesssim \theta_{13} \lesssim 6^\deg\,$.\,
 In addition, we reveal a new hidden symmetry that dictates the solar mixing angle
 $\,\theta_{12}\,$ by its group-parameter, and includes the conventional tri-bimaximal mixing
 as a special case, allowing deviations from it.
 \\[2mm]
 \\[2mm]
 Journal of Cosmology and Astroparticle Physics (2010), in Press.
 %{\bf Keywords:} neutrino properties, neutrino theory, baryon asymmetry\\
 %{\bf ArXiv ePrint:} 1001.0940
 \end{abstract}
 %\vspace*{1cm}

 \newpage
 \setcounter{page}{2}

 \tableofcontents

 \newpage
 \setcounter{footnote}{0}
 \renewcommand{\thefootnote}{\arabic{footnote}}
 \baselineskip 18.5pt

 \section{Introduction}
 \label{sec:introduction}

 Neutrino physics has come to a golden era for a decade. The discovery
 of oscillations of solar and atmospherical neutrinos, and the further
 confirmations by using terrestrial neutrino beams
 produced in the reactor and accelerator experiments, have
 established the massiveness of neutrinos as well as the large mixing angles
 in the leptonic mixing matrix\,\cite{review}\footnote{The leptonic mixing matrix
 is conventionally called Pontecorvo-Maki-Nakagawa-Sakata (PMNS) matrix\,\cite{PMNS}
 whose form will be discussed later in
 Sec.\,\ref{sec:model-independent-reconstruction-formalism}. \label{fn:pmns}}.
 The two mass-squared differences
 $\left( \Delta_s,\,\Delta_a \right)$ and two
 large mixing angles $(\theta_{12},\,\theta_{23})$
 are measured with good accuracy.
 A summary of the $3\nu$ global analysis on these
 parameters\,\cite{Fogli-08} is shown in Table-\ref{tab:1}, with the conventions of
 $\,\Delta_s = m_2^2-m_1^2\,$  and
 $\,\Delta_a =|m_3^2-(m_1^2+m_2^2)/2|\,$  for the fit\,\cite{Fogli-05,Lisi-09}.
\vspace*{2mm}
\begin{table*}[h]
\begin{center}
\label{tab:1} {\small
\begin{tabular}{c||ccccc}
\hline\hline
& & & \\
{\tt Parameters} & $\Delta_s\,(10^{-5}{\rm eV}^2)$ &
$\Delta_a\,(10^{-3}{\rm eV}^2)$ & $\sin^2\theta_{12}\,(\theta_{12})$
& $\sin^2\theta_{23}\,(\theta_{23})$ &
$\sin^2\theta_{13}\,(\theta_{13})$
\\
& & & & & \\[-2mm]
\hline
& & & & & \\[-2mm]
{\tt Best Fit} & $7.67$ & $2.39$ & $.312$~($34.0^\deg$) &
$.466$~($43.0^\deg$) & $.016$~($7.3^\deg$)
\\[1mm]
\hline
& & & & & \\[-2mm]
$\boldsymbol{1\sigma}$ {\tt Limits} & $7.48-7.83$ & $2.31-2.50$ &
$.294-.331$ & $.408-.539$ & $.006-.026$
\\[0.6mm]
& & & ($32.8^\deg -35.1^\deg$) & ($39.7^\deg -47.2^\deg$) &  ($4.4^\deg -9.3^\deg$)
\\[1mm]
\hline
& & & & & \\[-2mm]
$\boldsymbol{2\sigma}$ {\tt Limits} & $7.31-8.01$ & $2.19-2.66$ &
$.278-.352$ & $.366-.602$ & $<.036$
\\[0.6mm]
& & & ($31.8^\deg -36.4^\deg$) & ($37.2^\deg -50.9^\deg$) &  ($<10.9^\deg$)
\\[1mm]
\hline
& & & & & \\[-2mm]
$\boldsymbol{3\sigma}$ {\tt Limits} & $7.14-8.19$ & $2.06-2.81$ &
$.263-.375$ & $.331-.644$ & $<.046$
\\[0.6mm]
& & & ($30.9^\deg -37.8^\deg$) & ($35.1^\deg -53.4^\deg$) &  ($<12.4^\deg$)
\\[1mm]
\hline
& & & & & \\[-2mm]
90\%\,C.L. Limits  & $7.36-7.94$ & $2.24-2.60$ & $.283-.344$&
$.380-.582$ &  $<.032$
\\[0.6mm]
& & & ($32.2^\deg -35.9^\deg$) & ($38.1^\deg -49.7^\deg$) & ($<10.4^\deg$)
\\[1mm]
\hline
& & & & & \\[-2mm]
99\%\,C.L. Limits  & $7.23-8.13$ & $2.10-2.75$ & $.270-.365$ &
$.344-.627$  & $<.042$
\\[0.6mm]
& & & ($31.3^\deg -37.2^\deg$) & ($35.9^\deg -52.3^\deg$) &  ($<11.8^\deg$)
\\[1mm]
\hline\hline
\end{tabular}
} \caption{The global $3\nu$ fit\,\cite{Fogli-08}
 for the neutrino mass-squared differences and mixing
 angles including the available data
 from solar, atmospheric, reactor (KamLAND
 and Chooz) and accelerator (K2K and MINOS) experiments\,\cite{nu2008}.}
\label{tab:data}
\end{center}
\end{table*}
\vspace*{-5mm}

The Table-\ref{tab:1} shows that the neutrino mixings pose a striking pattern
with two large mixing angles
$(\theta_{12},\,\theta_{23})$ and a small mixing angle $\,\theta_{13}$\,,\,
very different from that of quark sector.
It is also intriguing to note that the central value of $\,\theta_{23}\,$ is already
somewhat below the maximal mixing angle\,\footnote{It was noted back in
2004\,\cite{D23-Smirnov} that the data have slightly favored a smaller $\theta_{23}$ than
its maximal value. Although this intriguing small deviation $\theta_{23}-45^\deg < 0$
is not yet statistically significant, the tendency of having a negative $\theta_{23}-45^\deg$
is found to be robust, due to the excess of $e$-like atmospheric neutrino events in
the sub-GeV sample\,\cite{D23-Smirnov}.\label{footnote-2}} ($45^\deg$)
and $\theta_{13}$ is favored to deviate from
zero at $95\%$\,C.L.\footnote{A very recent oscillation analysis on $\theta_{13}$ gives,
$\sin^2\theta_{13}=0.02\pm 0.01$ %[$\,5.7^\deg <\theta_{13}(8.1^\deg)<10^\deg\,$]
at $1\sigma$ level, with the central value slightly shifted upward\,\cite{Fogli-09}.}\,
To be specific, the data allow
two small deviations of the same order, at the $2\sigma$ level,
\beqa
\label{eq:da-dx-exp}
-7.8^\deg < (\theta_{23}-45^\deg) < 5.9^\deg\,,~~~~~~~~~~
0^\deg \leqq (\theta_{13}-0^\deg)<10.9^\deg \,,
\eeqa
with the best fitted values,
$\,(\theta_{23}-45^\deg)=-2.0^\deg$\, and \,$(\theta_{13}-0^\deg)=7.3^\deg$.\,
This naturally provides a fairly good {\it zeroth order approximation,}
\,$\theta_{23}=45^\deg$\, and \,$\theta_{13}=0^\deg$,\,
under which two exact discrete symmetries emerge,
namely, the $\mu\!-\!\tau$ symmetry\,\cite{mutauRev}
and the Dirac CP conservation\footnote{Here the possible Majorana CP-phases may still appear,
but in our theory construction (Sec.\,\ref{sec:common-origin}), they are not independent and will vanish as the
Dirac CP violation goes to zero in the $\mutau$ symmetric limit.} in the neutrino sector.
We stress that the $\mutau$ symmetry (as well as associated Dirac CP-invariance),
as a good {\it zeroth order approximation} reflected
by all neutrino data, has to appear in any viable theory for neutrino mass-generation.

 Since the $\mutau$ symmetry, as a good zeroth order flavor symmetry of
 the neutrino sector, leads to $\,\theta_{13}=0\,$ and thus the Dirac CP-conservation,
 the $\mutau$ symmetry breaking terms must be small and also serve as the source of
 the Dirac CP-violation.
 On the theory ground, it is natural and tempting to expect a common origin for all
 CP-violations, although the Dirac and Majorana CP-violations appear
 differently in the low energy effective theory of light neutrino mass-matrix.
 Given such a common origin for two kinds of
 CP-violations, then they must vanish together in the $\mutau$ symmetric limit.

 With these key observations, we conjecture that
 both discrete $\mutau$ and CP symmetries are fundamental
 symmetries of the seesaw Lagrangian {\it (respected by interaction terms),}
 and they are {\it only softly broken, arising from a common origin via
 a unique dimension-3 Majorana mass-term of the heavy right-handed singlet neutrinos.}
 The reason for the $\mutau$ and CP breaking terms being {\it soft} is because
 we consider both symmetries to be respected by interaction terms of the seesaw
 Lagrangian\,\cite{Mseesaw}, thus the unique place for such breakings is the
 dimension-3 singlet Majorana mass term of right-handed neutrinos.

 From the above conceptually attractive and simple construction with the soft breakings of two discrete
 symmetries from a common origin, we can predict the soft $\mutau$ breaking at low energies,
 leading to quantitative {\it correlations}
 between the apparently {\it two small deviations}
 \footnote{It is interesting to note that
 a nonzero $\tbc \!-45^\circ$ is further motivated by the quark-lepton complementarity\,\cite{QLC}
 via the relation $\theta_{23}^{q}+\theta_{23}^\ell = 45^\deg$, where experiments have measured
 the quark-sector CKM (Cabibbo-Kobayashi-Maskawa\,\cite{CKM}) mixing angle
 $\theta_{23}^{q}=(2.36\pm 0.06)^\deg$ \cite{PDG}, causing a deviation in the leptonic mixing
 angle \,$\theta_{23}^\ell = 45^\deg - \theta_{23}^{q} < 45^\deg$,\,
 consistent with the data in Table-\ref{tab:1}.},
 $\delta_a\equiv \tbc \!-45^\circ$ and $\delta_x\equiv\tac\!-0^\circ$.\,
 The mixing angle $\theta_{13}$ will be more precisely probed at the upcoming
 reactor experiments Double Chooz\,\cite{2CHOOZ}, Daya Bay\,\cite{DayaBay}, RENO\,\cite{RENO}
 as well as the accelerator experiments T2K\,\cite{T2K} and NO$\nu$A\,\cite{NOvA}, etc\,\footnote{For
 instance, the proposed LENA experiment\,\cite{LENA} will also probe $\theta_{13}$ and CP-violations
 with good precision via long baseline neutrino oscillations from CERN to the LENA detector
 at Pyhasalmi mine (2288\,km apart).},
 while improved measurements of the mixing angle $\theta_{23}$ are expected from
 the Minos\,\cite{MINOS} and T2K\,\cite{T2K} experiments etc.
 The future neutrino factory and super-beam facility\,\cite{nuFact} will further
 pin down these key parameters with high precision.
 Finally, we further derive the cosmological baryon asymmetry $\eta_B^{}$ via
 leptogenesis and analyze the interplay between the leptogenesis scale $M_1$
 and the low energy Jarlskog invariant $J$
 as well as neutrinoless double-beta decay observable $M_{ee}$.

Table-\ref{tab:1} also shows that
the solar angle $\theta_{12}$ is best measured among
the three mixing angles\footnote{Table-\ref{tab:1} shows that
the so-called tri-bimaximal (TBM) mixing ansatz,
$\,\sin^2\theta_{12}=\f{1}{3}~(\theta_{12}=35.3^\deg)$,\, is
above the $1\sigma$ upper bound on $\theta_{12}$.\,
A latest three-flavor oscillation analysis of
SNO Collaboration\,\cite{SNO} combined data from all solar experiments and
KamLAND reactor antineutrino experiment\,\cite{KamLAND}, this gave an even
tighter constraint on the mixing angle,
$\,\theta_{12}=34.38^{+1.16}_{-0.97}\,$(degrees),
at $1\sigma$ level, which also leaves the TBM mixing value around the $1\sigma$
upper boundary.},\,
so it is not our major concern.
The main goal here is to examine the least known angle $\theta_{13}$
(to be measured at Double-Chooz\,\cite{2CHOOZ},
Daya Bay\,\cite{DayaBay}, RENO\,\cite{RENO}, T2K\,\cite{T2K} and NO$\nu$A\,\cite{NOvA}
experiments) which is crucial for nonzero Dirac CP-violation,
as well as its quantitative correlation with the deviation of
$\theta_{23}-45^\deg$ from the naive maximal mixing (which will be further probed
by MINOS\,\cite{MINOS} and T2K\,\cite{T2K} experiments, etc).
We can predict nontrivial lower and upper limits on the mixing angle $\,\theta_{13}$\,.

This paper is organized as follows.
In Sec.\,\ref{sec:origin} we present a unique construction
of the soft $\mutau$ and CP breakings from a common origin
in the neutrino seesaw.
Then, in Sec.\,\ref{sec:general} we present a
low energy reconstruction of the light neutrino mass matrix with
$\mutau$ and CP violations including the small
parameters \,$\delta_a$\, and \,$\delta_x$\,.\,
With these, we will further derive, in Sec.\,\ref{sec:diagonalization},
the low energy $\mutau$ and CP violation observables
from the common soft breaking in the neutrino seesaw.
As the \,$\delta_a$\, and \,$\delta_x$\, arise from the common origin,
they are proportional to a (small) common soft breaking parameter
at the next-to-leading order of the well-defined expansion.
So we can make quantitative predictions for their correlations,
and discuss their implications for the upcoming experimental probes.
Then, in Sec.\,\ref{sec:solution} we analyze the cosmological CP violation via
leptogenesis\,\cite{lepG,lepGrev} in our model, which provides the origin of matter
--- the baryon asymmetry of the universe.  After inputting all the known neutrino data
and the observed baryon asymmetry\,\cite{WMAP08,PDG},
we can derive the direct link between the cosmological CP-violation
and the low energy Jarlskog invariant\,\cite{J}. We further place a lower bound on
the leptogenesis scale for producing the observed baryon asymmetry, and deduce new
lower and upper limits on the mixing angle $\,\theta_{13}$\,.
In Sec.\,\ref{sec:solar}, we generally prove why the solar mixing angle $\,\theta_{12}^{}\,$
does not receive any correction from the soft $\mutau$ and CP breakings. We show that
$\,\theta_{12}^{}\,$ is actually protected by a new hidden symmetry in the seesaw Lagrangian,
and its value is dictated by the group parameter of this hidden symmetry.
We finally conclude in Sec.\,\ref{sec:conclusions}.

\vspace*{3mm}
\section{Soft $\boldsymbol{\mutau}$ and CP Breaking Originated from Neutrino Seesaw}
 \label{sec:origin}

 We conjecture that both the discrete $\mutau$ and CP symmetries
 are fundamental symmetries of the neutrino seesaw Lagrangian
 {\it (respected by interaction terms),} and are only
 {\it softly broken from a common origin}. In fact, the {\it only place} in
 the seesaw Lagrangian which can provide such a common origin is the
 {\it dimension-3} singlet Majorana mass-term of right-handed neutrinos.
 The Dirac mass-term cannot provide such a soft breaking as it is generated by the
 {\it dimension-4} Yukawa interactions after spontaneous electroweak symmetry breaking.
 In Sec.\,\ref{sec:zeroth-seesaw}, we first consider the minimal neutrino
 seesaw Lagrangian with exact $\mutau$ and CP invariance, and derive
 the seesaw mass-matrix for the light neutrinos (from which we
 deduce the zeroth-order mass-eigenvalues and mixing angles).
 This $\mutau$ and CP symmetric limit predicts the mixing angles
 \,$(\theta_{23},\,\theta_{13})=(45^\deg,\,0^\deg)$.\,
 Then, in Sec.\,\ref{sec:common-origin}, we will construct
 a unique soft breaking term providing a common origin for both
 $\mutau$ and CP breakings. From this we will further derive predictions for
 the $\mutau$ and CP breakings in the low energy light neutrino mass-matrix,
 by treating the small soft-breaking as perturbation up to the first nontrivial order
 (Sec.\,\ref{sec:diagonalization}). The predictions for the seesaw-scale leptogenesis
 and its correlations with low energy observables will be analyzed in Sec.\,\ref{sec:solution}.

 \vspace*{3mm}
 \subsection{%\hspace*{-5mm}.\hspace*{-0.6mm}
 Neutrino Seesaw with $\boldsymbol{\mutau}$ and CP Symmetries}
 \label{sec:zeroth-seesaw}

 Seesaw mechanism\,\cite{Mseesaw} provides a natural explanation of the small Majorana
 masses for light neutrinos which violate lepton number by two units\footnote{It was
 shown\,\cite{Dseesaw}
 that the seesaw mechanism can also be realized for generating small Dirac masses for
 light neutrinos with lepton-number conservation. This direction will not be explored
 in the current study.}.
 For simplicity of demonstration,
 we consider the Lagrangian for the minimal neutrino seesaw\,\cite{MSS,He2003},
 with two right-handed singlet Majorana neutrinos
 besides the standard model (SM) particle content,
 \beqa
 {\cal L}_{\rm ss}
 & = &
     - \;\ov{L}\;Y_{\ell}\;\Phi\ell_R^{}
  \; - \;\ov{L}\;Y_\nu\Phit\;\N + \f{1}{2}\N^TM_R\widehat{C}\N + {\rm h.c.}
 \nn\\
 & = &
 - \;\ov{\ell_L}\;M_\ell\;\ell_R
              \;-\;\ov{\nuL}\;\mD\;\N + \f{1}{2}\N^TM_R\widehat{C}\N
 + {\rm h.c.} + (\textrm{interactions})
 \,,
 \label{eq:L-seesaw}
 \eeqa
 where $\,L\,$ denotes three left-handed neutrino-lepton weak doublets,
 $\,\ell=(e,\,\mu,\,\tau)^T\,$ contains charged leptons,
 $\,\nuL =(\nu_e^{},\,\nu_\mu^{},\,\nu_\tau^{})^T\,$ is for
 the light flavor neutrinos, and $\N =(N_\mu,\,N_\tau)^T$ represents two heavy
 right-handed singlet neutrinos.
 We have also denoted the SM Higgs doublet by $\Phi$ ($\Phit$) with hypercharge
 $\f{1}{2}$ ($-\f{1}{2}$), and the $3\!\times\!3$ lepton and neutrino Yukawa-coupling
 matrices by $Y_{\ell}$ and $Y_{\nu}$, respectively.
 The lepton Dirac-mass-matrix $\,M_\ell = v\,Y_\ell /\sqrt{2}\,$
 and the neutrino Dirac-mass-matrix $\,\mD = v\,Y_\nu /\sqrt{2}\,$ arise from the dimension-4 Yukawa
 interactions after the spontaneous electroweak symmetry breaking,
 $\,\left<\Phi\right> = (0,\,\f{v}{\sqrt{2}}\,)^T\neq 0\,$,
 and the dimension-3 Majorana mass-term for $\,M_R\,$ is a gauge-singlet.

 This minimal seesaw Lagrangian in Eq.\,(\ref{eq:L-seesaw}) can be regarded as
 an effective theory of the general three neutrino seesaw where the right-handed
 singlet $N_e$ is much heavier than the other two $(N_\mu,\,N_\tau)$ and
 thus can be integrated out at the mass-scales of $(N_\mu,\,N_\tau)$,
 giving rise to Eq.\,(\ref{eq:L-seesaw}).
 Consequently, the minimal seesaw generically predicts
 a massless light neutrino\,\cite{MSS},
 which is always a good approximation as long as
 the lightest left-handed neutrino has its mass much smaller than
 the other two (even if not exactly massless). As we will prove in Sec.\,6.3,
 the general three-neutrino-seesaw under the $\mutau$ symmetry ($\mathbb{Z}_2^{\mu\tau}$)
 and hidden symmetry ($\mathbb{Z}_2^s$) also predicts a massless light neutrino,
 sharing the same feature as the minimal seesaw we consider here.

 After integrating out the heavy neutrinos $(N_\mu,\,N_\tau)$
 from Eq.\,(\ref{eq:L-seesaw}), we can derive the seesaw formula for the
 $3\times 3$ symmetric Majorana mass-matrix of the light neutrinos,
 \begin{eqnarray}
  M_\nu ~\simeq~
  \mD M_R^{-1} \mDT \,.
  \label{eq:seesaw-formula}
 \end{eqnarray}

 In general it is expected that the lepton sector
 would obey a flavor symmetry $\mathbb{G}_\ell$ different from the
 $\mutau$ symmetry $\mathbb{Z}_2^{\mu\tau}$
 in the neutrino sector\footnote{The $\mutau$ symmetry $\mathbb{Z}_2^{\mu\tau}$
 will be defined shortly, in Eqs.\,(\ref{eq:mdtrans})-(\ref{eq:T3T2}) and
 the text just above them.},\,
 due to the large mass-splitting between $(\mu,\,\tau)$ leptons.
 The two symmetries $\mathbb{Z}_2^{\mu\tau}$ and $\mathbb{G}_\ell$ could
 result from spontaneous breaking of a larger flavor symmetry
 $\mathbb{G}_F$\,\cite{Gf}. The invariance of lepton mass matrix under
 the transformation of left-handed leptons $G_{\ell L}^{}\,\(\in\mathbb{G}_\ell^{}\)$
 is $\,G_{\ell L}^\dag M_\ell^{} M_\ell^\dag G_{\ell L}^{} = M_\ell^{} M_\ell^\dag\,$.
 The mass matrix $\,M_\ell M_\ell^\dag\,$
 can be ensured to be diagonal via proper model-buildings,
 {\it e.g.,} via making $G_{\ell L}^{}$ diagonal and nondegenerate.
 For an explicit realization, we can make the simplest choice of
 $\,\mathbb{G}_{\ell}=\ZZ_3\,$,\, with\footnote{Actually any discrete group $\mathbb{Z}_n$ with
 $\,n\geqq 3\,$ can provide a diagonal nondegenerate 3-dimensional representation and thus does
 the same job\,\cite{Gf}.}
 \beqa
 \label{eq:GL-lep}
 G_{\ell L} ~=~ \textrm{diag}(1,\,\eta,\,\eta^2)\,,
 \eeqa
 where $\,\eta =e^{i2\pi/3}\,$ and $\,G_{\ell L}^3=I\,$ with $I$ being unit matrix.
 Thus, the above invariance equation
 $\,G_{\ell L}^\dag M_\ell^{} M_\ell^\dag G_{\ell L}^{} = M_\ell^{} M_\ell^\dag\,$
 makes $\,M_\ell M_\ell^\dag$\, fully diagonal,
 and corresponds to the left-handed leptons in their mass-eigenbasis.
 This means that the PMNS matrix \,$V = U_{\ell L}^\dag U_{\nu L}^{} = U_{\nu L}^{}$\,.\,
 The right-handed leptons can be further rotated into their mass-eigenbasis,
 without affecting the PMNS matrix, except having a diagonal lepton-mass-matrix $M_{\ell}$
 in Eq.\,(\ref{eq:L-seesaw}).
 (Note that the relation $\,V = U_{\nu L}^{}$\, holds as long as
 $\,M_\ell M_\ell^\dag\,$ is ensured to be diagonal via the above symmetry transformation
 $\,G_{\ell L}\in \ZZ_3$\,,\, but this does not require the lepton mass-matrix $\,M_\ell\,$
 itself being generally diagonal because the right-handed leptons can still be
 in the non-mass-eigenbasis.)

 The Lagrangian (\ref{eq:L-seesaw}) is defined to
 respect both the $\mutau$ and CP symmetries.
 So the Dirac and Majorana mass matrices $\mD$ and $M_R^{}$
 are real, as well as symmetric under the transformations,
 \,$\nu_\mu \leftrightarrow p\nu_\tau$\, and
 \,$N_\mu \leftrightarrow p'N_\tau$\,,\, where
 \,$p,p'=\pm$\, denote the even/odd parity assignments of the light and heavy neutrinos
 under the $\mutau$ symmetry $\,\mathbb{Z}_2^{\mu\tau}$\,.\,
 This means that the mass matrices $\mD$ and $M_R^{}$ obey the following
 invariance equations,
 \begin{eqnarray}
 \label{eq:mdtrans}
 \widetilde{T}_L^{\dag} \tm_D^{} \widetilde{T}_R^{}
 ~=~ \tm_D^{}\,, &&~~~
 \widetilde{T}_R^T  \widetilde{M}_R^{} \widetilde{T}_R^{}
 ~=~ \widetilde{M}_R^{} \,,
 \end{eqnarray}
with
 \begin{eqnarray}
 \label{eq:T3T2}
\widetilde{T}_L^{} ~=~ \!\!\(
  \ba{ccc}
   1 & 0 & 0 \\
   0 & 0 & p \\
   0 & p & 0
  \ea \) \!,
 &&~~~
  \widetilde{T}_R^{} ~=~ \!\!
  \begin{pmatrix}
     0  & p' \\
     p' & 0
  \end{pmatrix} \!,
 \end{eqnarray}
 where we have added a ``tilde'' on $\mD$ and $M_R$
 to indicate their most general forms including possible $\mutau$ parities,
 and later we will remove the ``hat'' after some useful simplifications.
 Thus we can deduce $\widetilde{m}_D^{}$ and $\widetilde{M}_R^{}$ to have the following structure,
 \begin{eqnarray}
 \label{eq:mDMR}
  \tm_D ~= \(
  \ba{rr}
    a &~~ p'a \\[2mm]
    b &~~ c \\[2mm]
    pp'c & pp'b
  \ea \) \!,
 \qquad
  \widetilde{M}_R^{} ~=\,
  \pR\!
  \begin{pmatrix}
    M_{22}      &  p'_R M_{23} \\[2mm]
    p'_R M_{23} &  M_{22}
  \end{pmatrix} \!,~
 \qquad (\,p,p',\pR , p'_R = \pm\,)\,,
 \end{eqnarray}
 which are all real due to the CP conservation.
 Here, the hatted quantities denote the mass matrices before rephasing, while
 $M_{22}$ and $M_{23}$ denote the absolute values of the $\{22\}$ and $\{33\}$
 elements of $\widetilde{M}_R^{}$ with their signs defined as $\pR$ and $\pR p'_R$, respectively.
 Note that the Dirac mass-matrix
 $\widetilde{m}_D^{}$ contains only three independent real mass-parameters $(a,b,c)$,
 and the symmetric Majorana mass-matrix $\widetilde{M}_R$ has just two, $(M_{22}^{},\,M_{23}^{})$.
 Substituting the above $\widetilde{m}_D^{}$ and $\widetilde{M}_R^{}$
 into the seesaw equation (\ref{eq:seesaw-formula}),
 we deduce the mass matrix for light neutrinos,
 \begin{equation}
   \widetilde{M}_\nu
 ~=~
   \pR \!
  \begin{pmatrix}
    \frac{2a^2}{M_{22}+p' p'_R M_{23}}
  & \frac{a(b+p'c)}{M_{22}+p' p'_R M_{23}}
  & \frac{pa(b+p'c)}{M_{22}+p' p'_R M_{23}}
  \\[3mm]
    %\frac{a(b+p'c)}{M_{22}+p'M_{23}}
  & \f{1}{2}\left[\frac{(b+p'c)^2}{M_{22}+p' p'_R M_{23}}
                +\frac{(b-p'c)^2}{M_{22}-p' p'_R M_{23}} \right]
  & \f{\,p}{2}
    \[ \f{(b + p' c)^2}{M_{22} + p' p'_R M_{23}}
      -\f{(b - p' c)^2}{M_{22} - p' p'_R M_{23}} \]
  \\[3mm]
  & \frac{1}{2}
    \left[ \frac{(b+p'c)^2}{M_{22}+p' p'_R M_{23}}
          +\frac{(b-p'c)^2}{M_{22}-p' p'_R M_{23}} \right]
  \end{pmatrix} \!,
  \label{eq:Mnu-0b}
  \end{equation}
 which we call the {\it zeroth order mass-matrix} since we
 will further include the small soft-breaking effect in the next subsection.

 For the mass formula (\ref{eq:Mnu-0b}), some comments are in order.
 First,  it is clear that the sign factor $p'=\pm$ (due to the parity of the
 $\mutau$ transformations under $\Zmutau$)
 accompanies $c$ and $M_{23}$ everywhere in Eq.\,(\ref{eq:Mnu-0b}).
 Besides, $p'_R$ is always associated with $M_{23}$, as expected. It is convenient
 to simply rescale $c$ as $p'_R c$; then $p'$ and $p'_R$ always accompany
 each other in (\ref{eq:Mnu-0b}), and appear together in front of both $c$ and $M_{23}$.
 Second,  we note that the $\{12\}$ and $\{13\}$ elements of $M_\nu$
 only differ by an overall sign $p=\pm$ (due to the other parity factor under $\Zmutau$);
 the sign $p$ also determines the sign of
 the mixing angle $\theta_{23}$.
 As our sign-convention, we always define
 \,$\theta_{23}\in [0,\f{\pi}{2}] \geqq 0$,\, so we can make a simple rephasing
 for the light neutrino $\nu_\tau^{}$, i.e.,
 $\,(\nu_e^{},\,\nu_\mu^{},\,\nu_\tau^{})^T\to
  \mathcal{P}_L (\nu_e^{},\,\nu_\mu^{},\,\nu_\tau^{})^T\,$, where $\mathcal{P}_L$ is a $3\times3$
  diagonal matrix, $\mathcal{P}_L = {\rm diag}(1,1,p)\,$.
  This will not affect the leptonic PMNS mixing matrix since it only
  contributes to the rephasing matrix $U''$ as defined in
  Sec.\,\ref{sec:model-independent-reconstruction-formalism} (which is not related to the PMNS matrix).
  Under this rephasing,  the mass matrix (\ref{eq:Mnu-0b}) becomes
  $\mathcal{P}_L^T \widetilde{M}_\nu \mathcal{P}_L$.\,\footnote{From Eq.\,(\ref{eq:Mnu-0b})
  or Eq.\,(\ref{eq:Mnu-0}) below,
 we have \,$\det(\widetilde{M}_\nu)=\det(M_\nu)=0$,\, which
 actually holds in general minimal seesaw
 (even after including $\mutau$ and CP violations).}\,
 Third,
 we note that the overall sign $\pR$ in (\ref{eq:Mnu-0b}) originates from the overall sign
 in (\ref{eq:mDMR}) and can be rotated away by uniformly rephasing the Majorana neutrino fields
 $\,\mathcal{N} \rightarrow \sqrt{\pR}\mathcal{N}$\,;\, we can further remove the relative sign
 $p'_R$ in $\widetilde{M}_R$ [(\ref{eq:mDMR})] by a rephasing of $\mathcal{N}_\mu$ or
 $\mathcal{N}_\tau$.\, In summary, to simplify the sign-conventions
 we just need to make the following combined transformations
 for light and heavy neutrinos, in one step,
 \beqa
 \nuL \to \mathcal{P}_L\nuL \,,
 &&
 \mathcal{N} \to \mathcal{P}_R^{}\,\mathcal{N}\,,
 \eeqa
 with
 $~\mathcal{P}_L^{} = \sqrt{\pR}\,\textrm{diag}(1,\,1,\,p)\,,
  ~~
   \mathcal{P}_R^{} = \sqrt{\pR}\,\textrm{diag}(1,\,p_R') \,,
 $~
 where $\,\nu = (\nu_e^{},\,\nu_\mu^{},\,\nu_\tau^{})^T\,$.~
 Then the two mass matrices in (\ref{eq:mDMR}) transform accordingly,
\beqa
  \widetilde{m}_D^{} ~~\to~~
  m_D^{} ~\equiv~ \mathcal{P}_L^\dag \widetilde{m}_D^{} \mathcal{P}_R^{}\,,
&~~~&
  \widetilde{M}_R ~~\to~~
  M_R ~\equiv~  \mathcal{P}_R^T \widetilde{M}_R \mathcal{P}_R^{}\,,
\eeqa
with
\beqa
\label{eq:mD-MR-new}
  \mD ~=
  \begin{pmatrix}
    a & \pb a \\
    b & c \\
    \pb c & \pb b
  \end{pmatrix} \!,
&&~~~
  M_R ~= \begin{pmatrix}
    M_{22} & M_{23} \\[2mm]
    M_{23} & M_{22}
  \end{pmatrix}
  \equiv\,
  M_{22}\begin{pmatrix}
         1 & R \\[2mm]
         R & 1
        \end{pmatrix}\!,
\eeqa
where
\beqa
  \pb ~\equiv~ p' p'_R = \pm \,,
&~~~&
  R ~\equiv~ \f{M_{23}}{M_{22}} \,\equiv\, 1 - r \,>\, 0 \,.
  \label{eq:mDMR-new}
\eeqa
In the above we have simply rescaled $c$ as $\,p'_R c$\, (which does not matter since
we can fully eliminate the parameter $c$ in our final relations between physical observables).\,
Since both $M_{22}$ and $M_{23}$ denote absolute values, we always have \,$R\equiv 1-r >0$\,,\,
i.e., $\,r < 1\,$.\,

Furthermore, the $\mutau$ symmetry transformation matrices
$\widetilde{T}_L$ and $\widetilde{T}_R$ will change into
$T_L$ and $T_R$ accordingly,
\beqa
 \widetilde{T}_L ~=~ \mathcal{P}_L^{} T_L \mathcal{P}_L^{-1} \,,
 &~~~&
 \widetilde{T}_R ~=~ \mathcal{P}_R^{} T_R \mathcal{P}_R^{-1} \,,
\eeqa
with
\begin{eqnarray}
 \label{eq:T3new}
 T_L ~=~ \(
  \ba{ccc}
   1 & 0 & 0 \\
   0 & 0 & 1 \\
   0 & 1 & 0   \ea \)\!,
   &~~~&
  T_R ~=~
  \begin{pmatrix}
    0      & \pb \\[1.8mm]
    \pb    & 0
  \end{pmatrix}\!, ~~~~~~~~~~ (\pb =p'p'_R)\,.
 \end{eqnarray}
Thus, the $\mutau$ invariance equations for the Dirac and Majorana mass-matrices
in (\ref{eq:mD-MR-new}) become,
\beqa
\label{eq:TL-TR-mD-MR}
T_L^\dag \mD T_R^{} ~=~ \mD\,, &~~~&
T_R^T M_R T_R^{} ~=~ M_R \,,
\eeqa
which can be readily verified.
Finally, we rewrite the symmetric seesaw mass-matrix $\widetilde{M}_\nu$ in (\ref{eq:Mnu-0b}) as,
\begin{eqnarray}
  M_\nu
&=&
  \frac 1 {M_{22}}
  \begin{pmatrix}
    \frac{2a^2}{1 + \pb R}
  & \frac{a(b+ \pb c)}{1 + \pb R}
  & \frac{a(b+ \pb c)}{1 + \pb R}
  \\[3mm]
  & \f{1}{2}\left[\frac{(b+ \pb c)^2}{1 + \pb R}
                + \frac{(b- \pb c)^2}{1 - \pb R} \right]
  & \frac 1 2
    \left[
      \frac{(b + \pb c)^2}{1 + \pb R}
    - \frac{(b - \pb c)^2}{1 - \pb R}
    \right]
  \\[3mm]
  & &
  \frac{1}{2}
  \left[
    \frac{(b+ \pb c)^2}{1 + \pb R}
  + \frac{(b- \pb c)^2}{1 - \pb R}
  \right]
  \end{pmatrix}
  \!,
  \label{eq:Mnu-0}
\end{eqnarray}
 which is invariant under $\mutau$ symmetry,
 \beqa
 \label{eq:TL-Mnu}
 T_L^T M_\nu T_L^{} ~=~ M_\nu \,.
 \eeqa
 As mentioned earlier, we can readily verify $\,\det(M_\nu) = 0$,\, which actually holds for any
 two-neutrino seesaw.

 Diagonalizing the $\mutau$ and CP symmetric mass-matrix (\ref{eq:Mnu-0}),
 we derive the mass eigenvalues and mixing angles
\begin{subequations}
 \label{eq:LO-MassAngle}
 \begin{eqnarray}
 \label{eq:LO-Mass}
 & &
   m_{1}^{}
 \,=\,
   0,~~~~~
   m_{2}^{}
 \,=\,
   \frac{2a^2+(b + \pb c)^2}
    {M_{22} |1  + \pb R|},~~~~~
   m_{3}^{}
 \,=\,
   \frac{(b - \pb c)^2}{M_{22} |1 - \pb R|}\,,
 \\[3mm]
 \label{eq:LO-Angle}
 & &
   \tan \theta_{12} \,=\, \f{\,\sqrt{2}|a|\,}{|b + \pb c|}\,, ~~~~~
   \theta_{23} \,=\, 45^\deg\,, ~~~~~
   \theta_{13} \,=\, 0^\deg \,,
 \end{eqnarray}
\end{subequations}
where we have made all mass-eigenvalues positive and the three mixing angles within the range
$\[0,\,\f{\pi}{2}\]$ by properly defining the rotation matrix
(cf. Sec.\,\ref{sec:general} and Sec.\,\ref{subsec:solar}).
Here the mixing angles $\,(\theta_{23},\,\theta_{13})=(45^\deg,\,0^\deg)\,$ are direct consequence
of the $\mutau$ symmetry, while the value of $\theta_{12}$ does not depend on it.
The determination of $\theta_{12}$ from additional flavor symmetry will be further analyzed
in Sec.\,\ref{sec:solar}.
The Eq.\,(\ref{eq:LO-Mass}) shows that the mass-spectrum of light neutrinos falls into the
``normal hierarchy'' (NH) pattern ($\,m_1<m_2<m_3\,$).\,
From (\ref{eq:mD-MR-new}), we can diagonalize the $\mutau$ and CP symmetric mass-matrix $M_R$
for right-handed neutrinos, with two mass-eigenvalues given by
\beqa
\label{eq:LO-MR-M1M2}
M_{1} ~=~ |r|M_{22}\,,
&&
M_{2} ~=~ (2-r)M_{22} \,,
\eeqa
where $\,2-r>0\,$ is ensured because of $\,R\equiv 1-r > 0\,$ and thus $\,r<1\,$.\,

Finally, we comment on the sign choice for $\,\pb =\pm\,$.\,
From Table-\ref{tab:1}, we deduce, at $2\sigma$ level,
 \beqa
 \label{eq:ratio-m2m3}
 \f{m_2}{m_3} ~=~ \sqrt{\frac {\d m^2_{21}}{\d m^2_{31}}} ~\simeq~
 \sqrt{\f{\Delta_s}{\Delta_a}} ~=~ 0.17-0.19 ~\ll~ 1 \,.
 \eeqa
So we can derive the mass ratio,
  \beqa
  \f{m_2}{m_3}
  ~=~
  \f{2 a^2 + (b + \pb c)^2}{(b - \pb c)^2}
  \f{|1 - \pb R|}{|1 + \pb R|} ~\ll~ 1 \,,
  \label{eq:mass2-diff-ratio}
  \eeqa
 where we note $\,m_1^{}=0\,$ as given in (\ref{eq:LO-Mass})
(which holds even after including $\mutau$ and CP violations).
 Using the definition $\,R\equiv 1-r>0\,$ and the mass-eigenvalues in
 (\ref{eq:LO-MR-M1M2}), we can rewrite the ratio,
 \beqa
 \label{eq:q}
 \f{|1-\pb R|}{|1+\pb R|}
   ~= \left\{
   \ba{ll}
   =~ \dis\f{|r|}{2-r} ~=~ \f{M_{1}}{M_{2}}\equiv q  \,,
   &~~~ (\textrm{for}~\pb = +)\,,
   \\[5mm]
   =~ \dis\f{2-r}{|r|} ~=~ \f{M_{2}}{M_{1}} = q^{-1} \,,
   &~~~ (\textrm{for}~\pb = -)\,.
   \ea
   \right.
 \eeqa
 The Dirac mass-matrix $\mD$ in (\ref{eq:mD-MR-new})
 arises from the products of Yukawa couplings and Higgs vacuum expectation value.
 So for a natural seesaw without fine-tuning where Yukawa couplings are of $\O(1)$,
 we expect $\,\mD ={\cal O}(100\,\textrm{GeV})\,$, lying at the weak scale, so the ratio
 $\,(2 a^2 + (b + \pb c)^2)/(b - \pb c)^2={\cal O}(1) \,$.\,
 Hence, we deduce from (\ref{eq:mass2-diff-ratio}) and (\ref{eq:q}),
 \beqs
 \label{eq:q2}
 \beqa
 \label{eq:q2+}
 q &\!\!\equiv\!\!& \f{M_{1}}{M_{2}} ~\ll~ 1\,, ~~~~ (\textrm{for}~ \pb =+ )\,,
 \\[1mm]
 \label{eq:q2-}
 q^{-1} &\!\!\equiv\!\!& \f{M_{2}}{M_{1}} ~\ll~ 1\,, ~~~~ (\textrm{for}~ \pb =- )\,.
 \eeqa
 \eeqs
 This indicates that the mass-difference between the heavy neutrinos is
 large, which is also what we need for the leptogenesis analysis in Sec.\,\ref{sec:solution}.
 From (\ref{eq:q}) we can further resolve the parameter $\,r$\,,
 \beqa
 \label{eq:r}
 r ~=~  \left\{
 \ba{ll}
 \dis
 \f{2\pr q}{1+\pr q} ~=~ 2\pr q + \O(q^2) ~\ll~ 1\,,
 &~~~ (\textrm{for}~ \pb=+)\,,
 \\[5mm]
 \dis
 \f{2}{1+q^{-1}} ~=~ 2(1-\pr q^{-1}) + \O(q^{-2}) ~\simeq~ 2 \,,
 &~~~ (\textrm{for}~ \pb=-)\,,
 \ea
 \right.
 \eeqa
 where $\,\pr =\pm\,$ denotes the sign of \,$r$.\,
 We see that the case ~$\pb =-$~ leads to $\,r\simeq 2\,$ which contradicts with the
 original condition $\,R\equiv 1-r > 0\,$.\, Hence, the sign-choice of $\pb =-$ is excluded
 (or strongly disfavored), and we can focus on the case with $\,\pb =+\,$ from now on.

\vspace*{3mm}
\subsection{Common Origin of Soft $\boldsymbol{\mutau}$ and CP Breaking in
            Neutrino Seesaw} \label{sec:common-origin}

\vspace*{2mm}
\subsubsection{Unique Construction of Soft $\boldsymbol{\mutau}$ and CP Breaking in Neutrino Seesaw}
\label{sec:unique}

 As we pointed out in the introduction (Sec.\,\ref{sec:introduction}),
 the $\mutau$ symmetry serves as a good zeroth order flavor symmetry of the neutrino sector,
 it further leads to $\,\theta_{13}=0\,$ and thus the Dirac CP-conservation.
 This means that the $\mutau$ symmetry breaking is not only small,
 but also the source of the Dirac CP-violation.
 On the theory ground, it is natural and tempting to expect a common origin for all
 CP-violations, although the Dirac and Majorana CP-violations appear
 differently in the low energy effective theory of light neutrino mass-matrix.
 With such a common origin for two kinds of CP-violations,
 then they must vanish together in the $\mutau$ symmetric limit.

 With the above key observations, we conjecture
 that both discrete $\mutau$ and CP symmetries are fundamental symmetries of the
 seesaw Lagrangian {\it (respected by interaction terms),}  and
 are only {\it softly broken from a common origin in the seesaw sector.}
 We observe that the only place for such a soft breaking is the
 {\it dimension-3 singlet Majorana mass-term} of the right-handed heavy neutrinos $\N$
 in Eq.\,(\ref{eq:L-seesaw}).
 The Dirac mass term cannot provide such a soft breaking as it is generated by the
 {\it dimension-4} Yukawa interactions after spontaneous electroweak symmetry breaking.

 Under this conjecture, we can thus define the {\it common soft-breaking term}
 for $\mu-\tau$ and CP symmetries in the Majorana mass-matrix \,$M_R$\,,
 via a single complex parameter \,$\zeta e^{i\om}$\,,\, in a unique way,
 \beqa
 \label{eq:MR-SoftB-I}
  M_R &=&
  \begin{pmatrix}
    ~M_{22} & ~M_{23} \\[3mm]
    ~M_{23} & ~M_{22}\(1 - \zeta e^{i \omega}\)
  \end{pmatrix} ,
 \eeqa
where the module \,$0 < \zeta < 1$\, and the phase angle $\omega\in [0,2\pi)$.\,
Since the sign of $\zeta$ can always be absorbed into the phase $e^{i\omega}$,
we only need to consider $\,\zeta > 0\,$.\, Besides, $\zeta$ should be significantly smaller
than one as the $\mutau$ and CP violations are small.
As explained in Sec.\,(\ref{sec:zeroth-seesaw}), the elements $M_{22}$ and $M_{23}$
will be taken in the positive range without losing generality.
We note that the unique CP phase $e^{i\om}$ in (\ref{eq:MR-SoftB-I}) will not only generate
the low energy CP-violations (via both Dirac and Majorana phases) appearing in the
neutrino oscillations and neutrinoless double-$\beta$ decays,
but also provide the CP asymmetry needed
for the successful leptogenesis of baryon asymmetry (cf. Sec.\ref{sec:leptogenesis}).

 We also note that this soft-breaking term cannot be located in the $\{23\}$ and $\{32\}$
 elements of $M_R$ because the Majorana mass-matrix $M_R$ is symmetric and
 the equality between the two off-diagonal elements automatically respects the $\mutau$ symmetry.
 Hence, the only place to softly and simultaneously break both $\mu\tau$ and CP symmetries
 is in the diagonal elements of $M_R$.
 Then, we may wonder whether we could relocate the soft-breaking term in
 the $\{11\}$-element instead of $\{22\}$-element of $M_R$,
 \beqa
  \ov{M}_R &=&
  \begin{pmatrix}
    ~M_{22}\(1 - \zeta e^{i \omega}\) & ~M_{23}~ \\[3mm]
    ~M_{23} & ~M_{22}~
  \end{pmatrix}.
  \label{eq:MR-SoftB-II}
 \eeqa
 But we can readily prove that the
 definition (\ref{eq:MR-SoftB-II}) is actually equivalent to that in Eq.\,(\ref{eq:MR-SoftB-I})
 for our present study (where we are only interested in deriving correlations among the physical
 observables). To see this, we can simply rename the right-handed neutrinos
 $\,\NN=(N_\mu,\,N_\tau)\,$
 as $\,\NN'=(N_\tau',\,N_\mu')\,$, i.e.,
 \beqa
 \mathcal{N} ~=~  \mathcal{R} \mathcal{N}' \,,
 ~~~&&~~~
 \mbox{with} \qquad
   \mathcal{R} ~\equiv
  \begin{pmatrix}
    0 & 1 \\
    1 & 0
  \end{pmatrix} .
 \eeqa
 Then, the mass matrices $\,\mD\,$ in (\ref{eq:mD-MR-new})
 and $\,\ov{M}_R\,$ in (\ref{eq:MR-SoftB-II}) change accordingly,
 \beqs
 \label{eq:mD'-MR'}
 \beqa
 \label{eq:mmD'}
   \mD ~\to~
  m_D' = m_D \mathcal \RR  =
  \begin{pmatrix}
     \pb a & a \\
         c & b \\
     \pb b & \pb c
   \end{pmatrix}
  \equiv
   \begin{pmatrix}
     a'     & \pb a' \\
     b'     & c' \\
     \pb c' & \pb b'
   \end{pmatrix},
 \\[4mm]
 \label{eq:MMR'}
  \ov{M}_R ~\to~  \ov{M}_R' = \RR^T \ov{M}_R \RR
 =
  \begin{pmatrix}
    M_{22} & M_{23} \\[2mm]
    M_{23} & M_{22}(1 - \zeta e^{i \omega})
  \end{pmatrix},
 \eeqa
 \eeqs
 where we have simply renamed,
 $\,(\pb a,\,c,\,b) \equiv (a',\,b',\,c')\,$.\,
 From this we see that the Majorana mass-matrix $\ov{M}_R'$ is identical to $M_R$
 as defined in Eq.\,(\ref{eq:MR-SoftB-II}), while
 the Dirac mass-matrix $m_D'$
 takes exactly the same form as $\mD$ in Eq.\,(\ref{eq:mD-MR-new}) except all
 its elements are primed in the new notation.
 Since our final physical results in Sec.\,\ref{sec:diagonalization}-\ref{sec:solution} only concern the
 correlations among physical observables and do not explicitly rely
 on either $(a,\,b,\,c)$ or $(a',\,b',\,c')$\,\footnote{We note that under our present construction
 these parameters will all be eliminated from our final results
 since they are neither direct observable nor theoretically predicted (cf.\ Sec.\,\ref{sec:diagonalization}-\ref{sec:solution}).},\,
 we conclude that the parametrization in (\ref{eq:MR-SoftB-II}) is physically equivalent to
 (\ref{eq:MR-SoftB-I}) for our study, thus the formulation of the common origin for
 soft $\mutau$ and CP violations is indeed unique under our construction.
 In consequence, we only need to focus on the analysis of
 Eq.\,(\ref{eq:MR-SoftB-I}) for the rest of our paper.

 In summary, the Dirac mass-matrix $\mD$ and Majorana
 mass-matrix $M_R$ (with soft $\mutau$ and CP breakings from a common origin)
 can be uniquely parameterized as follows,
 \beqa
 \label{eq:mD-MR-FF}
  \mD ~=
  \begin{pmatrix}
    a & a \\[1mm]
    b & c \\[1mm]
    c & b
  \end{pmatrix} \!,
\qquad
  M_R ~=
  \begin{pmatrix}
    M_{22} & M_{23} \\[2.5mm]
    M_{23} & M_{33}
  \end{pmatrix}\!,~~~~~
  M_{33}~=~M_{22}\(1 - \zeta e^{i \omega}\),
 \eeqa
 where we have set the sign $\,\pb=+\,$ for $\mD$ as discussed
 by the end of Sec.\,\ref{sec:zeroth-seesaw}.

 Thus, we can explicitly derive the seesaw mass-matrix for light neutrinos,
  \begin{eqnarray}
  M_\nu &=&
  m_D M^{-1}_R m^T_D
  \nn\\[3mm]
 && \hspace{-14mm}
 =\frac{1}{\Vert M_R \Vert}
 \begin{pmatrix}
        a^2(M_{22} \!+\! M_{33} \!-\! 2M_{23})
      & a[cM_{22} \!+\! bM_{33}\!-\!(b\!+\!c)M_{23}]
      & a[bM_{22}\!+\!cM_{33}\!-\!(b+c)M_{23}]
      \\[2mm]
        & c^2M_{22} \!+\! b^2M_{33} \!-\! 2bcM_{23}
        & bc(M_{22} \!+\! M_{33}) \!-\! (b^2+c^2)M_{23}
        \\[2mm]
        & & b^2M_{22}\!+\!c^2M_{33}\!-\!2bcM_{23}
    \end{pmatrix}
    \nn\\[4mm]
    && \hspace{-14mm}
    \equiv
    \begin{pmatrix}
      A & B_1 & B_2 \\
        & C_1 & D \\
        &     & C_2
    \end{pmatrix} \!,
    \label{eq:Mnu-New}
  \end{eqnarray}
 where we have denoted, $\,\Vert M_R \Vert \equiv \det(M_R) = M_{22}M_{33}-M_{23}^2\,$.
 We find that it is useful to further decompose $M_\nu$ into $\mutau$ symmetric and
 anti-symmetric parts,
\beqa
\label{eq:Mnu-decomposition}
  M_\nu &\equiv&   M^{s}_\nu + M^{a}_\nu \,,
\eeqa
\vspace*{-4mm}
\begin{subequations}
  \begin{eqnarray}
  \label{eq:Mnu-decomposition-s}
    M^{s}_\nu
  & \equiv &
    \f{1}{2}
    \left(
      M_\nu + T_L^T M_\nu T_L
    \right)
  \equiv
    \begin{pmatrix}
      A & B_s & B_s \\
        & C_s & D \\
        &     & C_s
    \end{pmatrix} \!,
  \\[3mm]
  \label{eq:Mnu-decomposition-a}
    M^{a}_\nu
  & \equiv &
    \f{1}{2}
    \left(
      M_\nu - T_L^T M_\nu T_L
    \right)
  \equiv
    \begin{pmatrix}
      0 & B_a  &  -B_a \\
        & C_a  & 0     \\
        &      & -C_a
    \end{pmatrix} \!,
  \end{eqnarray}
  \label{eq:Mnu-decomposition-all}
\end{subequations}
with
\begin{subequations}
  \begin{eqnarray}
    A
  &=& \frac{\,M_{22} \!+\! M_{33} \!-\! 2 M_{23}\,}{\DetMR} a^2 \,,
  \\[3mm]
    B_s
  &=& \f{B_1+B_2}{2}
  ~=~  \frac{\,M_{22} \!+\! M_{33} \!-\! 2 M_{23}\,}{2\DetMR} a (b + c) \,,
  \\[3mm]
    C_s + D
  & = & \f{C_1+C_2}{2} + D
  ~=~
    \frac{\,M_{22} \!+\! M_{33} \!-\! 2 M_{23}\,}{2\DetMR} (b + c)^2 \,,
  \\[3mm]
    C_s - D
  & = & \f{C_1+C_2}{2} - D
  ~=~
    \frac{\,M_{22} \!+\! M_{33} \!+\! 2 M_{23}\,}{2\DetMR} (b - c)^2 \,,
  \end{eqnarray}
\end{subequations}
and
  \begin{subequations}
  \label{eq:Ba-Ca}
  \begin{eqnarray}
  \label{eq:Ba}
  B_a  &\equiv&  \d B_a ~=~ \f{B_1-B_2}{2}
       ~=~  \frac{\,M_{22} \!-\! M_{33}\,}{2\DetMR} a (c - b) \,,
  \\[3mm]
  \label{eq:Ca}
  C_a  &\equiv&  \d C_a ~=~  \f{C_1-C_2}{2}
  ~=~   \frac{\,M_{22} \!-\! M_{33}\,}{2\DetMR} (c^2 - b^2) \,,
  \end{eqnarray}
where
\beqa
\f{\,M_{22} \!-\! M_{33}\,}{M_{22}} ~=~ \zeta e^{i\om} ~=~ O(\zeta )\,,
\eeqa
\end{subequations}
showing that the parameters $\d B_a$ and $\d C_a$ must vanish in the $\mutau$ (and CP)
symmetric limit $\,\zeta\to 0$\,.\,  So the $\mutau$ anti-symmetric part is a net
measure of the $\mutau$ breaking. From (\ref{eq:Ba-Ca}) we further compute the ratio
of $\d B_a$ and $\d C_a$,
\beqa
\label{eq:dBa/dCa}
\f{\d B_a}{\d C_a} &=& \f{a}{\,b+c\,}\,,
\eeqa
which is a real number, independent of the $\mutau$ breaking
and depending only on the elements of the $\mutau$ symmetric $\,\mD\,$.\,
Note that in (\ref{eq:Ba-Ca}), all $\mutau$ breaking effects are contained in the
coefficient, ~$(M_{22}-M_{33})/\DetMR$\,,\,
which is exactly canceled in the ratio $\,\d B_a/\d C_a$.\,
In Sec.\,\ref{subsec:solar}-\ref{subsec:solar-mixing-not-affected},
we will further prove that the ratio (\ref{eq:dBa/dCa})
is directly connected to $\,\tan\ts\,$ for the solar mixing angle.

\vspace*{3mm}
\subsubsection{Soft $\boldsymbol{\mutau}$ and CP Breaking via Perturbative Expansion}

According to our construction in Sec.\,\ref{sec:unique},
the size of the soft $\mutau$ and CP breakings
are characterized by the small quantity $\,|\zeta|<1$\,,\, which can be treated as a
perturbative expansion .  As we noted by the end of Sec.\,\ref{sec:zeroth-seesaw},
there is another small $\mutau$ and CP symmetric parameter $\,|r| < 1\,$,\,
appearing in the Majorana mass-matrix of heavy right-handed neutrinos,
\beqa
  M_R ~=~ M_{22}
  \begin{pmatrix}
    1 & 1-r \\[3mm]
    1-r & 1 - \zeta e^{i \omega}
  \end{pmatrix}
 \,=~
  M_{10}
  \begin{pmatrix}
    \frac{1}{r}   & \frac{1}{r}-\!1 \\[3mm]
    \frac{1}{r}-\!1 & \frac{1}{r}-\! X
  \end{pmatrix}\!,
  \label{eq:MR-SoftB-Ix}
 \eeqa
where we have used the notations,
 \beqa
 r ~\equiv\, 1-R ~\equiv\, 1 -\f{M_{23}}{M_{22}}\,, ~~~~~
 M_{10} \,\equiv~ rM_{22}\,, ~~~~~
 X ~\equiv\, \frac{\zeta}{r}\,e^{i\omega} =\, \O(1) \,.
 \eeqa
With these we can explicitly write the light neutrino mass matrix (\ref{eq:Mnu-New}) as,
 \beqs
 \beqa
  M_\nu &\!\!=\!\!&
  \frac{1}{(2\!-\!r\!-\!X)M_{10}}
  \begin{pmatrix}
  ra^2(2\!-\!X) & ra(b\!+\!c\!-\!bX)  &  ra(b\!+\!c\!-\!cX)
  \\[2.5mm]
    & (b\!-\!c)^2\!+\!rb(2c\!-\!bX)     & -(b\!-\!c)^2\!+\!r(b^2\!+\!c^2\!-\!bcX)
  \\[2.5mm]
    & & (b\!-\!c)^2 \!+\! rc(2b\!-\!cX)
  \end{pmatrix}~~~~~~~~~~
  \label{eq:Mnu-all0}
  % \\[2mm]
  % &\!\!\equiv\!\!& M_\nu(r) + M_\nu(r,X)\,,
  % \label{eq:Mnu-all}
 \eeqa
 \eeqs
 Since the unique $\mutau$ and CP violation quantity
 $\,\zeta e^{i \omega}\,$ is now fully hidden in the ratio
 $\,X\equiv \f{\zeta}{r}e^{i \omega} = \O(1)\,$,\, we see that the only superficial small parameter
 for the effective perturbative expansion is $\,r\,$.\,
 So, we can reorganize the light neutrino mass-matrix
 $\,M_\nu\,$ in terms of the $r$-expansion alone, but with $X$ unexpanded,
 \begin{equation}
  M_\nu
  % ~\equiv~ M_{\nu}^{(0)} + \delta M_\nu
  ~=~
  M_{\nu}^{(0)} + \d M_{\nu}^{(1)} + O(r^2) \,,
  \label{eq:Mnu-expand}
 \end{equation}
 where
 \begin{subequations}
 \beqa
  M_{\nu}^{(0)}
 & = &
  % M^{(0)}_{\nu}(r) + M^{(0)}_{\nu}(r,X) ~=~
  \f{(b\!-\!c)^2}{(2\!-\!X)M_{10}}
  \(\ba{rrr}
    0 & 0 &  0\\
      & 1 & -1\\
      &   &  1
  \ea\) \!,
  \label{eq:Mnu0}
 \\[3mm]
 \d M_{\nu}^{(1)}
  % & = &
  % M^{(1)}_{\nu}(r) + M^{(1)}_{\nu}(r,X) \nonumber \\[3mm]
 & = &
  \f{r}{(2 \!-\! X)^2 M_{10}}
  \begin{pmatrix}
    (2 \!-\! X)^2 a^2
  & (2 \!-\! X)[(1 \!-\! X) b \!+\! c] a
  & (2 \!-\! X) [b \!+\! (1 \!-\! X)c] a
  \\[2.5mm]
  & [(1 \!-\! X) b \!+\! c]^2
  & (1 \!-\! X) (b \!+\! c)^2 \!+\! X^2 bc
  \\[2.5mm]
  & & [b \!+\! (1 \!-\! X) c]^2
  \end{pmatrix}\!. \hspace*{19mm}
\label{eq:Mnu1}
 \eeqa
 \label{eq:Mnu-expansion-r}
\end{subequations}
We note that the zeroth order mass-matrix $M_{\nu}^{(0)}$ is complex, but the CP-phase
only appears in an {\it overall} factor and thus causes no observable CP-violation.
The above decomposition can be symbolically denoted as,
 \beqa
 M_\nu &\!\!=\!\!&
 \(\ba{lll}
 A & B_1 & B_2 \\[1mm]
   & C_1 & D   \\[1mm]
   &     & C_2
 \ea \)
 ~\equiv~
 \(\ba{lll}
 A_0 & B_0 & B_0 \\[1mm]
     & C_0 & D_0   \\[1mm]
     &     & C_0
 \ea \) +
 \(\ba{lll}
 \d A & \d B_1 & \d B_2 \\[1mm]
      & \d C_1 & \d D   \\[1mm]
   &           & \d C_2
 \ea \)
\nn \\[3mm]
 &\!\!\equiv\!\!&
 M_\nu^{(0)} + \d M_\nu
 ~=~ M_\nu^{(0)} + \d M_\nu^{(1)} + \O(r^2) \,,
 \label{eq:Mu=Mu0+dMu}
 \eeqa
with its zeroth order matrix corresponding to (\ref{eq:Mnu0}) and the next-to-leading order (NLO)
matrix equal to (\ref{eq:Mnu1}) at $\O(r)$.

 For convenience, we further decompose $\d M_\nu$ into the
 $\mutau$ symmetric and anti-symmetric parts,
 \beqa
 \d M_\nu^{(1)} ~\equiv~ \d M_\nu^{s} + \d M_\nu^{a}
 &\equiv&
   \(\ba{lll}
 \d A & \d B_s & \d B_s   \\[1mm]
      & \d C_s & \d D   \\[1mm]
      &      & \d C_s
 \ea \) +
    \(\ba{llc}
    0 & \d B_a & -\d B_a   \\[1mm]
      & \d C_a & 0         \\[1mm]
      &        & -\d C_a
 \ea \)  \!,
 \label{eq:dMu=s+a}
 \eeqa
 with
 \beqs
 \beqa
 \d B_s  ~\equiv~ \f{\d B_1 + \d B_2}{2} \,, &&
 \d B_a  ~\equiv~ \f{\d B_1 - \d B_2}{2} \,,
 \\[2.5mm]
 \d C_s  ~\equiv~ \f{\d C_1 + \d C_2}{2} \,, &&
 \d C_a  ~\equiv~ \f{\d C_1 - \d C_2}{2} \,.
 \eeqa
 \label{eq:delta-BC-sa}
 \eeqs
 This decomposition is actually unique. For our current model with the expansion up to $\O(r)$,
 we deduce from (\ref{eq:Mnu0})-(\ref{eq:Mnu1}) and (\ref{eq:dMu=s+a})-(\ref{eq:delta-BC-sa}),
  \begin{eqnarray}
    A_0 ~=~ B_0 ~=~ 0\,, ~~~~~
    C_0 ~= -D_0 ~=~ \f{(b \!-\! c)^2}{\,(2 \!-\! X)M_{10}\,}\,.
  \end{eqnarray}
and
 \beq
 \label{eq:seesaw-Delta-m}
 \ba{ll}
 \dis\d A ~=~ \frac{a^2}{M_{10}}\;r \,,
 &~~~~~
 \dis\d D ~=~ \frac{\,[(b\!+\!c)^2(1\!-\!X)+bcX^2]\,}{(2\!-\!X)^2M_{10}}\;r \,,
 \\[4mm]
 \dis\d B_s  ~=~ \frac{\,a(b+c)\,}{2M_{10}}\;r \,,
 &~~~~~
 \dis\d C_s  ~=~ \f{\,[2(b\!+\!c)^2(1\!-\!X)+(b^2\!+\!c^2)X^2]\,}{\,2(2\!-\!X)^2M_{10}\,}\;r \,,~~~~
 \\[4mm]
 \dis\d B_a  ~=\, -\frac{a(b-c)X}{\,2(2\!-\!X)M_{10}\,}\;r \,,
 &~~~~~
 \dis\d C_a  ~=\, -\f{(b^2\!-\!c^2)X}{\,2(2\!-\!X)M_{10}\,}\;r \,.
 \ea
 \eeq
 We find it useful to further decompose the $\mutau$ symmetric part into,
 \begin{equation}
 \d M_\nu^{(s)} ~\equiv~
   \begin{pmatrix}
     \delta A   & \delta B_s & \delta B_s \\[1mm]
   & \delta C_s & \delta D \\[1mm]
   &            & \delta C_s
   \end{pmatrix}
 ~\equiv~
   \begin{pmatrix}
     \delta A   & \delta B_s & \delta B_s \\[1mm]
   & \delta E_s & \delta E_s \\[1mm]
   &            & \delta E_s
   \end{pmatrix}
 +
   \begin{pmatrix}
     0 & 0 & 0 \\[1mm]
       & \delta E'_s & - \delta E'_s \\[1mm]
       &             & \delta E'_s
   \end{pmatrix},
   \label{eq:decomposition-S}
 \end{equation}
with
\begin{subequations}
  \begin{eqnarray}
    \delta E_s  & \equiv &
    \frac 1 2 \( \delta C_s + \delta D \)
  ~=~ \f{\,(b \!+\! c)^2\,}{4 M_{10}}\; r \,,
    \label{eq:definition-E}
  \\[3mm]
    \delta E'_s
  & \equiv &
    \frac 1 2 \( \delta C_s - \delta D \)
  ~=~
    \frac {(b \!-\! c)^2 X^2}{\,4(2 \!-\! X)^2 M_{10}\,}\; r \,.
    \label{eq:definition-E1}
  \end{eqnarray}
    \label{eq:definition-EE1}
\end{subequations}

Finally, we note that from (\ref{eq:seesaw-Delta-m}) we can compute the ratio,
\beqa
\label{eq:dBa/dCa-2}
\f{\d B_a}{\d C_a} ~=~ \f{a}{\,b+c\,} \,,
\eeqa
which only depends on the elements of the $\mutau$ symmetric Dirac mass-matrix $\mD$,\,
and agrees to Eq.\,(\ref{eq:dBa/dCa}) at the end of Sec.\,\ref{sec:unique} where this ratio was
generally derived without invoking the explicit form of $M_R$ and without
making the $r$-expansion. As we will show in Sec.\,\ref{subsec:solar}-\ref{subsec:solar-mixing-not-affected},
this ratio is related to the solar mixing angle via $\tan\ts$
and is independent of the $\mutau$ breaking from the $M_R$.

\vspace{4mm}
Before concluding this section, for clarity we {\it summarize} the key features of our approach together
all assumptions we have used. As explicitly stressed in Sec.\,1 and at the beginning of Sec.\,2,
the oscillation data (Table-1) strongly support the $\mutau$ symmetry as a good approximate symmetry of
the neutrino sector, rather than the lepton sector (due to the large mass-splitting between the
$(\mu,\,\tau)$ leptons).  Thus the natural realization of $\mutau$ symmetry is in the diagonal
basis of {\it left-handed leptons} (corresponding to a diagonal mass-matrix product
$\,M_\ell M_\ell^\dag\,$), which can be easily realized by a proper lepton flavor symmetry like
$\,\mathbb{Z}_3\,$ [cf.\ Eq.\,(\ref{eq:GL-lep})],\,
so the measured PMNS mixing matrix $\,V=U_{\nu L}\,$ arises from
diagonalizing the light neutrino mass-matrix. Because the $\mutau$ symmetric limit enforces
$\,\theta_{13}=0\,$ and thus vanishing Dirac CP-violation in the PMNS matrix, the Dirac CP-violation
must originate from the $\mutau$ breaking. On the theory ground, it is tempting to expect a common origin
for all CP-violations (although the Dirac and Majorana CP-phases appear differently in the low energy
light neutrino mass-matrix). With these key observations,
{\it we have conjectured that both discrete $\mutau$
and CP symmetries are fundamental symmetries of the seesaw Lagrangian (respected by interaction terms),
and they are only softly broken, arising from a common origin; the only place for such a soft breaking
is the unique dimension-3 singlet Majorana mass-term for right-handed neutrinos.}
This means that Dirac mass-matrix $\mD$ naturally respects $\mutau$ and CP symmetries, and only the
singlet Majorana mass-matrix $M_R$ can softly break them.
This conjecture can hold for general neutrino seesaw with three right-handed neutrinos.
In our explicit formulation, we have chosen the minimal neutrino seesaw\,\cite{MSS,He2003}
(with two right-handed neutrinos added to the SM particle content) and proved that
the common origin of the soft $\mutau$ and CP breaking
is uniquely characterized by a single complex parameter in $\,M_R\,$
[cf. Eq.\,(\ref{eq:mD-MR-FF}) in Sec.\,2.2.1].
As explained above Eq.\,(\ref{eq:seesaw-formula}),
the minimal seesaw\,\cite{MSS,He2003} can be viewed as a useful effective theory
of the general three neutrino seesaw with one right-handed neutrino much heavier than
the other two and thus being integrated out of the seesaw Lagrangian (\ref{eq:L-seesaw}).
It serves as a good approximation as long as
the lightest left-handed neutrino has its mass much smaller than
the other two (even if not exactly massless).
We will further prove in Sec.\,6.3 that the general three-neutrino-seesaw under the
$\mutau$ symmetry ($\mathbb{Z}_2^{\mu\tau}$) and hidden symmetry ($\mathbb{Z}_2^s$)
also predicts a massless light neutrino, sharing the same feature as the minimal seesaw.

\vspace{4mm}
\section{Model-Independent Reconstruction of Light Neutrino Mass \\
\hspace{5mm} Matrix with $\boldsymbol{\mutau}$ and CP Violations from Low Energies}
 \label{sec:general}

 In this section, we present the {\it model-independent} reconstruction
 of the Majorana mass-matrix for light neutrinos, in terms of the
 low energy observables (mass-eigenvalues, mixings angles and CP
 phases). This reconstruction will be effectively expanded by
 experimentally well-justified small parameters up to the next-to-leading order.
 Then, in the next section we can readily apply this to
 our soft-breaking model (defined in Sec.\,\ref{sec:common-origin}), and systematically
 derive our model-predictions for the correlations among
 the low energy observables as well as for the link to the leptogensis
 at seesaw scale.

 \vspace*{3mm}
 \subsection{%\hspace*{-5mm}.\hspace*{-0.6mm}
 Model-Independent Reconstruction Formalism }
 \label{sec:model-independent-reconstruction-formalism}

 The $3\times3$ Majorana mass-matrix for the light neutrinos
 takes the following form,
 \beqa
 \label{eq:Mnu33}
  M_\nu
 &\equiv&
  \begin{pmatrix}
    m_{ee} & m_{e \mu} & m_{e \tau} \\[2mm]
    & m_{\mu \mu} & m_{\mu \tau} \\[2mm]
    && m_{\tau \tau}
  \end{pmatrix}
 ~\equiv~
  \begin{pmatrix}
    A & B_1 & B_2 \\[2mm]
      & C_1 & D   \\[2mm]
      &     & C_2
  \end{pmatrix} ,
 \eeqa
 which is generally symmetric and complex.
 The matrix (\ref{eq:Mnu33}) can be diagonalized by a unitary matrix \,$V$\,,
 \beqa
  V^T M_\nu V ~=~ D_\nu ~\equiv~ \textrm{diag}
  (m_1^{},\,m_2^{},\,m_3^{}) \,,
%  \begin{pmatrix}  m_1 \\  & m_2 \\  & & m_3  \end{pmatrix} \!,
  ~~~~& \textrm{or,} &~~~~
  M_\nu = V^* D_\nu V^\dagger\,,
  \label{eq:Mnu-V-D}
 \eeqa
 where the diagonal matrix $D_\nu$ contains three
 real mass-eigenvalues for the three light neutrinos.
 The mixing matrix $V$ can be generally expressed as a
 product of three unitary matrices including a CKM-type\footnote{Here,
 CKM stands for the Cabibbo-Kobayashi-Maskawa mixing matrix for the
 quark sector, including 3 mixing angles and 1 Dirac CP-phase\,\cite{CKM}.}
 mixing matrix $U$ plus two diagonal rephasing matrices $U'$ and $U''$,
 \beqa
 \label{eq:V}
 V ~\equiv~ U'' U U' \,,
 \eeqa
 with
 \begin{subequations}
 \label{eq:U'U"U}
 \beqa
 &&
   U' ~\equiv~ \textrm{diag}
   (e^{i \phi_1},\,e^{i \phi_2},\,e^{i \phi_3})\,, ~~~~~~
%   \begin{pmatrix}
%   e^{i \phi_1} \\   &  e^{i \phi_2} \\   && e^{i \phi_3}
%   \end{pmatrix} \!,  ~~~~~~
   U'' ~\equiv~ \textrm{diag}
   (e^{i \alpha_1},\,e^{i \alpha_2},\,e^{i \alpha_3}) \,,
%   \begin{pmatrix}
%     e^{i \alpha_1} \\   & e^{i \alpha_2} \\   &&e^{i \alpha_3}
%   \end{pmatrix} \!,
 \label{eq:U'U''}
 \\[3mm]
 && U ~\equiv~ U_{23} U_{13} U_{12} ~=~
  \begin{pmatrix}
    c_s c_x & - s_s c_x & - s_x e^{i \delta_D}
  \\[1.5mm]
    s_s c_a - c_s s_a s_x e^{-i\delta_D}
  & c_s c_a + s_s s_a s_x e^{-i\delta_D}
  & - s_a c_x
  \\[1.5mm]
    s_s s_a + c_s c_a s_x e^{-i\delta_D}
  & c_s s_a - s_s c_a s_x e^{-i\delta_D}
  & c_a c_x
  \end{pmatrix} \!,  \hspace*{20mm}
  \label{eq:U}
\\[3mm]
 &&  U_{23} \equiv
   \begin{pmatrix}
     1 & 0 & 0 \\
     0 & c_{a} & - s_{a} \\
     0 & s_{a} &   c_{a}
   \end{pmatrix}\!,
 \quad
   U_{13} \equiv
   \begin{pmatrix}
     c_{x} & 0 & - s_{x} e^{i\delta_D} \\
     0 & 1 & 0                          \\
     s_{x} e^{-i\delta_D} & 0 & c_{x}
   \end{pmatrix}\!,
 \quad
   U_{12} \equiv
   \begin{pmatrix}
     c_s & - s_s &  0 \\
     s_s &   c_s & 0 \\
     0 & 0 & 1
   \end{pmatrix}\!, ~~~~~~~~~
   \label{eq:U123}
 \eeqa
 \end{subequations}
 where $\d_D$ is the Dirac CP-phase. For notational convenience,
 we have denoted the three neutrino mixing angles of the PMNS matrix as,
 \beqs
 \beqa
 (\theta_{12},\,\theta_{23},\,\theta_{13})
 ~\equiv~ (\theta_s,\,\theta_a,\,\theta_x) ,\,
 \label{eq:theta-123}
 \eeqa
 (like in \cite{He2003}) in the above and in the rest of this paper.
 Consequently we further adopt the following notations,
 \beqa
 \label{eq:theta-123sc}
  (s_s,\,s_a,\,s_x) ~\equiv~ (\sin\ts ,\,\sin\ta ,\,\sin\tx )\,,
  &~~&
 (c_s,\,c_a,\,c_x) ~\equiv~ (\cos\ts ,\,\cos\ta ,\,\cos\tx )\,.~~~~
 \eeqa
 \eeqs

 The diagonal rephasing matrix $U'$ contains three Majorana phases, but after extracting
 an overall phase factor only two combinations of them persist
 (such as $\phi_1-\phi_3$ and $\phi_2-\phi_3$), which correspond to the two
 physical Majorana phases. The other three phases in $U''$ are associated with the
 flavor eigenbasis of light neutrinos
 and are needed for the consistency of diagonalizing a given
 mass-matrix \,$M_\nu$\,.

 From the second relation of (\ref{eq:Mnu-V-D}) we can fully reconstruct
 all the elements of the symmetric mass-matrix $M_\nu$
 in terms of the rephased mass-eigenvalues
 $\,(\tm_1,\,\tm_2,\,\tm_3)\equiv
    (m_1^{}e^{-i2\phi_1^{}},\,m_2^{}e^{-i2\phi_2^{}},\,m_3^{}e^{-i2\phi_3^{}})\,$,\,
 the mixing angles $(\ts ,\,\ta ,\,\tx )$, the Dirac phase $\d_D$,
 and the rephasing phases $\alpha_i$ (which do not appear in physical
 PMNS mixing matrix),
 \begin{subequations}
  \begin{eqnarray}
  \hspace*{-2mm}
    m_{ee}^{}
  & \!\!\!\!=\!\!\! &
    e^{-i2\alpha_1^{}}
  \[
    c^2_s c^2_x \widetilde m_1^{}
  + s^2_s c^2_x \widetilde m_2^{}
  + s^2_x e^{- 2 i \delta_D} \widetilde m_3^{} \] \!,
  \label{eq:Mnu-Reconstruct-ee}
  \\[1.5mm]
  \hspace*{-2mm}
    m_{\mu \mu}^{}
  & \!\!\!\!=\!\!\! &
    e^{-i2\alpha_2^{}}
  \[
    ( s_s c_a  - c_s s_a s_x e^{i \delta_D} )^2 \widetilde m_1^{}
  + ( c_s c_a  + s_s s_a s_x e^{i \delta_D} )^2 \widetilde m_2^{}
  + s^2_a c^2_x \widetilde m_3^{}
  \] \!,
  \label{eq:Mnu-Reconstruct-mm}
  \\[1.5mm]
  \hspace*{-2mm}
    m_{\tau\tau}^{}
  & \!\!\!\!=\!\!\! &
    e^{-i2\alpha_3^{}}
  \[
    ( s_s s_a + c_s c_a s_x e^{i \delta_D} )^2 \widetilde m_1^{}
  + ( c_s s_a - s_s c_a s_x e^{i \delta_D} )^2 \widetilde m_2^{}
  +   c^2_a c^2_x \widetilde m_3^{}
  \] \!,
  \label{eq:Mnu-Reconstruct-tt}
 \\[1.5mm]
 \hspace*{-2mm}
    m_{e\mu}^{}
  & \!\!\!\!=\!\!\! &
    e^{-i(\alpha_1^{}+\alpha_2^{})}
  \!\[
    c_s c_x (s_s c_a \!-\! c_s s_a s_x e^{i \delta_D}) \widetilde m_1^{}
  \!-\! s_s c_x (c_s c_a + s_s s_a s_x e^{i \delta_D}) \widetilde m_2^{}
  \!+\! s_a s_x c_x e^{- i\delta_D} \widetilde m_3^{}
  \] \!,~~~~~~~~
  \label{eq:Mnu-Reconstruct-em}
  \\[1.5mm]
  \hspace*{-2mm}
    m_{e\tau}^{}
  & \!\!\!\!=\!\!\! &
    e^{-i(\alpha_1^{}+\alpha_3^{})}
  \!\[
    c_s c_x (s_s s_a \!+\! c_s c_a s_x e^{i \delta_D}) \widetilde m_1^{}
  \!-\! s_s c_x (c_s s_a \!-\! s_s c_a s_x e^{i \delta_D}) \widetilde m_2^{}
  \!-\! c_a s_x c_x e^{- i \delta_D} \widetilde m_3^{}
  \] \!,~~~~~~~~
  \label{eq:Mnu-Reconstruct-et}
  \\[1.5mm]
  \hspace*{-2mm}
    m_{\mu\tau}^{}
  & \!\!\!\!=\!\!\! &
    e^{-i(\alpha_2^{}+\alpha_3^{})}
  \!\[
    ( s_s c_a - c_s s_a s_x e^{i \delta_D} )
    ( s_s s_a + c_s c_a s_x e^{i \delta_D} ) \widetilde m_1^{}
  \right.
    \nonumber
  \\
  \hspace*{-2mm}
  & \!\!\!\!\!\!\! &
  \hspace{16mm}
  \left.
  ~~ + ( c_s c_a + s_s s_a s_x e^{i \delta_D} )
    ( c_s s_a - s_s c_a s_x e^{i \delta_D} ) \widetilde m_2^{}
  -   s_a c_a c^2_x \widetilde m_3^{}
  \] \!,
  \label{eq:Mnu-Reconstruct-mt}
  \end{eqnarray}
  \label{eq:Mnu-Reconstruct}
\end{subequations}
where only two Majorana phases among $\phi_{1,2,3}^{}$
(hidden in the mass-parameters $\tm_{1,2,3}^{}$) are
independent since an overall phase factor (say $e^{i\phi_3^{}}$)
can be taken out of $U'$ and simply absorbed into the diagonal rephasing-matrix $U''$.\,
For the case with a vanishing mass-eigenvalue (such as $m_1^{}=0$ in our present model
defined in Sec.\,\ref{sec:origin}), only one independent phase combination
[say $e^{i(\phi_2^{}-\phi_3^{})}$] is left.

 If the light neutrino mass-matrix $M_\nu$ respects $\mutau$ symmetry, we should have
 \beqa
 \label{eq:mutau-cond}
  m_{e\mu}^{} ~=~ m_{e\tau}^{} \,,
  &~~~&
  m_{\mu\mu}^{} ~=~ m_{\tau \tau}^{} \,.
 \eeqa
 As discussed in Sec.\,\ref{sec:zeroth-seesaw},
 the $\mutau$ symmetry generally allows the relation $\,m_{e\mu}^{}=p\,m_{e\tau}^{}\,$
 (with $\,p=\pm\,$). But for $\,p=-$\, it can always be absorbed into the phase-matrix
 $U''$ by a simple shift $\,\a_3^{}\to \a_3^{}+\pi\,$
 [cf.\ (\ref{eq:Mnu-Reconstruct-em})-(\ref{eq:Mnu-Reconstruct-et})].
 (This shift also makes $m_{\mu\tau}^{}$ element flip a sign, but it does not affect
 the realization of $\mutau$ symmetry and we can simply denote $\,-m_{\mu\tau}^{}\,$ by
 $\,m_{\mu\tau}'\,$.)
 Taking the three mass-eigenvalues $(m_1^{},\,m_2^{},\,m_3^{})$
 as independent experimental inputs without extra tuning,
 we can derive six mass-independent equations
 from the general conditions in Eq.\,(\ref{eq:mutau-cond}),
 \beqs
\begin{eqnarray}
\label{eq:mutau-cond-1a}
  c_s c_x
\left[
  s_s c_a
- c_s s_a s_x e^{i \delta_D}
\right] e^{- i \alpha_2^{}}
& = &
  c_s c_x
\left[
  s_s s_a
+ c_s c_a s_x e^{i \delta_D}
\right] e^{- i \alpha_3^{}},
\\[1mm]
\label{eq:mutau-cond-1b}
  s_s c_x
\left[
  c_s c_a
+ s_s s_a s_x e^{i \delta_D}
\right] e^{- i \alpha_2^{}}
& = &
  s_s c_x
\left[
  c_s s_a
- s_s c_a s_x e^{i \delta_D}
\right] e^{- i \alpha_3^{}},
\\[1mm]
\label{eq:mutau-cond-1c}
  s_a s_x c_x e^{- i \delta_D} e^{- i \alpha_2^{}}
& = &
- c_a s_x c_x e^{- i \delta_D} e^{- i \alpha_3^{}},
\\[3mm]
\label{eq:mutau-cond-2a}
\left[
  s_s c_a
- c_s s_a s_x e^{i \delta_D}
\right]^2 e^{- i 2 \alpha_2^{}}
& = &
\left[
  s_s s_a
+ c_s c_a s_x e^{i \delta_D}
\right]^2 e^{- i 2 \alpha_3^{}},
\\[1mm]
\label{eq:mutau-cond-2b}
\left[
  c_s c_a
+ s_s s_a s_x e^{i \delta_D}
\right]^2 e^{- i 2 \alpha_2^{}}
& = &
\left[
  c_s s_a
- s_s c_a s_x e^{i \delta_D}
\right]^2 e^{- i 2 \alpha_3^{}},
\\[1mm]
\label{eq:mutau-cond-2c}
  s^2_a c^2_x e^{- i 2 \alpha_2^{}}
& = &
  c^2_a c^2_x e^{- i 2 \alpha_3^{}}\,.
\end{eqnarray}
\label{eq:mutau-six-eqs}
 \eeqs
 The last three conditions (\ref{eq:mutau-cond-2a})-(\ref{eq:mutau-cond-2c}) are not independent,
 since up to a trivial overall factor
 (\ref{eq:mutau-cond-2a})-(\ref{eq:mutau-cond-2b}) are just the squares of
 (\ref{eq:mutau-cond-1a})-(\ref{eq:mutau-cond-1b}), respectively, and (\ref{eq:mutau-cond-2c}) can be
 inferred from  (\ref{eq:mutau-cond-1a})-(\ref{eq:mutau-cond-1b}) (cf.\ below),
 where we note  $\,\theta_s \neq 0^\deg,\,90^\deg\,$ and $\,\theta_x \neq 90^\deg$\, due to
 the requirement of neutrino data in Table-\ref{tab:1}.
 So we only need to solve  (\ref{eq:mutau-cond-1a})-(\ref{eq:mutau-cond-1c}).
 Since the data in Table-\ref{tab:1} already enforce
 $\,\theta_s \neq 0^\deg,\,90^\deg\,$ and $\,\theta_x \neq 90^\deg$,\,
 we can deduce two equations from (\ref{eq:mutau-cond-1a})-(\ref{eq:mutau-cond-1b}),
 \beqs
 \beqa
 \label{eq:ca-sa-a23-1}
  c_a \,-\, s_a e^{i(\a_2^{}-\a_3^{})}  &\!=\!& 0 \,,
 \\
 \[s_a \,+\, c_a e^{i(\a_2^{}-\a_3^{})}\]s_x &\!=\!& 0 \,,
 \label{eq:ca-sa-a23-2}
 \eeqa
 \label{eq:ca-sa-a23}
 \eeqs
 which together give,
 \beqa
 \sin(\a_3^{}-\a_2^{})\,=\,0\,,  ~~~~~
 \tan\theta_a \,=\, \cos(\a_3^{}-\a_2^{})\,, ~~~~~
 s_x \,=\, 0 \,.
 \eeqa
 The solution $\,s_x=0\,$ also holds Eq.\,(\ref{eq:mutau-cond-1c}).
 Noting that $\,\tan\ta \geqq 0\,$ for $\,\ta\in [0,\,\f{\pi}{2}]\,$,\,
 we find that the $\mutau$ symmetry results in the unique consistent solution,
 \beqa
  \(\ta ,\,\tx \)
 \,=\, (45^\deg ,\, 0^\deg ) \,,
 &~~~~~& \a_2^{} \,=\, \a_3^{} \,,
  \label{eq:mutau-solution}
 \eeqa
 which automatically holds (\ref{eq:mutau-cond-2c}).
 The solution (\ref{eq:mutau-solution}) is also the sufficient condition of
 realizing $\mutau$ symmetry in $M_\nu$ since we can readily deduce the
 condition (\ref{eq:mutau-cond}) by substituting Eq.\,(\ref{eq:mutau-solution}) into the
 general reconstruction formulas (\ref{eq:Mnu-Reconstruct-mm})-(\ref{eq:Mnu-Reconstruct-et}).

 Finally, we note that, as shown in the $\mutau$ symmetric solution (\ref{eq:mutau-solution}),
 the solar mixing angle $\ts$ is independent of the $\mutau$ symmetry
 and thus can take any value in the $\mutau$ symmetric limit. So we need to go beyond
 the $\mutau$ symmetry for predicting $\ts$ (cf. Sec.\,\ref{sec:extra-z2}), though this is not the main
 concern of our present study.

\vspace*{3mm}
\subsection{Reconstruction with Normal Hierarchy Mass-Spectrum of Light Neutrinos }

In this subsection we apply the general reconstruction formalism in Sec.\,\ref{sec:model-independent-reconstruction-formalism} to
to our model defined in Sec.\,\ref{sec:origin} where the light neutrino mass-spectrum exhibits the
normal hierarchy (NH) pattern, $\,\ma<\mb<\mc\,$, and has $\,\ma=0\,$
[cf. Eq.\,(\ref{eq:LO-Mass})].
As noted in Eqs.\,(\ref{eq:ratio-m2m3})-(\ref{eq:mass2-diff-ratio}), the
neutrino data (Table-\ref{tab:1}) requires the small mass-ratio,
\begin{equation}
  y ~\equiv~
  \frac{m_2}{m_3}
  ~=~ \sqrt{\f{\d m^2_{21}}{\d m^2_{31}}}
  ~\simeq~ \sqrt{\f{\Delta_s}{\Delta_a}}
  ~=~ 0.17-0.19 \ll 1 \,,
  \label{eq:definition-y}
\end{equation}
at $2\sigma$ level.
From the relations in Eqs.\,(\ref{eq:mass2-diff-ratio})-(\ref{eq:r}) (with $\pb=+$),
we see that the small parameter $y$ is related to another small quantity $r$
in the seesaw sector and we have $\,y=\O(r) \ll 1\,$.\,
So, with $\,\ma =0\,$ and $\,y\equiv \f{\mb}{\mc} \ll 1\,$
for the NH mass-spectrum, we can further simplify
the general reconstruction formula (\ref{eq:Mnu-Reconstruct}) as,
\begin{subequations}
  \begin{eqnarray}
  \label{eq:NH-m-ee}
    m_{ee}^{}  &\!\!\! = \!\!\!&
    m_3^{} e^{-i2\ab_1^{}}
  \[
    s^2_s c^2_x y e^{- i2\phi_{23}^{}}
  + s^2_x e^{-i2\delta_D}
  \] ,
  \\[1.5mm]
  \label{eq:NH-m-mm}
     m_{\mu\mu}^{}
  &\!\!\! = \!\!\!&
    m_3^{} e^{-i2\ab_{2}^{}}
  \[
    y( c_s c_a + s_s s_a s_x e^{i \delta_D} )^2 e^{-i2\phi_{23}^{}}
    + s^2_a c^2_x
  \],
  \\[1.5mm]
  \label{eq:NH-m-tt}
      m_{\tau\tau}^{}
  &\!\!\! = \!\!\!&
    m_3^{} e^{-i2\ab_3^{}}
  \[
    y( c_s s_a - s_s c_a s_x e^{i \delta_D} )^2 e^{-i2\phi_{23}^{}}
    + c^2_a c^2_x
  \].
  \\[1.5mm]
  \label{eq:NH-m-em}
    m_{e\mu}^{}
  &\!\!\! = \!\!\!&
    m_3^{} e^{-i(\ab_1^{}+\ab_2^{})}
  \[
  - y s_s c_x (c_s c_a + s_s s_a s_x e^{i \delta_D}) e^{- i2\phi_{23}^{}}
  +     s_a c_x s_x e^{- i \delta_D}
  \],
  \\[1.5mm]
  \label{eq:NH-m-et}
    m_{e\tau}^{}
  & = &
    m_3^{} e^{-i(\ab_1^{}+\ab_3^{})}
  \[
  - y s_s c_x (c_s s_a - s_s c_a s_x e^{i \delta_D}) e^{- i2\phi_{23}^{}}
  - c_a c_x s_x e^{- i \delta_D}
  \],
  \\[1.5mm]
  \label{eq:NH-m-mt}
    m_{\mu\tau}^{}
  &\!\!\! = \!\!\!&
    m_3^{} e^{-i(\ab_2^{}+\ab_3^{})}
  \[
   y( c_s c_a + s_s s_a s_x e^{i \delta_D} )
    ( c_s s_a - s_s c_a s_x e^{i \delta_D} ) e^{- i2\phi_{23}^{}}
    - s_a c_a c^2_x
  \] ,~~~~~~~~~
  \end{eqnarray}
  \label{eq:Mnu-Reconstruct-NH}
\end{subequations}
where the Majorana phase $\phi_1^{}$ disappears due to \,$\ma=0\,$.\,
We have also absorbed an overall redundant Majorana phase $\phi_3^{}$ (from $U'$)
into the redefinition of $\,\a_j^{}$ (in $U''$),\,
and the only remaining independent Majorana phase is $\,\phi_{23}^{}\,$,
%\beqs
\beqa
\ab_j^{} ~\equiv~ \a_j^{} + \phi_3^{} \,,
&~~~~&
%\\[0.9mm]
\phi_{23}^{} ~\equiv~  \phi_2^{}-\phi_3^{} \,,
\eeqa
%\eeqs
%
where $\,j=1,2,3$.\,
In fact, for any minimal seesaw model the low energy physical PMNS mixing matrix $V_{\textrm{PMNS}}$
only contains one independent Dirac phase $\d_D$ and one independent Majorana phase (denoted as
$\phi_{23}^{}$ here for NH mass-spectrum).

With these notations we may slightly rewrite the general $\mutau$ symmetric solution as follows,
 \beqa
  \(\ta ,\,\tx \)_0^{}
 \,=\, \(\f{\pi}{4} ,\, 0 \) ,
 &~~~~~& \ab_{20}^{} \,=\, \ab_{30}^{} \,,
  \label{eq:mutau-solution-NH}
 \eeqa
 where the subscript ``0'' denotes the leading order (LO) results
 from the $\mutau$ symmetric limit.

Next, we will explicitly identify the $\mutau$ and CP violations from the
reconstruction formula (\ref{eq:Mnu-Reconstruct-NH}).
Since the existing neutrino data support the $\mutau$ symmetry prediction
(\ref{eq:mutau-solution-NH}) as an excellent approximation
[cf. Table-\ref{tab:1} and Eq.\,(\ref{eq:da-dx-exp})], we are strongly
motivated to treat the small $\mutau$ breakings,
\beqa
\da ~\equiv~ \ta -\f{\pi}{4} \,,
&~~~~~&
\dx ~\equiv~ \tx - 0\,,
\eeqa
as the parameters of perturbative expansion. So we can systematically analyze
these breakings under the expansion of $(\d_a,\,\d_x)$.\, For practical analysis
it is enough to keep the expansion up to the next-to-leading order (NLO), i.e.,
the linear order of $(\d_a,\,\d_x)$.\,
Besides $(\d_a,\,\d_x)$, the other NLO parameters for our current reconstruction analysis
will include $\,(\d m_2^{},\,\d m_3^{})\,$ and
$\,(\d \ab_1^{},\,\d\ab_2^{},\,\d\ab_3^{})\,$.\,
Since $\,m_1^{}\equiv 0\,$ in our minimal seesaw model with NH mass-spectrum, there is no
NLO correction to it. Also, we will explicitly show in Sec.\,\ref{sec:prediction-low-energy-parameter} that for our expansion
for the current model, the LO mass for $m_2^{}$
vanishes ($m_{20}^{}=0$) as well, so we have,
$\,\dis y\equiv \f{m_2^{}}{m_3^{}}=\f{\d m_2^{}}{m_3^{}}\,$,\,
where $\,m_{2}^{}=m_{20}^{}+\d m_{2}^{} = \d m_{2}^{}\,$.\,
We can introduce another small ratio $\dis\,z\equiv\f{\d m_3^{}}{m_{30}^{}}=\O(y)\,$.\,
In addition, as will be generally proven in Sec.\,\ref{subsec:solar-mixing-not-affected}, the solar angle
$\,\theta_s\,(\equiv\theta_{12})$\,
is independent of the soft $\mutau$ breaking and thus does not receive any NLO correction.
Furthermore, we note that the Dirac phase $e^{i\d_D}$ is always associated by the small
mixing parameter $\,s_x\,(\equiv\d_x)\,$ and the remaining Majorana phase $e^{i\phi_{23}^{}}$
in our present minimal seesaw model with NH mass-spectrum is always suppressed by the
small mass-ratio $\,y\,(\equiv \f{m_2^{}}{m_3^{}})\,$, so both phases only appear at NLO and
thus receive no more correction at this order of expansion.
Finally, we can summarize all independent
NLO parameters in our reconstruction analysis as follows,
\beqa
 (\, y,\,
     z,\,
     \d_a,\,
     \d_x,\,
     \d\ab_1^{},\,
     \d\ab_2^{},\,
     \d\ab_3^{} \,)\,,
  \label{eq:NLO-parameters}
\eeqa
with $\,\dis y\equiv \f{m_2^{}}{m_3^{}}=\f{\d m_2^{}}{m_3^{}}\simeq\f{\d m_2^{}}{m_{30}^{}}\,$
and  $\,\dis z\equiv\f{\d m_3^{}}{m_{30}^{}}=\O(y)\,$.\,
Each of these parameters is defined as the difference between its full value and zeroth-order
value. In Sec.\,\ref{sec:diagonalization} we will use our soft $\mutau$ and CP breaking model (defined in Sec.\,\ref{sec:common-origin})
to predict these deviations, especially the correlation between $\da$ and $\dx$ which will
be probed at the upcoming neutrino experiments.

Under the perturbation of (\ref{eq:NLO-parameters}), we expand the light neutrino mass-matrix
(\ref{eq:Mnu33}) up to NLO,
 \beqa
 \label{eq:Mnu33-NH}
  M_\nu
 &=& M_\nu^{(0)} + \d M_\nu^{(1)} + (\textrm{higher orders})
 \nn\\[4mm]
 &\equiv&
  \begin{pmatrix}
    m_{ee}^{(0)} & m_{e \mu}^{(0)} & m_{e \tau}^{(0)} \\[2mm]
    & m_{\mu \mu}^{(0)} & m_{\mu \tau}^{(0)} \\[2mm]
    && m_{\tau \tau}^{(0)}
  \end{pmatrix}
 +
   \begin{pmatrix}
    \d m_{ee}^{(1)} & \d m_{e\mu}^{(1)} & \d m_{e\tau}^{(1)} \\[2mm]
    & \d m_{\mu\mu}^{(1)} & \d m_{\mu\tau}^{(1)} \\[2mm]
    & & \d m_{\tau\tau}^{(1)}
  \end{pmatrix} \!.
 \eeqa
 The LO matrix $M_\nu^{(0)}$ and the NLO matrix
 $M_\nu^{(1)}$ can be explicitly derived
 from our reconstruction formula (\ref{eq:Mnu-Reconstruct-NH}).
 Let us first deduce the LO mass-matrix  $M_\nu^{(0)}$
 from (\ref{eq:Mnu-Reconstruct-NH}), under the $\mutau$ symmetric solution
 (\ref{eq:mutau-solution-NH}),
 \beqa
 \label{eq:Mnu-00}
 M_\nu^{(0)}
 ~\equiv~
 \begin{pmatrix}
    m_{ee}^{(0)} & m_{e \mu}^{(0)} & m_{e \tau}^{(0)} \\[2mm]
    & m_{\mu \mu}^{(0)} & m_{\mu \tau}^{(0)} \\[2mm]
    && m_{\tau \tau}^{(0)}
  \end{pmatrix}
 ~\equiv~
 \begin{pmatrix}
    A_0 & B_0 & B_0 \\[2mm]
        & C_0 & D_0 \\[2mm]
        &     & C_0
  \end{pmatrix}
 ~=~
 \f{1}{2}m_{30}^{}e^{-i2\ab_{20}^{}}
 \(\ba{ccr}
    0 & 0 &  0 \\[1mm]
      & 1 & -1 \\[1mm]
      &   &  1
   \ea\) \!,
 \eeqa
 where we have also matched to our notation of $M_\nu^{(0)}$ in (\ref{eq:Mu=Mu0+dMu}).
 So, as expected, the zeroth order $\mutau$ symmetric mass-matrix $M_\nu^{(0)}$ only
 contains an overall CP-phase $e^{-i2\ab_{20}^{}}$ which has no observable effect at this order
 and is needed only for consistency of the mass-diagonalization procedure (cf. Sec.\,\ref{sec:prediction-low-energy-parameter}).
 From (\ref{eq:Mnu-00}), it is clear that $M_\nu^{(0)}$ has the LO mass-eigenvalues
 $\,(0,\,0,\,m_{30}^{})\,$,\, while the mixing angle $\theta_s$ is not fixed at this order.

 Then, we derive elements of the NLO matrix $M_\nu^{(1)}$ from (\ref{eq:Mnu-Reconstruct-NH}),
\begin{subequations}
  \begin{eqnarray}
    \d m_{ee}^{(1)}  & \!\!\equiv\!\! &  \d A ~=~
    m_{30}^{}ys^2_s\, e^{-i2(\ab_{10} + \phi_{23}^{})}  \,,
\label{eq:dMnu-1-ee}
  \\[2mm]
    \d m_{\mu\mu}^{(1)}  & \!\!\equiv\!\! &  \d C_1 ~=~
    \hf m_{30}^{} e^{- i2\ab_{20}^{}}
    \[ yc_s^2e^{-i2\phi_{23}^{}} +z +2\d_a -i2\d\ab_2^{}\] ,
\label{eq:dMnu-1-mm}
  \\[2mm]
    \d m_{\tau\tau}^{(1)}  & \!\!\equiv\!\! &  \d C_2 ~=~
    \hf m_{30}^{} e^{- i2\ab_{20}^{}}
    \[ yc_s^2e^{-i2\phi_{23}^{}} +z -2\d_a -i2\d\ab_3^{}\] ,
\label{eq:dMnu-1-tt}
  \\[2mm]
    \d m_{e\mu}^{(1)}  & \!\!\equiv\!\! &  \d B_1 ~=~
    \f{1}{\sqrt{2}\,}m_{30}^{} e^{-i(\ab_{10} + \ab_{20})}
    \[ -y s_sc_s e^{-i2\phi_{23}^{}} +\d_x e^{-i\d_D}\] ,
\label{eq:dMnu-1-em}
  \\[2mm]
\d m_{e\tau}^{(1)}  & \!\!\equiv\!\! &  \d B_2 ~=~
    \f{1}{\sqrt{2}\,}m_{30}^{} e^{-i(\ab_{10} + \ab_{20})}
    \[ - y s_sc_s e^{-i2\phi_{23}^{}} - \d_x e^{-i\d_D}\] ,
\label{eq:dMnu-1-et}
  \\[2mm]
\d m_{\mu\tau}^{(1)}  & \!\!\equiv\!\! &  \d D ~=~
    \hf m_{30}^{} e^{-i2\ab_{20}^{}}
    \[ y c_s^2 e^{-i2\phi_{23}^{}} - z + i(\d\ab_2^{}+\d\ab_3^{})\] ,
\label{eq:dMnu-1-mt}
  \end{eqnarray}
  \label{eq:dMnu-1}
\end{subequations}
where we have matched to our notation of $\d M_\nu^{(1)}$ defined in (\ref{eq:Mu=Mu0+dMu}).
In the above formulas we have input the $\mutau$ symmetric LO parameters or
relations,
$\,m_1\equiv 0\,$,\, $\,(\theta_{a0},\,\theta_{x0})=(\f{\pi}{4},\,0)\,$
and $\,\ab_{20}^{}=\ab_{30}^{}\,$ [Eq.\,(\ref{eq:mutau-solution-NH})].

According to the definitions in (\ref{eq:dMu=s+a}),
we can uniquely decompose the elements of $\d M_\nu^{(1)}$
in (\ref{eq:dMnu-1}) as the $\mutau$ symmetric and anti-symmetric parts,
$\,\d M_\nu^{(1)} \equiv \d M_\nu^{s}+\d M_\nu^{a}\,$,\,
with their elements given by,
\beq
\label{eq:Reconstruct-dMnu-sa}
\ba{l}
\d B_s ~\equiv~ \dis\f{\d B_1+\d B_2}{2}
~=~ -\f{1}{\sqrt{2}\,}m_{30}^{}y s_s c_s \,
    e^{-i(\ab_{10}^{}+\ab_{20}^{}+2\phi_{23}^{})} \,,
\\[4mm]
\d B_a ~\equiv~ \dis\f{\d B_1-\d B_2}{2}
~=~ \f{1}{\sqrt{2}\,} m_{30}^{}\d_x\, e^{-i(\ab_{10}^{}+\ab_{20}^{}+\d_D)} \,,
\\[4mm]
\d C_s ~\equiv~ \dis\f{\d C_1+\d C_2}{2}
~=~ \hf m_{30}^{}e^{-i2\ab_{20}^{}}
\[yc_s^2e^{-i2\phi_{23}^{}} + z - i(\d\ab_2^{}+\d\ab_3^{})
\] ,
\\[4mm]
\d C_a ~\equiv~ \dis\f{\d C_1-\d C_2}{2}
~=~ \hf m_{30}^{}e^{-i2\ab_{20}^{}}
\[ 2\d_a -i(\d\ab_2^{}-\d\ab_3^{})\] ,
\ea
\eeq
where we may further decompose $\d M_\nu^{(s)}$ according to (\ref{eq:decomposition-S}),
\beq
\label{eq:Reconstruct-dEs-dEs'}
\ba{l}
\d E_s ~\equiv~ \dis\hf (\d C_s+\d D)
~=~ \hf m_{30}^{}yc_s^2\, e^{-i2(\ab_{20}^{}+\phi_{23}^{})} \,,
\\[4mm]
\d E_s'~\equiv~ \dis\hf (\d C_s-\d D)
~=~ \hf m_{30}^{}e^{-i2\ab_{20}^{}}\[
z - i(\d\ab_2^{}+\d\ab_3^{})
\].
\ea
\eeq

In the next section, we will apply the above reconstruction formulas
(\ref{eq:Mnu33-NH})-(\ref{eq:Reconstruct-dEs-dEs'})
to match with (\ref{eq:Mu=Mu0+dMu})-(\ref{eq:definition-EE1})
in our soft breaking model at the LO and NLO, respectively.
From this, we can connect the seesaw parameters with the low energy neutrino
observables and derive quantitative predictions of our soft breaking model.

\vspace*{5mm}
Before concluding this section, we note that so far we have presented our
reconstruction formalism at the low energy scale. There are possible renormalization
group (RG) running effects for the low energy neutrino parameters
when connecting them to the model predictions at the seesaw scale.
Such RG effects were extensively discussed in the literature\,\cite{nuRG}.
The application to our present analysis is straightforward.
Below the seesaw scale, heavy right-handed neutrinos can be integrated out from the
effective theory and the seesaw mass-matrix $\,M_\nu\,$ for light neutrinos obey
the one-loop RG equation (RGE)\,\cite{nuRG},
\beqa
\label{eq:RGE-Mnu}
\f{dM_\nu}{dt}~=~ \f{1}{16\pi^2}
\[\ahat \,M_\nu +
  {\cal C}\(M_\nu (Y_\ell^\dag Y_\ell) + (Y_\ell^\dag Y_\ell)^TM_\nu\)\],
\eeqa
where $\,t=\ln(\mu/\mu_0^{})\,$ with $\,\mu\,$ the renormalization scale,
$\,{\cal C} = -\f{3}{2}\,$ for the SM, and
$\,Y_\ell\,$ denotes the diagonal lepton Yukawa coupling matrix in the lepton
mass-eigenbasis. For the SM, the coupling parameter $\,\ahat\,$ consists of
\beqa
\label{eq:RG-alpha}
\ahat &\!=\!& -3g_2^2 + 2(y_\tau^2+y_\mu^2+y_e^2)
       +6(y_t^2+y_b^2+y_c^2+y_s^2+y_d^2+y_u^2) + \lambda
\nn\\
&\!\simeq\!& -3g_2^2 +6y_t^2 +\lambda \,,
\eeqa
where $\,(g_2^{},\,y_t^{},\,\lambda)\,$ stand for the $SU(2)_L$ weak gauge coupling,
the top Yukawa coupling and Higgs self-coupling, respectively, which are all
functions of $\,t$\, at loop-level.
Thus, the RGE for mass-eigenvalues $\,m_j^{}\,$($j=1,2,3$) can be derived as\,\cite{nuRG},
\beqa
\label{eq:RGE-mj}
\f{dm_j^{}}{dt} &\!=\!& \f{1}{16\pi^2}
\[\,\ahat \,+\, 2\,{\cal C}\,y_\tau^2\,{\cal F}_j\,\]m_j^{} \,,
\\[2.5mm]
{\cal F}_1 &\!=\!& 2s_s^2s_a^2 - 4s_xs_sc_ss_ac_a\cos\d_D +2s_x^2c_s^2c_a^2 \,,
\nn\\
{\cal F}_2 &\!=\!& 2c_s^2s_a^2 + 4s_xs_sc_ss_ac_a\cos\d_D +2s_x^2s_s^2c_a^2 \,,
\nn\\
{\cal F}_3 &\!=\!& 2c_x^2c_a^2 \,,
\nn
\eeqa
where $\,y_\tau^{} = \O(10^{-2})\,$ is the largest lepton Yukawa coupling.
We see that the flavor-dependent term on the RHS of (\ref{eq:RGE-mj}) is suppressed by
$\,y_\tau^2 =\O(10^{-4})\,$ relative to the universal $\,\ahat$-term, and thus completely
negligible for the present study.
Hence, from the RGE (\ref{eq:RGE-mj}) we can express the running of the mass-parameter
$\,m_j^{}\,$ from the scale $\mu_0^{}$ to $\,\mu\,$, to good accuracy,
\beqs
\label{eq:RG-Run-mj}
\beqa
m_j^{}(\mu) & = & \chi (\mu,\mu_0^{})\,m_j^{}(\mu_0^{})\,,
\\[2mm]
\chi (\mu,\,\mu_0^{}) & \simeq & \exp\[\f{1}{16\pi^2}\int_{0}^t \ahat(t')\,dt'\],
\eeqa
\eeqs
where $\,t=\ln(\mu/\mu_0^{})\,$.\,
For current study we will set,
$\,(\mu_0^{},\,\mu) = (M_Z,\,M_1)\,$,\,
where $Z$ boson mass $M_Z$ lies at the weak scale and the heavy mass $M_1$
represents the seesaw scale.

For the minimal neutrino seesaw with NH mass-spectrum, $\,0=\ma < \mb \ll \mc\,$,\,
we note that the zero-eigenvalue $\,\ma\,$ and the ratio $\,y\equiv\f{\mb}{\mc}\,$
do not depend on the RG running scale $\,\mu\,$.\,  So we can derive the running
of the two nonzero mass-parameters from weak scale to seesaw scale,
\beqs
\label{eq:RG-Run-m123}
\beqa
&& \mh_2^{} ~\equiv~ m_2^{}(M_1) ~=~ \chi_1^{}\, m_2^{}(M_Z) ~=~ y\,\mh_3^{}\,,
\\[0mm]
&& \mh_3^{} ~\equiv~ m_3^{}(M_1) ~=~ \chi_1^{}\, m_3^{}(M_Z)\,,
\\[0mm]
&& \chi_1^{} ~\equiv~ \chi (M_1,M_Z) \,,
\eeqa
\eeqs
and accordingly,
$\,\Delta_s(M_1)=\chi_1^2\Delta_s(M_Z)\,$ and $\,\Delta_a(M_1)=\chi_1^2\Delta_a(M_Z)\,$.\,
The RG running factor $\,\chi_1^{}\equiv \chi (M_1,M_Z)\,$ can be evaluated numerically,
and depends on the inputs of initial values for the weak gauge coupling
$\,\alpha_2^{}=g_2^2/(4\pi)\,$,\,
the top-quark Yukawa coupling $\,y_t^{}\,$ and the Higgs boson mass $\,M_H\,$, via the combination
$\,\ahat\,$ in Eq.\,(\ref{eq:RG-alpha}).  The existing electroweak data gives,
$\,\a_2^{-1}(M_Z)=29.57\pm 0.02\,$,\, $\,m_t^{}=173.1\pm 1.4\,$GeV, and the Higgs mass
range $\,115\leqq M_H^{} \leqq 149\,$GeV\, for the SM at $90\%$\,C.L. \cite{PDG,mH-fit}.
So the universal running factor $\,\chi (M_1,M_Z)\,$ is found to be around
$\,1.3-1.4\,$ for $\,M_1 = 10^{13}-10^{16}\,$GeV\, and $\,M_H = 115-149\,$GeV.
In Sec.\,\ref{sec:diagonalization}-\ref{sec:solution}
this running effect will be numerically computed for our analyses.
Other running effects due to the leptonic mixing angles and CP-phases are all negligible for the
present study since their RGEs contain only flavor-dependent terms and are all suppressed by
$\,y_\tau^2 =\O(10^{-4})\,$ at least\,\cite{nuRG}.

For the analyses  in Sec.\,\ref{sec:diagonalization}-\ref{sec:solution},
we will first run the low energy neutrino parameters in the current
reconstruction formalism up to the seesaw scale $\,M_1\,$, and then match them
with our model-predictions.  The relevant places for including such RG effects are just to replace
the light neutrino mass-eigenvalues $\,(m_2^{},\,m_3^{})\,$ at the low scale by
$\,(\mh_2^{},\,\mh_3^{})\,$ at the seesaw scale $\,M_1\,$.\,

\newpage
%\vspace{4mm}
\section{%\hspace*{-5.7mm}.\hspace*{-0.6mm}
Predictions of Neutrino Seesaw with Common Soft $\boldsymbol{\mutau}$ \\
\hspace{5mm} and CP Breaking} \label{sec:diagonalization}

 In this section we systematically derive the predictions of our common soft $\mutau$ and CP breaking
 model (Sec.\ref{sec:common-origin}) for the low energy neutrino observables by using the reconstruction
 formalism (including the RG running effects) in Sec.\,\ref{sec:general}.
 Especially, we will analyze the {\it correlation} between the two small $\mutau$ breaking parameters
 $\,\d_x \(\equiv \theta_{13} - 0\)\,$ and $\,\d_a \(\equiv\theta_{23}-\f{\pi}{4}\)\,$.\,
 Other correlations of $\d_x$ with the Jarlskog invariant $J$ or the $0\nu\beta\beta$ decays
 parameter $|m_{ee}^{}|$ are also analyzed.

\vspace*{3mm}
\subsection{Correlation between
            $\boldsymbol{\theta_{13}}$ and $\boldsymbol{\theta_{23}\!-\!45^\deg}$}
%\label{sec:dx-da}
\label{sec:prediction-low-energy-parameter}

In this subsection, we first analyze a general feature of the NLO mass-matrix $M_\nu^{(1)}$,
which predicts $\,\dx\propto\da\,$ and imposes a {\it lower bound} on $\,\dx\,$
for any nonzero $\,\da\,$, independent of all details of the $\mutau$ and CP breakings.

Inspecting the reconstruction formula (\ref{eq:Reconstruct-dMnu-sa}),
we see that $\,\d_x$\, and $\,\d_a\,$ are only contained in the
$\mutau$ anti-symmetric parts $\d B_a$ and $\d C_a$, respectively.
Using these two formulas we compute their ratio,
\beqa
\label{eq:dBa-dCa-abc}
  \f{\delta B_a}{\delta C_a} &=&
  \f{\sqrt{2}\d_x e^{-i(\ab_{10}^{}-\ab_{20}^{}+\d_D)}}
    {2\d_a  - i(\d\ab_2^{} - \d\ab_3^{})} \,.
\eeqa
On the other hand, from Eq.\,(\ref{eq:dBa/dCa-2}) [cf.\ also Eq.\,(\ref{eq:dBa/dCa})],
our soft breaking model predicts this ratio to be real,
\beqa
\label{eq:dBa-dCa-k}
\f{\delta B_a}{\delta C_a}
~=~ \f{a}{\,b+c\,} ~\equiv~ \f{1}{k} \,,
\eeqa
which does not depend on any detail of the $\mutau$ and CP breakings.
According to our general proof (cf.\ Sec.\,\ref{subsec:solar-mixing-not-affected} for detail),
this ratio $\,k\,$ is related to the solar mixing angle $\,\ts\,$
via Eq.\,(\ref{eq:theta12-dd}).  So we can rewrite (\ref{eq:dBa-dCa-k}),
\beqa
\label{eq:dBa-dCa-th12}
\f{\delta B_a}{\delta C_a}
~=~ \f{p_k^{}}{\,\sqrt{2}\,}\tan\ts \,,
\eeqa
where $\,p_k^{}=\pm\,$ denotes the sign of $\,k\,$.\,
So, from (\ref{eq:dBa-dCa-k}) and (\ref{eq:dBa-dCa-th12})
we deduce two conditions,
\beqs
\beqa
\da  &=&  \dx\,\cot\ts\cos(\ab_{10}^{}-\ab_{20}^{}+\d_D)p_k^{} \,,
\label{eq:da-dx-1}
\\[2mm]
\d\ab_2^{}-\d\ab_3^{} &=& 2\tan(\ab_{10}^{}-\ab_{20}^{}+\d_D)\,\da \,.
\label{eq:dalpha23-da}
\eeqa
\eeqs
Eq.\,(\ref{eq:da-dx-1}) shows that at this order the two small $\mutau$ breaking parameters
are {\it proportional to each other,} $\,\dx \propto \da\,$.\,
Strikingly, due to \,$|\cos(\ab_{10}^{}-\ab_{20}^{}+\d_D)| \leqq 1$\,,\,
we can further derive from (\ref{eq:da-dx-1}) a generic {\it lower bound}
on $\,\delta_x$\,,\, for any nonzero $\da$,
\beqa
\label{eq:Lbound-dx}
  \delta_x &\geqq&
  |\delta_a|\tan\theta_s  \,,
\eeqa
where $\,\dx \equiv \theta_{13}\in [0,\f{\pi}{2}]\,$ in our convention.

We stress that the above derivation of (\ref{eq:da-dx-1})-(\ref{eq:Lbound-dx})
depends only on the ratio of two $\mutau$ anti-symmetric elements $\,\d B_a/\d C_a\,$
in (\ref{eq:dBa-dCa-abc}),
and our model predicts this ratio as $\,1/k\,$ in (\ref{eq:dBa-dCa-k}),
which originates from the Dirac mass-matrix $\mD$ alone and does not depend on
any detail of the $\mutau$ and CP breakings.
Hence, Eqs.\,(\ref{eq:da-dx-1}) and (\ref{eq:Lbound-dx}) reflect the general feature
of our model.

\vspace*{3mm}
\subsection{Full NLO Analysis for Low Energy Neutrino Observables}
\label{sec:full-NLO}

In our model, both $\mutau$ and CP violations arise from
a common origin in the seesaw Lagrangian, characterized by the unique soft breaking
parameter $\,\zeta e^{i\om}\,$ (appearing at the NLO of our perturbative analysis).
Thus, the small $\mutau$ breaking parameters $(\d_a,\,\d_x)$ as well as
the low energy Dirac and Majorana CP phases $(\d_D,\,\phi_{23}^{})$
in the light neutrino mass-matrix are controlled
by $\,\zeta\,$ and $\,\om\,$.\,
In this subsection, using the general reconstruction formalism (Sec.\,\ref{sec:general})
for diagonalizing the light neutrino mass-matrix at the NLO,
we will systematically derive predictions of our model for these
low energy observables and their correlations.

Let us first have an examination on the reconstructed LO mass-matrix $M_\nu^{(0)}$
in (\ref{eq:Mnu-00}).  Comparing the (\ref{eq:Mnu-00}) with our model prediction
(\ref{eq:Mnu0}) at the same order, we can deduce,
\begin{subequations}
\beqa
&&
   m_{10}^{} \,=\, 0 \,, \qquad
   m_{20}^{} \,=\, 0 \,, \qquad
  \mh_{30}^{} \,=\, \frac{2(b-c)^2}{|(2 - X) M_{10}|} \,,
  \label{eq:m123-0}
\\[2mm]
&&
  e^{i2\ab_{30}^{}}
~=~ p_r^{}  \frac{2-X}{|2 - X|} \,,
\label{eq:alpha-20}
\eeqa
 \label{eq:zeroth-b}
\end{subequations}
where we have evolved the relevant low energy mass-parameter in (\ref{eq:Mnu-00})
up to its value at the seesaw scale via (\ref{eq:RG-Run-m123}),
the symbol $\,p_r^{}=\pm\,$ denotes the sign of $r$ as before,
and $\,\ab_{30}^{}=\ab_{20}^{}\,$ due to
the $\mutau$ symmetric solution (\ref{eq:mutau-solution}).
The fact that $\,\mh_{30}^{}\,$ is the largest mass-eigenvalue at the LO
shows that light neutrinos pose NH mass-spectrum in our model.
Also, in this expansion the solar mixing angle $\theta_s$
is not fixed at the zeroth order, and it will be determined at the NLO below.

We then inspect the light neutrino mass-matrix $\d M_\nu^{(1)}$ at the NLO, as given by
our soft breaking model in Eqs.\,(\ref{eq:dMu=s+a})-(\ref{eq:definition-EE1}) and
by our reconstruction formulas in Eqs.\,(\ref{eq:dMnu-1})-(\ref{eq:Reconstruct-dEs-dEs'}).
We can match the two sets of formulas for the $\mutau$ symmetric elements,
\begin{subequations}
  \begin{eqnarray}
   \d A ~=~ \f{a^2}{M_{10}}\, r  & = &
           \mh_{30}^{}\,y\,s_s^2\,
            e^{-i2(\ab_{10}^{} + \phi_{23}^{})} \,,
    \label{eq:re-delta-equality-A}
  \\[2mm]
   \d B_s ~=~ \f{\,a(b\!+\!c)\,}{2 M_{10}}\,r  & = &
  - \f{1}{\sqrt{2}\,} \mh_{30}^{} \,y\,s_sc_s\,
    e^{-i(\ab_{10}^{} + \ab_{20}^{} + 2\phi_{23}^{})} \,,
    \label{eq:re-delta-equality-Bs}
  \\[2mm]
  \d E_s ~=~  \frac {\,(b \!+\! c)^2\,}{4 M_{10}}\, r
  & = &
    \f{1}{2} \mh_{30}^{}\,y\,c_s^2\,
    e^{-i2(\ab_{20}^{} + \phi_{23}^{})} \,,
    \label{eq:re-delta-equality-Es}
  \\[2mm]
  \d E_s' ~=~ \f{(b \!-\! c)^2 X^2}{\,4(2 \!-\! X)^2 M_{10}\,}\, r
  & = &
    \f{1}{2}\mh_{30}^{} e^{-i2\ab_{20}^{}}
    \[ z - i(\d\ab_2^{} + \d\ab_3^{})
    \],
    \label{eq:re-delta-equality-Es'}
  \end{eqnarray}
  \label{eq:re-delta-equality-s}
\end{subequations}
and for $\mutau$ anti-symmetric elements,
\beqs
\beqa
 \dis\d B_a  ~=\, -\frac{a(b-c)X}{\,2(2\!-\!X)M_{10}\,}\;r
 &=&
 \f{1}{\sqrt{2}\,} \mh_{30}^{}\d_x\, e^{-i(\ab_{10}^{}+\ab_{20}^{}+\d_D)} \,,
 \label{eq:dBa}
 \\[2mm]
 \dis\d C_a  ~=\, -\f{(b^2\!-\!c^2)X}{\,2(2\!-\!X)M_{10}\,}\;r
 &=&
 \hf \mh_{30}^{}e^{-i2\ab_{20}^{}}
 \[ 2\d_a -i(\d\ab_2^{}-\d\ab_3^{})\] ,
 \label{eq:dCa}
\eeqa
\label{eq:dBa-dCa}
\eeqs
where we have run the low energy mass-parameter $\,m_{30}^{}\,$
up to $\,\mh_{30}^{}\,$ at the seesaw scale $\,M_1\,$
via Eq.\,(\ref{eq:RG-Run-m123}).

We first analyze the $\mutau$ symmetric
Eqs.\,(\ref{eq:re-delta-equality-A})-(\ref{eq:re-delta-equality-Es'}).
Inspecting the model predictions for $\,(\d A,\,\d B_s,\,\d E_s)\,$
on the left-hand-sides (LHS's)
of (\ref{eq:re-delta-equality-A})-(\ref{eq:re-delta-equality-Es}),
we observe that $\,(\d A,\,\d B_s,\,\d E_s)\,$ are all real.
This requires the RHS's of (\ref{eq:re-delta-equality-A})-(\ref{eq:re-delta-equality-Es})
to have vanishing imaginary parts. We thus derive,
\beqa
\label{eq:a10a20ph23-1}
2(\ab_{10}^{} + \phi_{23}^{}) ~=~ n_1^{}\pi\,,~~~~~
\ab_{10}^{} + \ab_{20}^{} + 2\phi_{23}^{} ~=~ n_2^{}\pi\,,
\eeqa
where \,$n_1^{}\pi,\,n_2^{}\pi=0,\pi$\,(mod $2\pi$).\,
From these we have,
%
%\beqs
\beqa
\ab_{10}^{} + \phi_{23}^{} ~=~ \f{n_1^{}\pi}{2} \,,~~~~~
%\label{eq:a20+phi23}
%\\[1.5mm]
\ab_{10}^{} - \ab_{20}^{} ~=~ n\pi \,,
%\label{eq:a10-a20}
\label{eq:a10a20ph23-2}
\eeqa
%\eeqs
%
where $\,n\pi=(n_1^{}\!-\!n_2^{})\pi=0,\,\pi$~(mod $2\pi$).
Then, we compute the ratio of (\ref{eq:re-delta-equality-A})
and (\ref{eq:re-delta-equality-Bs}), and arrive at
\begin{equation}
  \tan\ts ~=~ (-)^{n+1}\f{\,\sqrt{2}a\,}{\,b + c\,} \,,
  \label{eq:t12}
\end{equation}
where we can always ensure $\,\tan\ts >0\,$, i.e., we set $\,n=1\,$ for $\,a/(b\!+\!c)>0\,$,\,
and $\,n=0\,$ for $\,a/(b+c)<0\,$.\,
We note that due to the NH mass-spectrum, we have a small parameter $\,r=\O(m_2^{}/m_3^{})\,$
in our present model [cf. (\ref{eq:LO-Mass}) with $\pb =+$]
and a corresponding small parameter $\,y\equiv m_2^{}/m_3^{}\,$ in our reconstruction
formalism [cf. (\ref{eq:definition-y})]. As a result of expanding $r$ and $y$ in the NH scheme,
it is clear that the solar mixing angle $\theta_s$ is not fixed from the LO mass-matrix
$M_\nu^{(0)}$ in (\ref{eq:Mnu0}) or (\ref{eq:Mnu-00}), but determined at the NLO as above.
(In Sec.\,\ref{subsec:solar}-\ref{subsec:solar-mixing-not-affected}
we will generally prove that the $\ts$ formula (\ref{eq:t12}) actually does not
depend on the inclusion of $\mutau$ and CP violations at the NLO.)

Comparing (\ref{eq:t12}) with the $\,\tan\ts\,$
in (\ref{eq:dBa-dCa-k})-(\ref{eq:dBa-dCa-th12}),
we can fix the sign of $\,k\,$ as $\,p_k^{}=(-)^{n+1}\,$.\,
With the second relation of (\ref{eq:a10a20ph23-2}),
we can further simplify (\ref{eq:da-dx-1})-(\ref{eq:dalpha23-da}) as
\beqs
\beqa
\f{\da}{\dx}   &=& \!\! -\cot\ts \cos\d_D \,,
\label{eq:da-dx-2}
\\[2mm]
\d\ab_2^{}-\d\ab_3^{} &=& 2\tan\d_D\,\da \,,
\label{eq:dalpha23-da-f}
\eeqa
\eeqs
where the sign of (\ref{eq:dalpha23-da-f}) does not depend on the choice of $\,n$\,.\,
Since our convention already ensures $\,\cot\ts\,$ and $\,\dx\,(\equiv\sin\tx)\,$
to be both positive, the sign of the deviation $\da$ just equals the sign of $\,-\cos\d_D\,$.\,

With Eqs.\,(\ref{eq:m123-0}) and (\ref{eq:a10a20ph23-1})-(\ref{eq:a10a20ph23-2}),
we can further solve from
Eqs.\,(\ref{eq:re-delta-equality-A})-(\ref{eq:re-delta-equality-Es'}),
\beqs
  \beqa
    a^2  & = &
    (-)^{n_1^{}}s_s^2 \mh_{30}^{} M_{10}Y \,,
    \label{eq:expression-abc-1}
  \\[2mm]
    a(b+c)  & = &
  (-)^{n_2^{}\!+\!1} \sqrt{2}\, s_s c_s \mh_{30}^{} M_{10}Y \,,
    \label{eq:expression-abc-3}
  \\[2mm]
    (b+c)^2  & = &
  (-)^{n_1^{}} 2c_s^2 \mh_{30}^{}M_{10}Y \,,
    \label{eq:expression-abc-2}
  \\[2mm]
    (b - c)^2  & = &
    \f{1}{2} |(2 - X)M_{10}| \mh_{30}^{} \,,
    \label{eq:expression-abc-4}
  \eeqa
  \label{eq:expression-abc}
\eeqs
where for convenience we have defined a ratio,
\begin{eqnarray}
\label{eq:Y-def}
Y \,\equiv\, \dis\f{y}{r} \,=\, \O(1)
\,.
\label{eq:definition-Y}
\end{eqnarray}
Since $\,M_{10}Y=yM_{22} > 0\,$ and $\,a^2,\,(b+c)^2>0\,$,\, we deduce,
$\,n_1^{}=0\,$, which further leads to $\,n_2^{}=-n\,$
[cf.\ the comment below (\ref{eq:a10a20ph23-2})].
With these we can solve the expressions for $\,(a,\,b,\,c)$\,,
\begin{subequations}
\begin{eqnarray}
  a     &=&
(-)^{n_2^{}+1}p_+^{} s_s \sqrt{\mh_{30}^{} M_{10}Y\,}  \,,
  \label{eq:expression-a-new}
\\[2mm]
  b + c &=&
  p_+^{}  c_s \sqrt{2 \mh_{30}^{} M_{10}Y\,} \,,
  \label{eq:expression-b+c-new}
\\[2mm]
  b - c &=&
  p_-^{} \sqrt{\hf\left|(2 - X)M_{10}\right|\mh_{30}^{}\,} \,,
  \label{eq:expression-b-c-new}
\end{eqnarray}
\label{eq:expression-abc-new}
\end{subequations}
where $\,p_\pm =\pm\,$ denotes the signs of $\,b \pm c\,$.\,
Combining Eqs.\,(\ref{eq:zeroth-b}) and (\ref{eq:re-delta-equality-Es'}), we derive,
\begin{equation}
\f{X^2}{\,4(2 \!-\! X)\,}\,r
~=~
z - i \,\( \d\ab_2^{} + \d\ab_3^{} \) \,,
\end{equation}
which can be decomposed into real and imaginary parts,
\begin{subequations}
  \beqa
  z  & = &
    \f{|X|^2}{\,4|2 \!-\! X|^2\,}
    \( 2\cos 2\omega - |X|\cos\omega \)\, r \,,
  \\[1.5mm]
    \d\ab_2^{} + \d\ab_3^{}
  & = &
  -
    \f{|X|^2}{\,4|2 \!-\! X|^2\,}
    \( 2\sin 2\omega - |X|\sin\omega \)\, r \,.
  \eeqa
\end{subequations}

Next, we analyze the $\mutau$ anti-symmetric equations (\ref{eq:dBa})-(\ref{eq:dCa})
for $\,\d M_\nu^{(1)}$.\,
With Eqs.\,(\ref{eq:alpha-20}), (\ref{eq:expression-abc-new})
and (\ref{eq:a10a20ph23-1})-(\ref{eq:a10a20ph23-2}),
we can rewrite (\ref{eq:dBa})-(\ref{eq:dCa}) as
\begin{subequations}
  \begin{eqnarray}
    \f{\,p_+^{}p_-^{}(-)^{n_2^{}-n}\,}{2} \left|\f{X^2Y}{2\!-\!X}\right|^{\hf}
    e^{i(\om + \d_D)}
    s_s\, r
  & = &
    \delta_x \,,
    \label{eq:dB'-new}
  \\[3mm]
  - p_+^{}p_-^{}\left|\f{X^2Y}{2\!-\!X}\right|^{\hf}e^{i\omega}\, c_s\, r
  & = &
      2 \da - i (\d\ab_2^{} - \d\ab_3^{}) \,.
    \label{eq:dC'-new}
  \end{eqnarray}
  \label{eq:dB'C'-new}
\end{subequations}
Taking the ratio of the two sides of (\ref{eq:dB'-new})-(\ref{eq:dC'-new}) and
making use of (\ref{eq:dalpha23-da-f}), we deduce
\beqa
\f{\da}{\dx} &=& (-)^{n-n_2^{}+1}\cot\ts\cos\d_D \,,
\eeqa
which is consistent with Eq.\,(\ref{eq:da-dx-2}) due to
$\,n_2^{}=-n\,$ as we derived below Eq.\,(\ref{eq:expression-abc}).
Then, we may rewrite (\ref{eq:a10a20ph23-2}) as
\beqa
\ab_{10}^{} ~=~ -\phi_{23}^{}\,, &~~&
\ab_{20}^{} ~=~ -n\pi -\phi_{23}^{} \,,
\label{eq:a10-a20-phi23-solve}
\eeqa
where we have used $\,n_1^{}=0\,$ derived below (\ref{eq:expression-abc}).
With this we can derive, from (\ref{eq:alpha-20}), the Majorana phase angle $\,\phi_{23}^{}\,$
in terms of the original soft CP-breaking phase angle $\,\om$\,,
\beqa
\tan 2\phi_{23}^{} &=& \f{\zeta\sin\om}{\,2r-\zeta\cos\om\,} \,.
\label{eq:phi23}
\eeqa

Then, from Eq.\,(\ref{eq:dB'C'-new}) we finally solve,
\begin{subequations}
  \begin{eqnarray}
    \delta_D  & = &
  2\pi - \omega  \,,
    \label{eq:sol-deltaD'}
  \\[2mm]
    \dx  & = &
    \hf
    \left|\f{X^2Y}{2\!-\!X}\right|^{\hf}\! s_s|r| \,,
    \label{eq:sol-deltax'}
  \\
    \da & = &
   - \hf\left|\f{X^2Y}{2\!-\!X}\right|^{\hf}\!\cos\d_D\,c_s |r| \,,
    \label{eq:sol-deltaa'}
  \\
    \d\ab_2^{} - \d\ab_3^{}  & = &
    -\left|\f{X^2Y}{2\!-\!X}\right|^{\hf}\!\sin\d_D\,c_s |r| \,,
  \end{eqnarray}
  \label{eq:NLO-Solution'}
\end{subequations}
where we have defined the phase angles $\,\om,\,\d_D\in [0,\,2\pi)\,$ and the mixing angles
$\,(\ts,\,\ta,\,\tx )\in [0,\,\f{\pi}{2}]$,\, which requires the sign product
$\,p_+^{}p_-^{}p_r^{} = +\,$ for consistency.
From (\ref{eq:NLO-Solution'}) we can reproduce the relations
(\ref{eq:da-dx-2})-(\ref{eq:dalpha23-da-f}) in Sec.\,\ref{sec:prediction-low-energy-parameter}.
Furthermore, using (\ref{eq:sol-deltax'})-(\ref{eq:sol-deltaa'})
we can explicitly express $\,\dx\,$ and $\,\da\,$ in terms of $\,r\,$ and $\,\zeta$\,,
\beqs
\beqa
  \dx  &=&
  \f{\sqrt{y}\,s_s\,\zeta}{2\[\zeta^2 - 4 r \zeta \cos\d_D + 4 r^2\]^{\f{1}{4}}} \,,
  \label{eq:predict-dx-zeta}
\\[2mm]
 \da   &=&
  \f{-\sqrt{y}\,c_s\,\cos\d_D\,\zeta}
    {2\[\zeta^2 - 4 r \zeta \cos\d_D + 4 r^2\]^{\f{1}{4}}} \,,
  \label{eq:predict-da-zeta}
\eeqa
\label{eq:predict-da-zeta-all}
\eeqs
where  $\,y \equiv \f{m_2}{m_3} \simeq \sqrt{{\Delta_s}/{\Delta_a}}\,$,\,
$\,\d_D\in [0,\,2\pi)\,$ and $\,0 < \zeta <1\,$ in our convention.
Since
$\,\zeta^2 - 4 r \zeta\cos\d_D + 4r^2 = (2r-\zeta\cos\d_D)^2+(\zeta\sin\d_D)^2 >0\,$,\,
so the denominators in (\ref{eq:predict-da-zeta-all}) are well-defined.

It is useful to note that reversing (\ref{eq:predict-dx-zeta})
we may resolve the seesaw parameter $r$ as a function of $\,\zeta$\, and $\,\d_D$\,,
\beqa
\label{eq:rr}
r ~=~ \f{\zeta}{2}\[
\cos\d_D \pm \sqrt{\f{s_s^4}{16}\f{y^2\zeta^2}{\dx^4} - \sin^2\d_D\,} \,
\] ,
\eeqa
where the parameters
$\,s_s \,(=\sin\theta_{12})\,$,\,
$\,y\,\(=\f{m_2^{}}{m_3^{}}\simeq \sqrt{\Delta_s/\Delta_a}\,\)\,$ and $\,\dx\,(=\theta_{13})\,$
have been measured by the oscillation experiments (Table-\ref{tab:1}).
The square-root in (\ref{eq:rr}) has to be real, so we have the condition,
\beqa
\label{eq:LowerBound-zeta}
\zeta ~\geqq~ \f{4\,\d_x^2}{\,s_s^2\,y\,}|\sin\d_D| \,,
\eeqa
which will put a lower bound on \,$\zeta$\, for nonzero \,$\dx$\, and \,$\sin\d_D$\,.

\begin{figure}[t]
  \centering
  \includegraphics[width=8.1cm,clip=true]{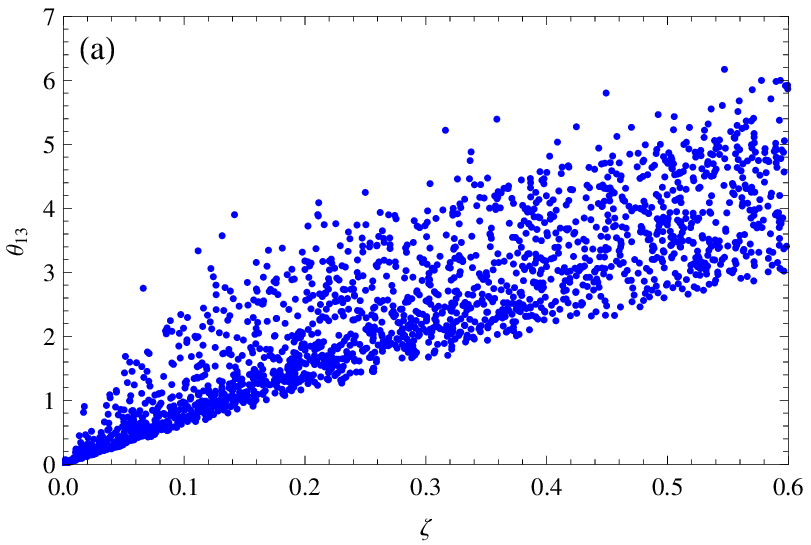}
  \includegraphics[width=8.2cm,clip=true]{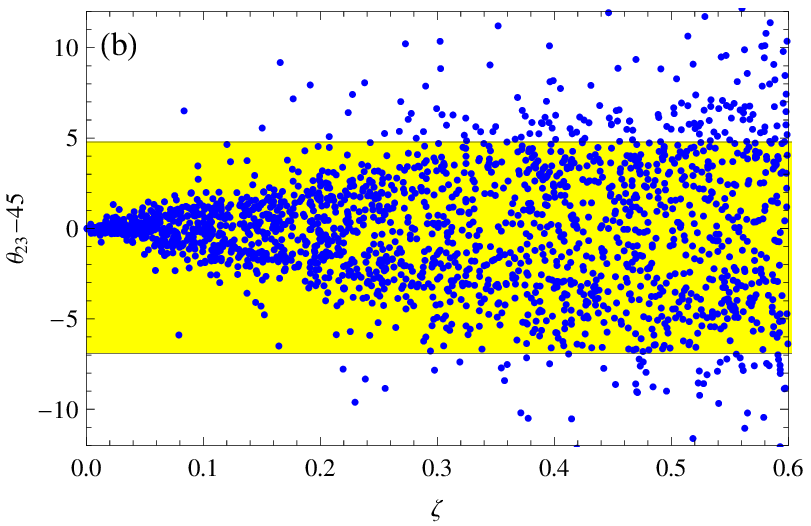}
  \includegraphics[width=8.1cm,clip=true]{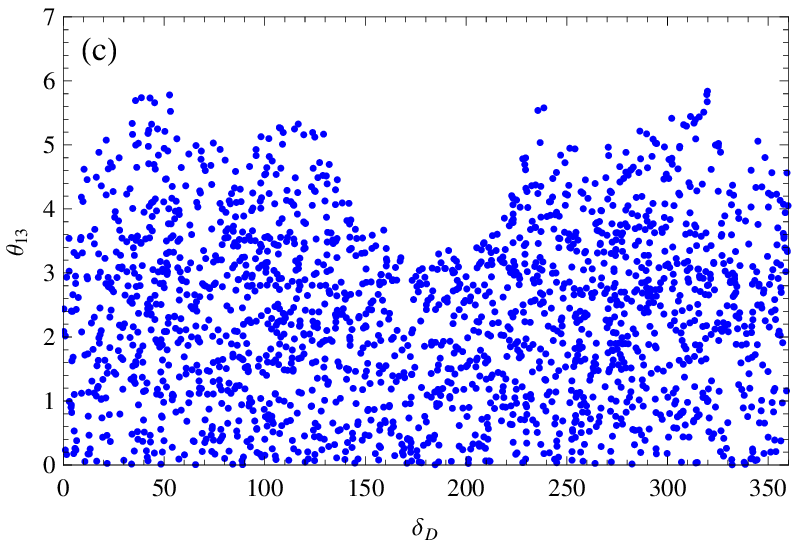}
  \includegraphics[width=8.2cm,clip=true]{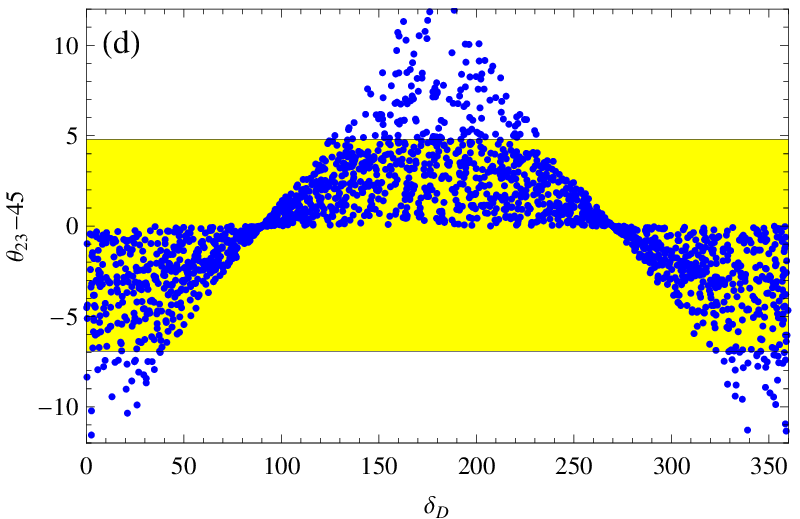}
  \caption{The predictions of $\theta_{13}$ and $\theta_{23}-45^\deg$
  as functions of the soft $\mutau$ breaking parameter
  $\zeta$ and CP-phase angle $\d_D$,
  are shown in the upper plots (a)-(b) and lower plots (c)-(d), respectively.
  The experimental inputs are scanned within 90\%\,C.L. ranges
  and the Dirac phase angle $\,\d_D\in [0^\deg,\,360^\deg)$\,,\, with 2000 samples.
  The shaded region (yellow) denotes the current bounds ($90\%$\,C.L.) on
   $\theta_{23}-45^\deg$ according to Table-\ref{tab:1}.}
  \label{fig:dx-ta-zeta}
\end{figure}

Inputting the neutrino data for $\ts$ and $(\Delta_s,\,\Delta_a)$ in
Table-\ref{tab:1} and scanning the phase angle
$\,\d_D \in [0^\deg ,\, 360^\deg )$\,,\,
we can plot $\,\theta_{13}\,(\equiv\dx)$\,
and \,$\theta_{23} -45^\deg\,(\equiv\da)$,\,
from (\ref{eq:predict-dx-zeta})-(\ref{eq:predict-da-zeta}),
as functions of the parameters $\,\zeta\,$ and \,$\d_D$\,.
This is shown in Fig.\,\ref{fig:dx-ta-zeta} with the experimental inputs varied
within 90\%\,C.L. ranges and with the natural regions of $\,|r|,\zeta\in [0,\,0.6]$\,.\,
Inspecting Fig.\,\ref{fig:dx-ta-zeta}a and Fig.\,\ref{fig:dx-ta-zeta}c,
we see that for $\,\zeta\in [0,\,0.6]\,$,\,  an upper bound
$\,\theta_{13} \lesssim 6^\deg\,$ holds for most of the parameter space.
Fig.\,\ref{fig:dx-ta-zeta}b  depicts how
the deviation $\,\theta_{23} -45^\deg\,$ behaves as a function of $\,\zeta\,$
via Eq.\,(\ref{eq:predict-da-zeta}) in which all measured observables are
constrained by the $90\%$\,C.L. data and
the variable $\,\zeta\,$ needs to ensure that $\,\theta_{13}\,(\dx)\,$
obeys the current oscillation limits (Table-\ref{tab:1}) via Eq.\,(\ref{eq:predict-dx-zeta}).
As a result, we see that both Fig.\,\ref{fig:dx-ta-zeta}a and Fig.\,\ref{fig:dx-ta-zeta}b
confined our parameter space into $\,\zeta \gtrsim 0.1\,$ region.
We further note that in Fig.\,\ref{fig:dx-ta-zeta}d there is a positive peak-region of
$\,\theta_{23} -45^\deg\,$ for the parameter space
$\,120^\deg\lesssim\d_D\lesssim 220^\deg\,$;\,
but Fig.\,\ref{fig:dx-ta-zeta}c requires that around the same range of $\,\d_D\,$
the $\,\theta_{13}\,$ is much suppressed, and the two peak-regions of $\,\theta_{13}\,$
in Fig.\,\ref{fig:dx-ta-zeta}c around $\,\d_D\sim 50^\deg\,$ and $\,\d_D\sim 310^\deg\,$
just correspond to the negative deviation, $\,\theta_{23} -45^\deg <0\,$.\,

\begin{figure}[t]
\centering
\includegraphics[width=14cm,clip=true]{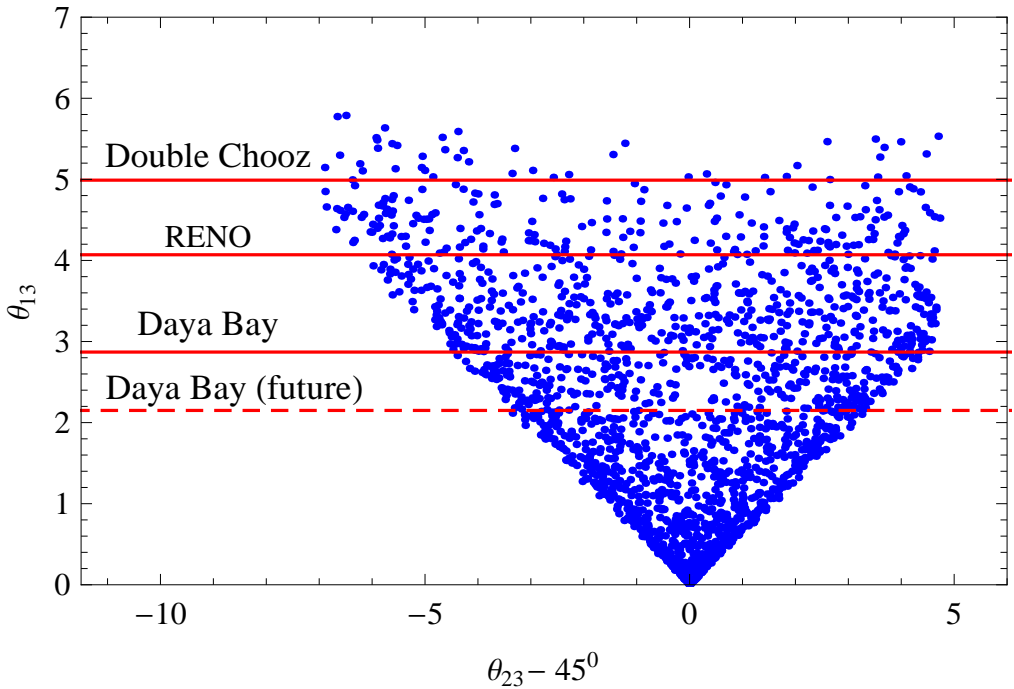}
  \caption{The correlation between $\theta_{13}$ and
  $\theta_{23}-45^\deg$, based on Eqs.\,(\ref{eq:predict-dx-zeta})-(\ref{eq:predict-da-zeta}),
  where we have scanned the experimental inputs within 90\%\,C.L. ranges
  and the Dirac-phase angle $\,\d_D\in [0^\deg,\,360^\deg)$\,,\, with 2000 samples.
  The sensitivities of
  Double Chooz\,\cite{2CHOOZ}, RENO\,\cite{RENO} and Daya Bay\,\cite{DayaBay} experiments
  to $\,\theta_{13}$\, are shown by the three solid horizontal lines at 90\%\,C.L.,
  as $5.0^\deg$, $4.1^\deg$ and $2.9^\deg$
  (from top to bottom), based on three years of data-taking; the dashed horizontal line gives
  Daya Bay's future sensitivity ($2.15^\deg$) after six years of running\,\cite{Lindner}.
  }
  \label{fig:dx-da}
\end{figure}

With (\ref{eq:predict-dx-zeta})-(\ref{eq:predict-da-zeta}) we can plot the {\it correlation}
between the two $\mutau$ breaking mixing angles $\theta_{13}$ and $\,\theta_{23} -45^\deg\,$.\,
This is shown in Fig.\,\ref{fig:dx-da}, where we have varied the measured parameters within
their 90\%\,C.L. ranges and the Dirac-phase angle $\,\d_D\in [0^\deg,\,360^\deg)$\,, as well as
$\,\zeta,r\in [0,\,0.6]\,$.\,
The current 90\%\,C.L. limits on $\theta_{13}$ are shown by the shaded region (yellow),
while the $\theta_{13}$ sensitivities of the upcoming
Double Chooz\,\cite{2CHOOZ}, RENO\,\cite{RENO} and Daya Bay\,\cite{DayaBay} experiments to
are shown by the three horizontal (red) lines at 90\%\,C.L.,
as $5.0^\deg$, $4.1^\deg$ and $2.9^\deg$
(from top to bottom), based on three years of data-taking.
From Fig.\,\ref{fig:dx-da}, we see that our model generally predicts
$\,\theta_{13}\lesssim 6^\deg\,$;\, the Double Chooz experiment\,\cite{2CHOOZ} could marginally
probe the predicted $\theta_{13}$ and the RENO experiment\,\cite{RENO} can be a better job.
The NO$\nu$A experiment\,\cite{NOvA} is also expected to probe $\theta_{13}$, with a sensitivity
of $\,\theta_{13}\simeq 4.4^\deg\,$, slightly lower than RENO.
It is clear that {\it the Daya Bay experiment\,\cite{DayaBay} holds the best sensitivity and can probe
the large central region of our predicted parameter space.}
(For more details of the experimental sensitivities
to $\,\theta_{13}\,$,\, see a nice comparative analysis in \cite{Lindner}.)

We note that the sharp edges on the two sides of the allowed parameter space
in Fig.\,\ref{fig:dx-da} are essentially determined by the lower bound given in
(\ref{eq:Lbound-dx}), $\,\dx \geqq |\da|\tan\ts\,$,\,
where the measured parameter $\,\tan\ts\sim 0.67\,$ (Table-\ref{tab:1}) just corresponds to the slopes
of the sharp edges which are nearly straight lines.
This means that for {\it any measured value of
$\,\theta_{23}-45^\deg \neq 0\,$,\, the Fig.\,\ref{fig:dx-da}
imposes a lower bound on $\theta_{13}$,} which will be probed
by the $\,\theta_{13}\,$ experiments such as Daya Bay and RENO.
This is a really interesting and encouraging
prediction for the upcoming neutrino oscillation experiments which
will probe the $\mutau$ violating observables $\,\theta_{13}\!-\!0^\deg\,$
and $\,\theta_{23}\!-\!45^\deg\,$ to much higher precision. Note that the current
oscillation data already favor the central values of $\theta_{23}$
to be smaller than $45^\deg$ (Table-\ref{tab:1}) and this feature
is quite robust (cf. footnote-\ref{footnote-2} and Ref.\,\cite{D23-Smirnov}).
Hence, {\it our findings strongly encourage the experimental efforts (such as
MINOS\,\cite{MINOS} and T2K\,\cite{T2K}) to further probe
$\,\theta_{23}\,$ as precise as possible}, hopefully to the similar level of
accuracy as the present $\,\theta_{12}\,$ measurement.

\vspace*{2mm}
Next, we analyze our model predictions for the low energy CP-violation
(via Jarlskog invariant) and the neutrinoless double beta decays (via
the element $|m_{ee}|$ of $M_\nu$).
In our construction (Sec.\,\ref{sec:common-origin}), the original soft breaking CP-phase $\,e^{i\om}\,$
is the source of both low energy Dirac and Majorana CP-violations via the
phase angles $\,\d_D\,$ and $\,\phi_{23}^{}$\,.

The Dirac CP-phase $\d_D$ will manifest itself in Jarlskog invariant
$J$ while both $\d_D$ and the Majorana CP-phase $\phi_{23}^{}$ will appear
in the neutrinoless double beta decay observable $\,|m_{ee}^{}|$\,.\,
The Jarlskog invariant $J$ can be written as\,\cite{J},
\beqa
\label{eq:J}
J  ~\equiv~
\f{1}{8}\sin2\theta_s \sin2\theta_a \sin2\theta_x \cos2\theta_x \sin \delta_D
~=~  \frac{\dx}{4}\sin2\theta_s\sin\delta_D +\O(\dx^2,\da^2)\,,~~~~
\eeqa
where as defined earlier, $\,\dx\equiv\theta_x\,$ and $\,\da\equiv\ta -\f{\pi}{4}\,$.

Then we analyze the neutrinoless double-beta decay. Our model predicts the NH
mass-spectrum with $m_1^{}=0$, so from (\ref{eq:NH-m-ee})
we can derive the mass-matrix element $|m_{ee}^{}|$
for neutrinoless double beta decays,
\begin{eqnarray}
  M_{ee} &\equiv&
  \left| m_{ee}^{} \right| ~=~
  \left| \sum {V_{ej}^*}^2 m_j^{} \right|
 \nn\\[2mm]
  &=&
  m_3^{}\left|
    y\,s^2_s c^2_x  e^{-i2\phi_{23}^{}} + s^2_x e^{-i2\delta_D} \right|
 \nn\\[2mm]
 & \simeq &
  m_3^{}  \sqrt{
    y^2 s_s^4
  + 2 y\d_x^2 s_s^2\cos2(\d_D\!-\!\phi_{23}^{})
  + (\d_x^4 - 2s_s^4y^2\d_x^2)\,}  ~.
  \label{eq:Mee}
\end{eqnarray}
\begin{figure}[t]
\hspace*{-4mm}
\includegraphics[width=8.3cm,clip=true]{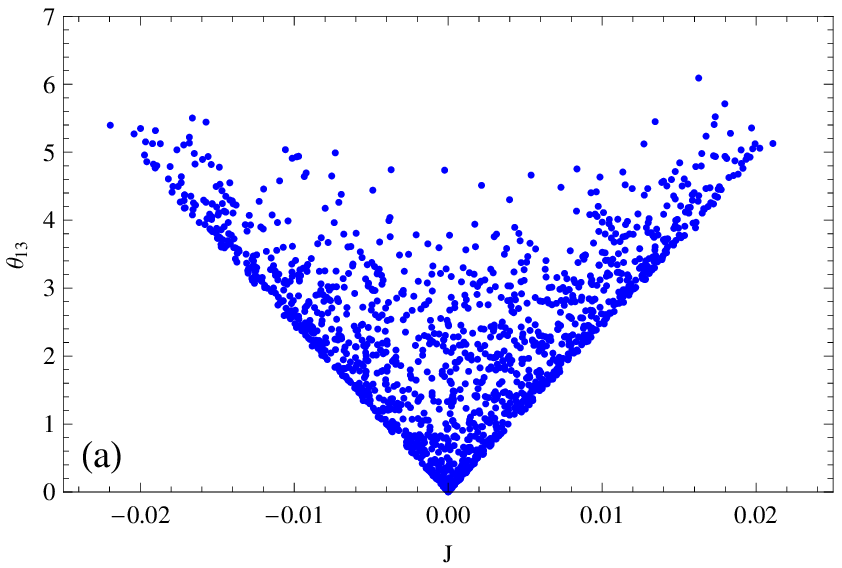}
\hspace*{2mm}
\includegraphics[width=8.2cm,clip=true]{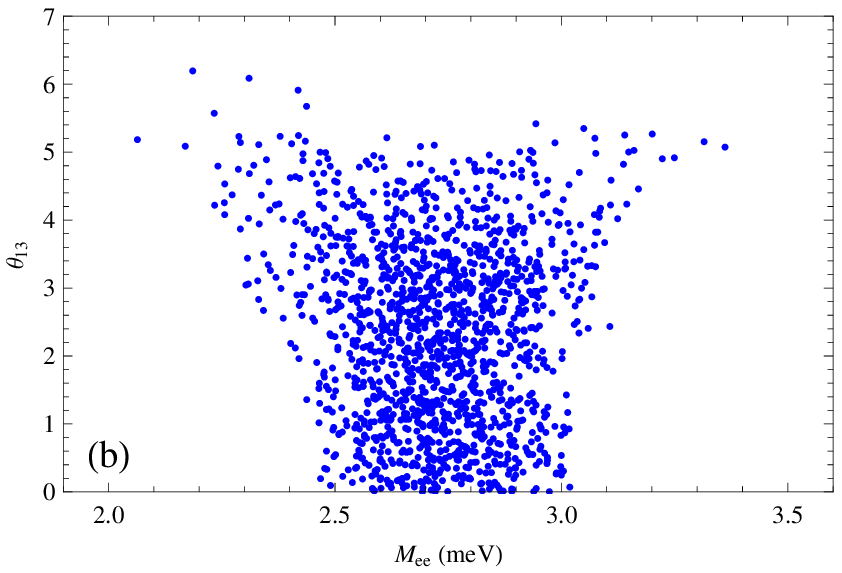}
  \caption{Correlations of  $\theta_{13}$ (in degree) with the Jarlskog invariant
           $J$ [plot-(a)] and with the neutrinoless-doublet-beta-decay observable
           $M_{ee}$ [plot-(b)]. Each plot has computed 1500 samples.
          % The shaded region (yellow) is allowed by the
          % current data at 90\%\,C.L.
          }
  \label{fig:dx-J-Mee}
\end{figure}

From Eqs.\,(\ref{eq:a10-a20-phi23-solve}),(\ref{eq:sol-deltaD'}) and
noting $\,\ab_{20}^{}=\ab_{30}^{}$\,,\,
we derive the phase-angle combination,
\beqa
(\d_D - \phi_{23}^{}) &=& (\ab_{30}^{}-\om) +(2+n)\pi \,.
\eeqa
Thus, we can compute, by using Eq.\,(\ref{eq:alpha-20}),
%
% \beqa
% e^{i2(\d_D-\phi_{23}^{})}
% ~=~ p_r^{}e^{-i2\om}\f{2\!-\!X}{\,|2\!-\!X|\,}
% \nn\\[3mm]
% ~=~ \f{(2r\cos2\om -\zeta\cos\om)-i(2r\sin 2\om -\zeta\sin\om)}
%      {\sqrt{\zeta^2 - 4 r \zeta \cos\d_D + 4 r^2\,}}\,,~~~~~
% \eeqa
%
%which results in
%
\beqa
\cos 2(\d_D\!-\!\phi_{23}^{}) &=&
\f{\,\Re\mathfrak{e}\[p_r^{}e^{-i2\om}(2\!-\!X)\]\,}{\,|2\!-\!X|\,}
\non\\[2mm]
&=&
\f{\,2r\cos 2\d_D \!-\!\zeta\cos\d_D\,}
  {\sqrt{\zeta^2 \!-\! 4 r \zeta \cos\d_D \!+\! 4 r^2\,}\,} \,,
\eeqa
where we have used the solution $\,\om = 2\pi-\d_D\,$ in (\ref{eq:sol-deltaD'}).

In Fig.\,\ref{fig:dx-J-Mee}a we show the correlation between
$\theta_{13}$ and the Jarlskog invariant $J$, and in Fig.\,\ref{fig:dx-J-Mee}b the
neutrinoless doublet beta decay observable $M_{ee}\,(\equiv |m_{ee}^{}|)$ depicted.
In plotting this figure, we have used Eq.\,(\ref{eq:predict-dx-zeta}) where
$\theta_{13}$ is taken as a function of $r$ and $\zeta$.  All other measured
parameters in the formulas of $\theta_{13}$, $J$ and $M_{ee}$ are scanned within
their 90\%C.L. ranges, while the Dirac CP-phase $\d_D$ is varied in $[0,\,2\pi)$.
The shaded region (yellow) in this figure is allowed by the current data at 90\%\,C.L.
From Fig.\,\ref{fig:dx-J-Mee}(a), we see that a nonzero $J$ can place
a lower bound on $\theta_{13}$ in our model.
This is because $\,|\sin\d_D|\leqq 1\,$ and thus we can deduce from Eq.\,(\ref{eq:J}),
\beqa
\label{eq:Lbound-t13-J}
\dx ~\geqq~ \f{4|J|}{~\sin 2\ts\,} \,.
\eeqa
Fig.\,\ref{fig:dx-J-Mee}(b) shows an upper bound
$\,M_{ee} \lesssim 3.4\,$meV which is quite small
(as commonly expected for schemes with NH mass-spectrum) and
poses a challenge to the future neutrinoless double beta decay experiments\,\cite{0nu2beta}.

\vspace*{3mm}
\subsection{Analysis of Seesaw Parameter Space}
\label{sec:Seesaw-Space}

In this subsection, we further analyze the allowed seesaw parameter space, which will also
be needed for our leptogenesis analysis in Sec.\,\ref{sec:solution}.

\begin{figure}[t]
\hspace*{-4mm}
\includegraphics[width=8.3cm,clip=true]{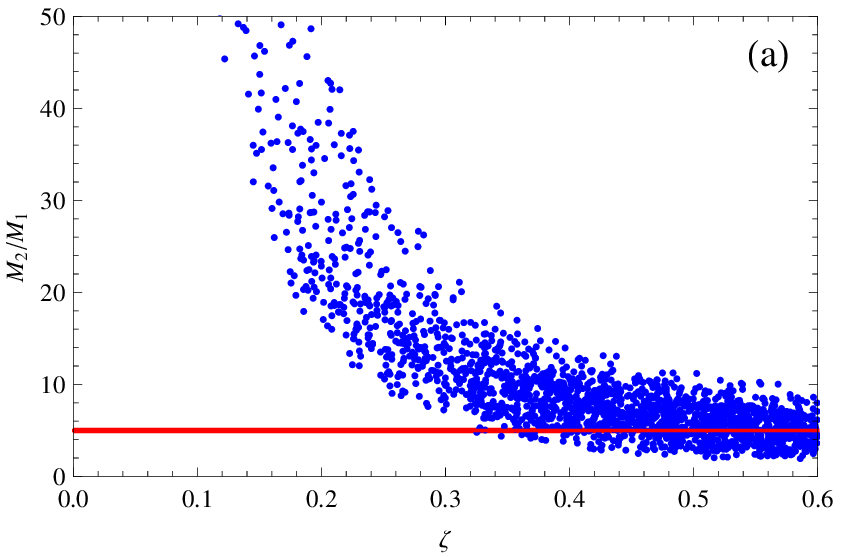}
\includegraphics[width=8.3cm,clip=true]{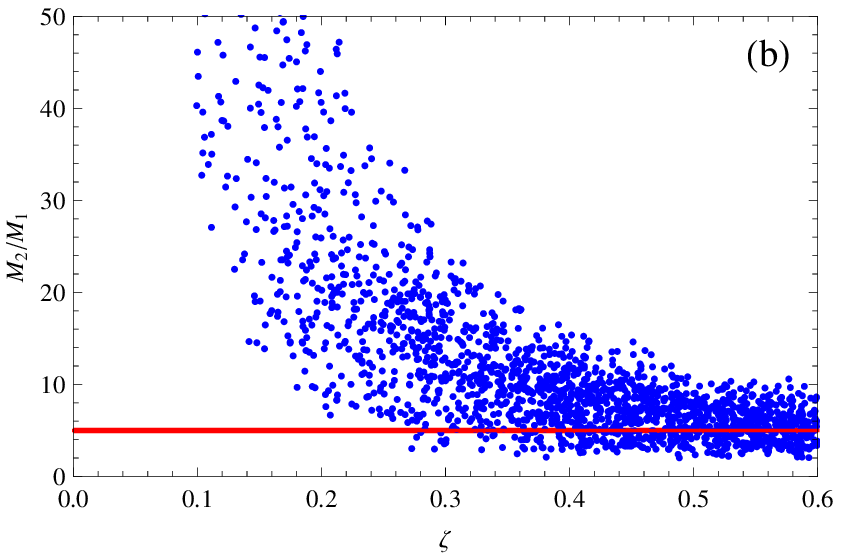}
  \caption{The mass ratio $M_2/M_1$ of right-handed heavy neutrinos as
   a function of the soft $\mutau$ breaking parameter $\,\zeta\,$,
   with the experimental inputs varied within 90\%\,C.L. ranges [plot-(a)]
   and $2\sigma$ ranges [plot-(b)], respectively.
   Each plot has computed 2000 samples.}
  \label{fig:M1M2-ratio}
  \vspace*{2mm}
\end{figure}

The mass-eigenvalues of the right-handed neutrinos can be directly derived from
diagonalizing the Hermitian matrix $M_R^{}M_R^\dag$ or $M_R^\dag M_R^{}$,
under the expansion of $\zeta$ and $r$\,,\footnote{We can also diagonalize
the complex matrix $M_R$ by using the unitary transformation matrix $\,V_R$\,,\,
as shown in Appendix-A.}
\begin{subequations}
  \label{eq:MR-M1M2}
  \begin{eqnarray}
    M_1  & \simeq &
    M_{22}
    \[r^2 - r \zeta \cos \omega + \f{1}{4}\zeta^2\,\]^{\hf}  \!,
    \label{eq:MR-M1}
  \\
    M_2  & \simeq &
    2M_{22}
    \[ 1 - \f{1}{4}(2 r + \zeta \cos \omega) \] \!.
    \label{eq:MR-M2}
  \end{eqnarray}
\end{subequations}
So we derive the mass ratio,
\beqa
\label{eq:Ratio-M2/M1}
\f{M_2}{M_1} ~\simeq~
\f{4-(2r+\zeta\cos\om)}{\sqrt{(2r-\zeta\cos\om)^2+\zeta^2\sin^2\om\,}\,} \,.
\eeqa
As shown in Eq.\,(\ref{eq:rr}), the seesaw parameter $r$ can be resolved as a function of
$\zeta$ and $\omega$ with $\,(s_s,\,y,\,\dx)\,$ from the oscillation data.
In Fig.\,\ref{fig:M1M2-ratio} we plot the ratio $M_2/M_1$ as a function of the $\mutau$ breaking
parameter $\,\zeta$\,,\, where we have scanned the parameter space $\,\om\in [0,\,2\pi)\,$.
As we noted earlier in Eq.\,(\ref{eq:q2+}), the neutrino data already
favors $\,M_2/M_1\gg 1\,$.\,   For our leptogenesis study in Sec.\,\ref{sec:solution} we will also require
$\,M_2/M_1 \geqq 5\,$ (as indicated by the (red) horizontal line in Fig.\,\ref{fig:M1M2-ratio}),
so that lepton asymmetry is mainly produced by the decays of the
lighter right-handed Majorana neutrino with mass $M_1$.

\begin{figure}[h]
   \centering
  \vspace*{3mm}
  \hspace*{-3mm}
  \includegraphics[width=8.3cm,clip=true]{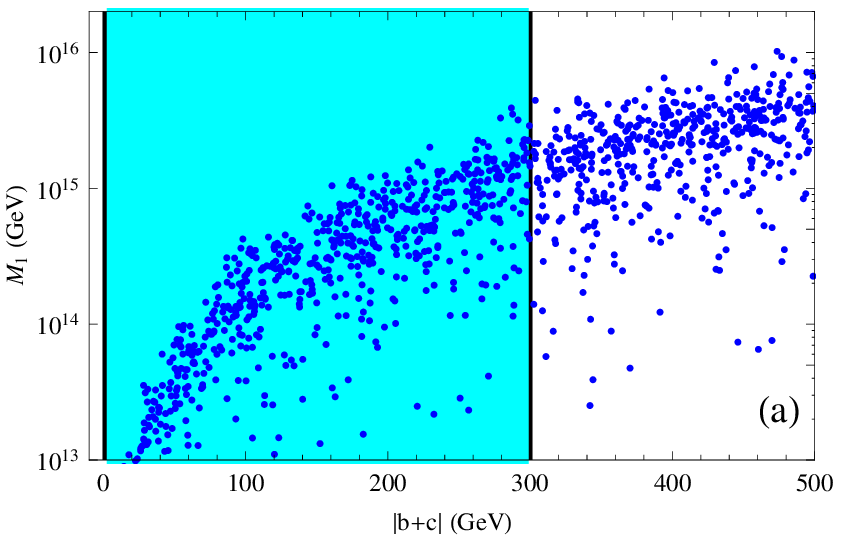}
  \includegraphics[width=8.3cm,clip=true]{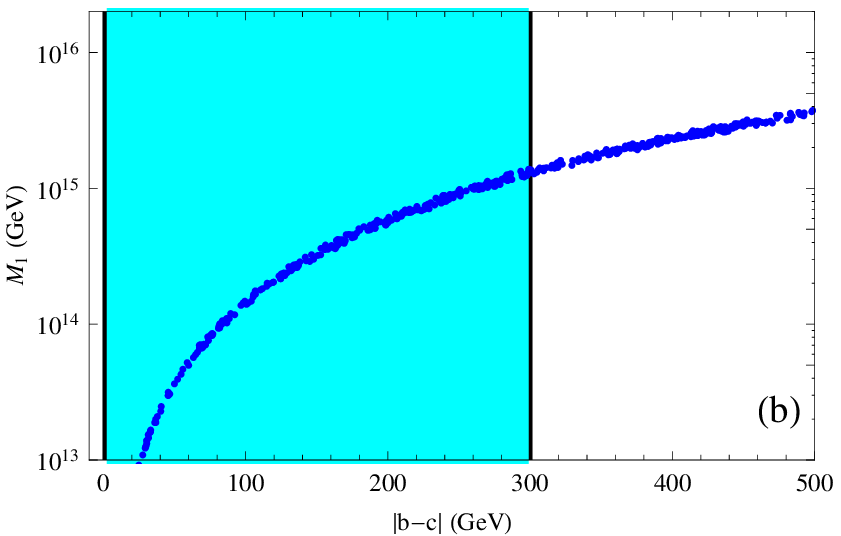}\\
 \includegraphics[width=8.3cm,clip=true]{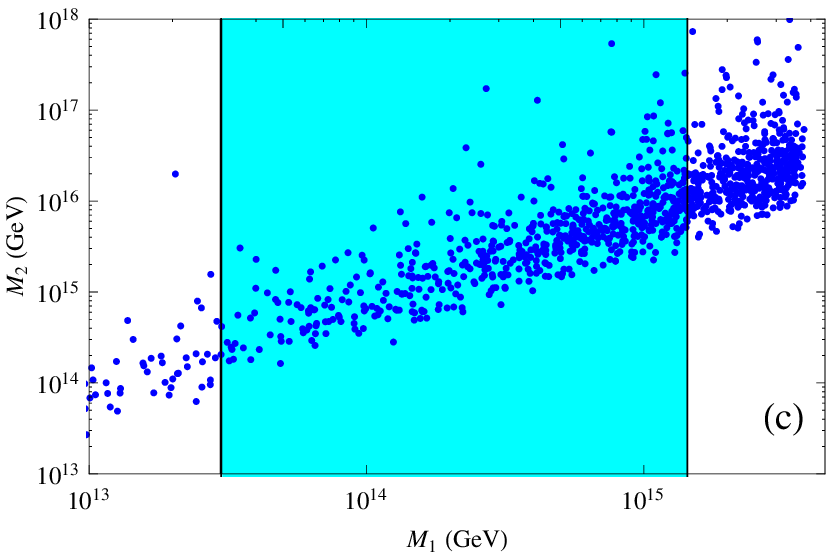}
  \caption{Seesaw scale $M_1$ as a function of the elements $|b\pm c|$ in the Dirac mass-matrix $\mD$
  is shown in plot-(a) and (b), where the shaded regions correspond to
  the natural perturbative region $\,|b\pm c|\in [1,\,300]$\,GeV.
  This puts an upper bound, $\,M_1\leqq 1.4\times 10^{15}\,$GeV, from plot-(b).
  The natural parameter-space in the $\,M_1-M_2\,$ plane is depicted in plot-(c), where the shaded
  region is bounded from the right by the upper limit of $M_1$ as inferred from plot-(b) and
  from the left by the lower limit of $M_1$
  as imposed by (\ref{eq:M1-LowerBound}) via the leptogenesis in Sec.\,\ref{sec:model-prediction-leptogenesis}.
  }
\label{fig:M1-|bc|}
\end{figure}

From Eqs.\,(\ref{eq:expression-abc-2}), (\ref{eq:expression-abc-4}),
(\ref{eq:MR-M1}) and (\ref{eq:sol-deltaD'}), we can connect the seesaw scale $M_1$ to the element
of Dirac mass-matrix \,$\mD$\,,
\beqs \beqa \label{eq:M1-b+c} M_1 &\simeq& \frac{\,(b+c)^2\,}{2c^2_s
\mh_2^{}}
\[r^2-r\zeta\cos\d_D+\f{1}{4}\zeta^2\]^{\f{1}{2}}
\\[1mm]
\label{eq:M1-b-c}
    &\simeq& \frac{\,(b - c)^2\,}{\mh_3^{}} \,,
\eeqa \label{eq:M1-bc} \eeqs
where the Dirac mass-parameters,
$\,\,b \pm c = (y_b^{}\pm y_c^{})v/\sqrt{2}\,$,\,
arise from the Yukawa interactions (with couplings $\,y_b^{}\,$ and $\,y_c^{}\,$
for $\,b\,$ and $\,c\,$,\, respectively),
and the seesaw parameter $\,r\,$ is given by Eq.\,(\ref{eq:rr}).
Thus, we can plot $\,M_1\,$ as a function of the magnitude of the Dirac
mass-parameters $\,|b \pm c|\,$ in Figs.\,\ref{fig:M1-|bc|}a and
\ref{fig:M1-|bc|}b, where we have input the measured
quantities in their 90\%\,C.L. ranges and scanned the allowed
parameter space for \,$(\zeta,\,\d_D)$,\, with 1200 samples.

We note that the Yukawa couplings $\,y_b^{}\,$ and $\,y_c^{}\,$ cannot be too small
(to avoid excessive fine-tuning) or too large (to keep valid perturbation).
So, we will take the Dirac mass-parameters $\,|b \pm c|\,$ in the natural range
$[1,\,300]$\,GeV, corresponding to the Yukawa coupling $\,y_j^{}\,$'s no
smaller than $\,\O(10^{-2})\,$ and no larger than $\,\O(y_t^{})$,  where
$\,y_t^{}=\sqrt{2}m_t^{}/v\simeq 1\,$ is the top-quark Yukawa coupling in the SM.\,
This natural perturbative range of $\,|b\pm c|\,$ is indicated by
the shaded area in \,Figs.\,\ref{fig:M1-|bc|}a-b,\, which results in an
upper limit on the seesaw scale $\,M_1$\, due to the perturbativity requirement.
We see that Fig.\,\ref{fig:M1-|bc|}b puts a much stronger bound on $\,M_1\,$ than
Fig.\,\ref{fig:M1-|bc|}a because Eq.\,(\ref{eq:M1-b+c}) has much larger uncertainties
due to scanning the parameter-space of $\,\zeta\,$ and $\,\d_D\,$.\,
We may also resolve $\,M_1\,$ from Eq.\,(\ref{eq:expression-abc-1}), so
$\,M_1\,$ is expressed in terms of the Dirac mass-parameter $\,a\,$,
\beqa \label{eq:M1-a} M_1 &\simeq& \frac{\,a^2\,}{s^2_s \mh_2^{}}
\[r^2-r\zeta\cos\d_D+\f{1}{4}\zeta^2\]^{\f{1}{2}} ,
\eeqa
which has similar uncertainties to Eq.\,(\ref{eq:M1-b+c}) and thus does not
provide better constraint on $\,M_1$\,.\,

Using the eigenvalue formulas
(\ref{eq:MR-M1})-(\ref{eq:MR-M2}), we have further plotted the
viable parameter-space in the $\,M_1-M_2\,$ plane, as shown in
Fig.\,\ref{fig:M1-|bc|}c with 1200 samples, where the shaded region
is bounded from the right by the upper limit of $M_1$ as inferred
from Fig.\,\ref{fig:M1-|bc|}b and from the left by the lower limit
of $M_1$ as imposed by (\ref{eq:M1-LowerBound}) via leptogenesis
(which will be derived later in
Sec.\,\ref{sec:model-prediction-leptogenesis}).

\pagebreak
%\vspace*{4mm}
\section{Origin of Matter from Soft $\bd{\mutau}$ and CP Breaking}
\label{sec:solution}

 In this section, we derive the predictions of our soft breaking seesaw model
 for cosmological matter-antimatter asymmetry (baryon asymmetry) via
 thermal leptogenesis\,\cite{lepG,lepGrev}.  We fully reconstruct the
 leptogenesis CP-asymmetry from the low energy Dirac CP phase and establish the
 direct link between the cosmological CP-violation and the low energy Jarlskog invariant.
 We analyze the {\it correlations} of the leptogenesis
 scale with the low energy observables such as the leptonic Jarlskog-invariant
 $J$ \cite{J} and neutrinoless double-beta decay parameter $M_{ee}$ \cite{0nu2beta}.
 We also deduce a lower bound on the leptogenesis scale for producing the observed
 matter-antimatter asymmetry.

\vspace*{3mm}
\subsection{%\hspace*{-4.7mm}.\hspace*{-0.6mm}
Matter-Antimatter Asymmetry via Leptogenesis in Neutrino Seesaw}
\label{sec:leptogenesis}

The universe is populated exclusively with matter instead of antimatter,
and the baryon density $\,n_B^{}\,(\,\gg n_{\ov B}^{}\simeq 0\,)$\,  relative to
the photon density $n_\gamma^{}$ is found
to be small but nonzero\,\cite{WMAP08},
\begin{equation}
\label{eq:etaB-exp}
  \eta_B^{}
 ~\equiv~
  \frac {n_B^{} - n_{\ov B}^{}}{n_\gamma^{}}
 ~=~
  (6.21\pm 0.16) \times 10^{-10} \,.
\end{equation}
Generating a net baryon asymmetry requires three conditions \'{a} la
Sakharov\,\cite{Sakharov}: (i) baryon number violating interactions,
(ii) C and CP violations, and (iii) departure from the thermal equilibrium.
The standard model (SM) could not provide the observed baryon asymmetry
due to having too small CP-violations from the CKM matrix and the lack of
sufficiently strong first-order electroweak phase transition\,\cite{EW-BG}.
But, in the seesaw extension of the SM,
the thermal leptogenesis\,\cite{lepG} has CP-violation phases arisen from the
neutrino sector and the lepton number asymmetry produced during
out-of-equilibrium decays of
heavy Majorana neutrino $N_j$ into the lepton-Higgs pair $\ell H$
and its CP-conjugate $\bar{\ell}H^*$.\,
Since the nonperturbative electroweak sphaleron\,\cite{sphaleron}
interactions violate $B+L$ \cite{tHooft} but preserve $B-L$,
the lepton asymmetry is partially converted to a baryon asymmetry\,\cite{HT,kappa-BDP},
\beqa
  \eta_B^{} ~=~
  \f{\xi}{\,f\,} N_{B-L}^f \,=\, -\f{\xi}{\,f\,} N_{L}^f
  \,,
  \label{eq:etaB}
\eeqa
where $\xi$ is the fraction of $B\!-\!L$ asymmetry
converted to baryon asymmetry via sphaleron process\,\cite{HT},
$\,\xi \equiv (8 N_F \!+\! 4 N_H) / (22 N_F \!+\! 13 N_H)\,$ with $\,N_F~(N_H)\,$ being
the number of fermion generations (Higgs doublets).
The SM has $\,(N_F,\,N_H)=(3,\,1)\,$ and thus $\,\xi=28/79\,$.\,
The parameter $\,f =N_\gamma^{\rm rec}/N_\gamma^* ={2387}/{86}\,$ is the dilution factor
calculated by assuming standard photon production from the onset of leptogenesis till
recombination\,\cite{kappa-BDP}.
The contribution from decays of the heavier right-handed neutrino ($N_2$)
will be washed out through the thermal equilibrium, only the lightest one
($N_1$) contributes effectively to the net lepton asymmetry
so long as $\,M_1\ll M_2$.\,
This is practically realized by requiring $\,{M_2}/{M_1} \geqq 5\,$ as indicated
by the (red) horizontal line in Fig.\,\ref{fig:M1M2-ratio}.
Note that $\,M_2/M_1\gg 1\,$ is also consistent with the condition (\ref{eq:q2+})
we derived in Sec.\,\ref{sec:zeroth-seesaw}.
Then, the net contribution to lepton asymmetry $N_L^f$ is
given by\,\cite{kappa-BDP},
\beqa
  N_L^f ~=~ \f{3}{4}\kappa_f^{}\epsilon_1^{} \,,
\label{eq:YL}
\eeqa
and thus we have the final baryon asymmetry,
\beqa
\label{eq:etaB-f}
\eta_B^{} ~=\, -\f{3\,\xi}{\,4f\,} \kappa_f^{}\ep_1^{}
~=\, - d \,\kappa_f^{}\ep_1^{} \,,
\eeqa
with
$\,d \equiv 3\xi/(4f) \simeq 0.96\!\times\! 10^{-2}\,$.\,
Here $\,\kappa_f^{}\,$ is an efficiency factor that depends on how much
out-of-equilibrium $N_1$-decays are. It is deduced from
numerically solving the Boltzmann equation\,\cite{kappa-BDP,kappa-Strumia}.
Practically, one can derive useful analytical formulas for $\,\kappa_f^{}\,$
by fitting the numerical solution of the Boltzmann equation.
For convenience, we will use the following fitting formula for $\,\kappa_f^{}$
\cite{kappa-Strumia},\footnote{Other fitting formulas to the exact solution
of $\kappa_f^{}$ also exist in the literature\,\cite{kappa-BDP} and the resulted
$\kappa_f^{}$ values are quite close in the relevant range of $\ov{m}_1^{}$.}
\beqa
  \kappa_f^{-1} ~\simeq~
  \(\f{\ov{m}_1^{}}{\,0.55\!\times\! 10^{-3}\,\textrm{eV}\,}\)^{1.16}
  +\f{\,3.3\!\times\! 10^{-3}\,\textrm{eV}\,}{\ov{m}_1^{}}  \,,
  \label{eq:kappa1}
\eeqa
where
\begin{equation}
  \ov{m}_1^{}  ~\equiv~
  \f{(\widetilde{m}^\dagger_D \widetilde{m}_D^{})_{11}^{}}{M_1}  \,,
  \label{eq:tilde-m1}
\end{equation}
and $\,\mt_D^{}\equiv \mD V_R\,$,\, with $V_R$ determined from the
diagonalization of $M_R$ (cf.\ Appendix-A).
The CP asymmetry parameter $\,\epsilon_1^{}\,$ can be expressed as
%
%\beqs
\begin{eqnarray}
  \epsilon_1^{}
~\equiv~
  \f {~\Gamma[N_1\to\ell H]-\Gamma[N_1\to\ov{\ell}H^*]~}
     {~\Gamma[N_1\to\ell H]+\Gamma[N_1\to\ov{\ell}H^*]~}
%\label{eq:ep1-def}
%\\[3mm]
%&\!\! = \!\!&
 ~=~ \f{1}{\,4\pi v^2\,} F\!\(\f{M_2}{M_1}\)
  \f{\Im\mathfrak{m}
  \left\{ [ (\widetilde{m}^\dagger_D \widetilde{m}_D^{})_{12}^{}]^2
  \right\}}
  {( \widetilde m^\dagger_D \widetilde m_D^{})_{11}^{}} \,,
\label{eq:ep1}
\end{eqnarray}
%\label{eq:ep1-all}
%\eeqs
%
where $v$ is the vacuum expectation value of the SM Higgs boson.
Note that in the mass-eigenbasis of $\,(N_1,\,N_2)\,$ their Yukawa couplings
and the corresponding effective Dirac mass-matrix $\,\mt_D^{}\,(\equiv \mD V_R)\,$ violate CP
due to the complex $M_R$ and thus its diagonalization matrix $V_R$ in our soft breaking model,
even though the original weak-basis Yukawa couplings and $\mD$ are real. It is this
complex $\,\mt_D^{}\,$ that differs the decay width $\,\Gamma[N_1\to\ell H]\,$
from $\,\Gamma[N_1\to\ov{\ell} H^*]\,$.\,
For the SM, the function $F(x)$ in (\ref{eq:ep1}) takes the form,
\beqs
\beqa
\label{eq:F1}
  F(x)
&\!\!\equiv\!\!&
  x \left[
    1 - (1 + x^2) \ln \frac {1 + x^2}{x^2}
  + \frac 1 {1 - x^2} \right]
\\[2mm]
&\!\!=\!\!&
- \f{3}{\,2x\,}
+ \O\!\( \f{1}{x^3} \),
\quad \textrm{for}~\,  x \gg 1\,.
\label{eq:F2}
\eeqa
\label{eq:F}
\eeqs
As explained earlier, in our numerical analysis we will require
the mass ratio $\,M_2/M_1\geqq 5\,$,\,
so we see that (\ref{eq:F2}) is a rather accurate
approximation to (\ref{eq:F1}).

\vspace*{3mm}
\subsection{Model Predictions for Leptogenesis}
\label{sec:model-prediction-leptogenesis}

In the formulas (\ref{eq:kappa1}) and (\ref{eq:ep1}), we have defined a mass matrix
$\,\widetilde{m}_D^{}\,$ which is connected to the original Dirac mass matrix $\,\mD\,$,
\beqa
  \widetilde{m}_D^{} ~\equiv~
  \mD V_R ~=
  \begin{pmatrix}
    e^{i \gamma_1^{}} \left[ \cR a + \sR e^{i \beta} a \right]
  & e^{i \gamma_2^{}} \left[ \cR e^{i \beta} a - \sR a \right] \\[2mm]
    e^{i \gamma_1^{}} \left[ \cR b + \sR e^{i \beta} c \right]
  & e^{i \gamma_2^{}} \left[ \cR e^{i \beta} c - \sR b \right] \\[2mm]
    e^{i \gamma_1^{}} \left[ \cR c + \sR e^{i \beta} b \right]
  & e^{i \gamma_2^{}} \left[ \cR e^{i \beta} b - \sR c \right] \\[2mm]
  \end{pmatrix} ,
\eeqa
where the unitary matrix $\,V_R\,$ is given by (\ref{eq:reconstruction-VR})
in Appendix-A from the diagonalization of $\,M_R$.\,
With this we compute,
\begin{subequations}
  \begin{eqnarray}
    ( \widetilde{m}^\dag_D \widetilde{m}_D^{} )_{11}^{}
  & \simeq &   (b - c)^2
  ~\simeq~     \mh_{30}^{} M_1 \,,
  \label{eq:md11}
  \\[2mm]
    ( \widetilde{m}^\dag_D \widetilde{m}_D^{} )_{12}^{}
  & = &
    ( a^2 + 2 b c )
    ( c^2_R e^{i \beta} - s^2_R e^{- i \beta} )
    e^{-i( \gamma_1^{} - \gamma_2^{})}
    \nonumber
  \\
  & = &
  - \frac{\zeta}{2} \mh_{3}^{} M_{10}
    \[ |Y| - \f{1}{4} |2 \!-\! X| \]
    e^{- i (\omega + \gamma_1 - \gamma_2)} \,.
  \end{eqnarray}
\end{subequations}
Then, we can derive the parameter $\ov{m}_1^{}$ in (\ref{eq:tilde-m1}),
\beqa
\label{eq:m1bar}
  \ov{m}_1^{} ~\simeq~
  \frac {(b - c)^2}{M_1} ~\simeq~ \mh_{30}^{} ~\simeq\, \chi_1^{}\sqrt{\Delta_a} ~,
\eeqa
as well as the imaginary part,
\begin{equation}
  \Im\mathfrak{m}\!
  \left\{ [ ( \widetilde{m}^\dag_D \widetilde{m}_D^{} )_{12}^{} ]^2 \right\}
~\,\simeq~ - \f{\zeta^2}{4} \mh_{3}^2 M_{10}^2
  \[ |Y| - \f{1}{4} |2 \!-\! X| \]^2
  \frac {\sin \omega \left( 4 r \cos \omega - \zeta \right)}
    {\sqrt{\zeta^2 - 4 r \zeta \cos \omega + 4 r^2}\,} \,.
\end{equation}
With the data of Table-\ref{tab:1} and using Eq.\,(\ref{eq:m1bar}),
we see that the light neutrino mass-parameter
$\,\ov{m}_1^{}\,$ falls into the range,
$\,0.045< \ov{m}_1^{} <0.053\,$eV, at $3\sigma$ level.
This means that in the formula (\ref{eq:kappa1})
the first term dominates $\kappa_f^{}$
and the second term is negligible.

Then, we further derive the CP-asymmetry parameter $\epsilon_1^{}$ from (\ref{eq:ep1}),
%
%\beqs
\beqa
  \epsilon_1^{} &\!\!\simeq\!\!&
%  \f{\mh_3^{}M_1}{4\pi v^2}
%  \f{\,3 (4 y - \sqrt{\zeta^2 \!-\! 4 r \zeta \cos \omega \!+\! 4 r^2}\,)^2\,}
%    {128 (\zeta^2 \!-\! 4 r \zeta \cos \omega \!+\! 4 r^2)}
%  \left( 4 r \cos \omega \!-\! \zeta \right)\sin\omega\,\zeta^2
% \label{eq:ep1-a}
% \\[3mm]
%  &\!\!=\!\!&
  -\f{\mh_3^{}M_1}{4\pi v^2}
  \f{\,3 (4 y - \sqrt{\zeta^2 \!-\! 4 r \zeta \cos\d_D \!+\! 4 r^2}\,)^2\,}
    {128 (\zeta^2 \!-\! 4 r \zeta \cos\d_D \!+\! 4 r^2)}
  \left( 4 r\cos\d_D \!-\! \zeta \right)\sin\d_D\,\zeta^2 \,,
\label{eq:ep1-b}
\eeqa
%\eeqs
%
where the solution (\ref{eq:sol-deltaD'}), $\,\om = 2\pi -\d_D$\,,\,
is used to replace $\omega$ by  $\,\d_D$\,.\,
As expected, Eq.\,(\ref{eq:ep1-b}) proves that
{\it in our model the leptogenesis CP-asymmetry $\ep_1^{}$ is
completely reconstructed from the the low energy Dirac CP phase $\,\d_D$\,.}\,

Combining the above formulas (\ref{eq:etaB-f}), (\ref{eq:ep1}),
(\ref{eq:F}) and (\ref{eq:ep1-b}), we finally derive,
\begin{eqnarray}
  \f{\eta_B^{}}{M_1} ~=~
  d \,\kappa_f^{}
  \f{\mh_3^{}}{\,4\pi v^2\,}
  \f{\,3\[4 y \!-\! \sqrt{\zeta^2 \!-\! 4 r \zeta\cos\delta_D \!+\! 4 r^2}\,\]^2\,}
    {128 (\zeta^2 \!-\! 4 r \zeta\cos\delta_D \!+\! 4 r^2)}
  \(4r\cos\delta_D \!-\! \zeta\) \sin\delta_D\,\zeta^2 \,,
\label{eq:etaB-M1}
\end{eqnarray}
where we note
$\,\zeta^2 \!-\! 4 r \zeta\cos\delta_D \!+\! 4 r^2
   = (2r\!-\!\zeta\cos\d_D)^2+\zeta^2\sin^2\!\d_D > 0\,$.\,
As WMAP data (\ref{eq:etaB-exp}) requires
the baryon asymmetry $\,\eta_B^{}\,$  to be positive, we have the constraint,
\beqa
  \(4 r \cos\d_D \!-\! \zeta \)\sin \d_D ~>~  0 ~ .
  \label{eq:Cond-etaB>0}
\eeqa
Substituting the expression (\ref{eq:rr}) into
the positivity conditions (\ref{eq:Cond-etaB>0}), we derive the following
the allowed regions of $\,\d_D\,$ with possible constraints
after a systematical analysis (cf.\ Appendix-B),
\begin{subequations}
\begin{eqnarray}
\label{eq:bound-deltaD-0}
&& \hspace*{-28mm}
\textrm{for} ~~r\,=\,r_+^{}\,=\,r_-\,,
\nonumber\\[2mm]
&& \hspace*{-19mm}
  \d_D \in
    \( 0,\,\f{\pi}{4} \)
  \bd{\cup}
    \( \f{3\pi}{4},\, \pi\)
  \bd{\cup}
    \( \f{5\pi}{4},\,\f{7\pi}{4} \)  ;
\\[3mm]
&& \hspace*{-28mm}
\textrm{for} ~~r \,=\, r_+^{} \,\neq\, r_- \,,
\nonumber\\[2mm]
 && \hspace*{-19mm}
\d_D \in
    \( 0,\,\f{\pi}{4} \]
  \bd{\cup}
    \( \f{\pi}{4},\, \f{\pi}2\)_>
  \bd{\cup}
    \( \f{3\pi}{4},\,\pi \)_<  \cup
    \( \pi,\,\f{5\pi}{4} \)_>  \cup
    \[\frac{5\pi}{4},\,\frac{3\pi}{2}\] \cup
    \( \frac{3\pi}{2},\,\f{7\pi}{4}
    \)_< \,;
\\
\label{eq:bound-deltaD-a}
&& \hspace*{-28mm}
\textrm{for} ~~r=r_-^{}\neq r_+
\nonumber \\[2mm]
&& \hspace*{-19mm}
\d_D \in
    \( 0,\,\f{\pi}{4} \)_<
  \bd{\cup}    \( \f{\pi}{2},\, \f{3\pi}4\)_>
  \bd{\cup}    \[ \f{3\pi}{4},\,\pi \)
  \bd{\cup}    \( \frac{5\pi}{4},\,\f{3\pi}{2} \)_<
  \bd{\cup}    \[\frac{3\pi}{2},\,\frac{7\pi}{4}\]
  \bd{\cup}    \( \frac{7\pi}{4},\,2\pi \)_> \,;
 \label{eq:bound-deltaD-b}
  \end{eqnarray}
\label{eq:bound-deltaD}
\end{subequations}
where $\,r_+ ~ (r_-)\,$ corresponds to taking the plus (minus) sign in the
formula (\ref{eq:rr}).
Also, the regions with subscript ``$>$"  require imposing the condition
$\,\cos^2\delta_D > \f{4}{s^4_s}\frac{\delta^4_x}{y^2 \zeta^2}\,$,\,
and those with subscript ``$<$" are constrained by the condition
$\,\cos^2\delta_D<\frac {4}{s^4_s}\f{\delta^4_x}{y^2 \zeta^2}\,$.\,

\begin{figure}[t]
  \centering
  \includegraphics[width=15cm,clip=true]{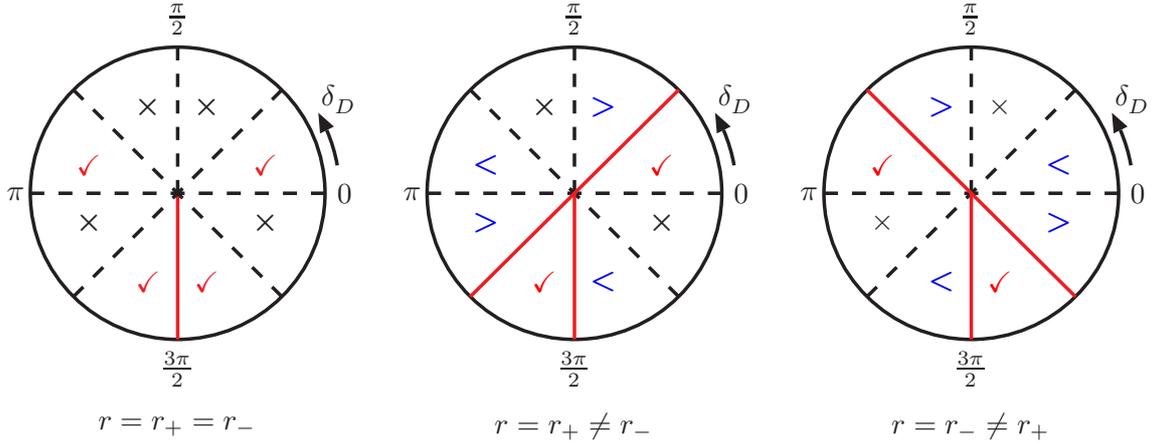}
  \caption{Pictorial summary of all constraints
(\ref{eq:bound-deltaD-0})-(\ref{eq:bound-deltaD-b}) from the
positivity condition (\ref{eq:Cond-etaB>0}) is depicted in each octant
of the CP-phase $\,\d_D\in [0,\,2\pi )\,$,\, where the octants
checked by $\,\checkmark$\,  automatically hold this
condition, the octants checked by $\,\times\,$  have
no solution, the octants checked by $\,>\,$ subjects to the condition
$\,\cos^2\delta_D > \f{4}{s^4_s} \f{\delta^4_x}{y^2\zeta^2}\,$,\,
and those checked by $\,<\,$ are constrained by
$\,\cos^2\delta_D < \f{4}{s^4_s}\f{\delta^4_x}{y^2 \zeta^2}\,$.\,
The radial solid-lines in each circle denote the corresponding $\,\d_D\,$
values which automatically hold (\ref{eq:Cond-etaB>0}) and the
radial dashed-lines correspond to the  $\,\d_D\,$ values excluded by
the positivity condition (\ref{eq:Cond-etaB>0}).}
\label{fig:rmp}
\end{figure}

\begin{figure}[h]
\vspace*{4mm}
\hspace*{-2mm}
\includegraphics[width=8.3cm,clip=true]{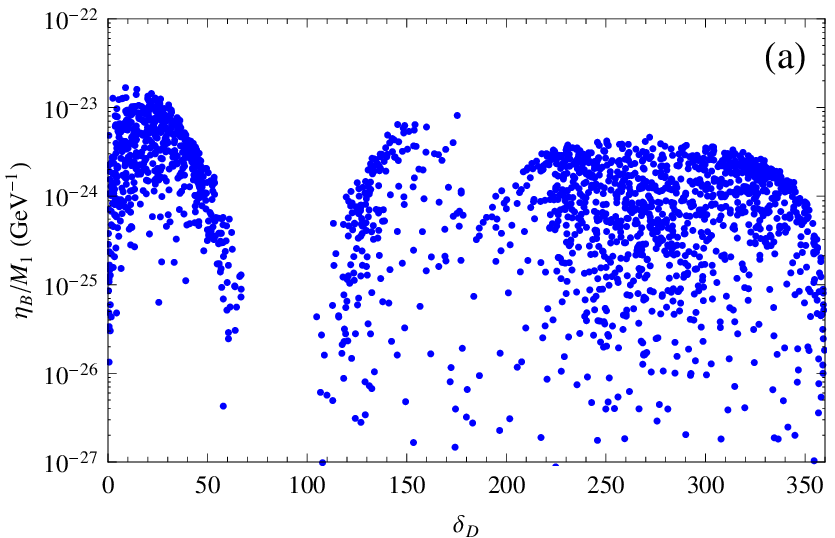}
\hspace*{2mm}
\includegraphics[width=8.2cm,clip=true]{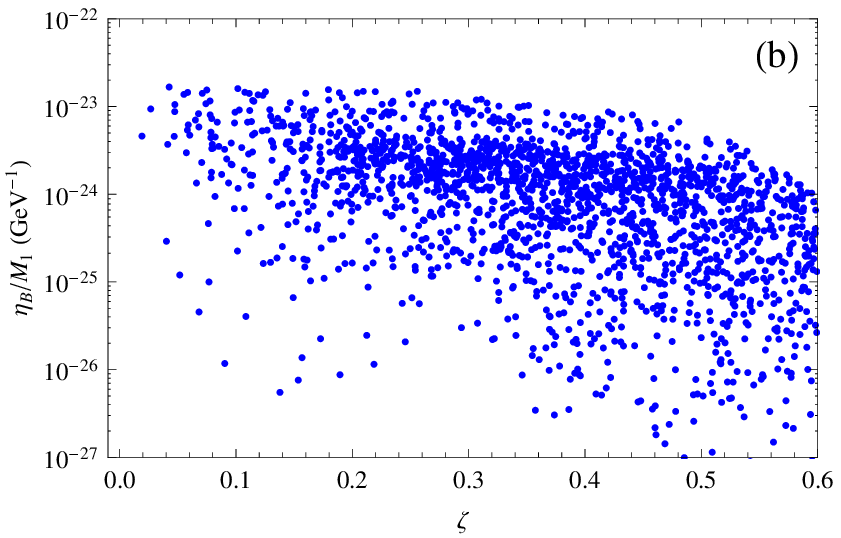}
\caption{Ratio $\,\eta_B^{}/M_1\,$ is shown
           as a function of soft breaking parameter $\,\zeta\,$ in plot-(a) and
           as a function of Dirac CP-phase $\,\delta_D\,$ in plot-(b), respectively,
           where all experimental inputs are scanned within
           their $90\%$\,C.L.\ ranges, for 2000 samples.}
\vspace*{2mm}
\label{fig-etaB/M1-omega}
\end{figure}

The positivity solution (\ref{eq:bound-deltaD}) is pictorially summarized
in Fig.\,\ref{fig:rmp}.
We see that the values $\,\d_D=\f{\pi}{2}\,$ and $\,\d_D=\pi\,$
are disallowed in all three cases (represented by dashed lines in Fig.\,\ref{fig:rmp}).
This feature is reflected in Fig.\,\ref{fig-etaB/M1-omega}a,
in which a small region around $\,\d_D\sim \f{\pi}{2}\,$
or $\,\d_D\sim \pi\,$ is excluded. The similar feature will be exhibited
more clearly by Fig.\,\ref{fig-J-D-new} in Sec.\,\ref{sec:model-prediction-leptogenesis} later.

Using Eq.(\ref{eq:rr}) we can eliminate $\,r\,$ in terms of
$\,(\zeta,\,\om)$\,.\, Finally we can plot, in
Fig.\,\ref{fig-etaB/M1-omega}, the ratio $\,\eta_B^{}/M_1\,$ as a function
of Dirac CP-phase $\,\delta_D$\, via Eq.\,(\ref{eq:etaB-M1}), where
all experimentally measured quantities are scanned within their
$90\%$\,C.L.\ range, for 2000 samples.
We see that Fig.\,\ref{fig-etaB/M1-omega} shows a robust upper bound,
\beqa
  \f{\eta_B^{}}{M_1} ~<~
  2 \times 10^{-23} \,\textrm{GeV}^{-1} \,.
\eeqa
Given the observed value of $\eta_B^{}$ in (\ref{eq:etaB-exp}), we infer a lower bound
for the leptogenesis scale \,$M_1$\,,
\beqa
\label{eq:M1-LowerBound}
  M_1 ~>~ (2.9-3.3) \times 10^{13} \,\textrm{GeV}\,,
\eeqa
where the experimental value of $\,\eta_B^{}\,$
is varied within its $90\%$\,C.L.\ range.

\vspace*{3mm}
\subsection{Direct Link of Leptogenesis with Low Energy Observables}

\begin{figure}[t]
\centering
\includegraphics[width=14cm,clip=true]{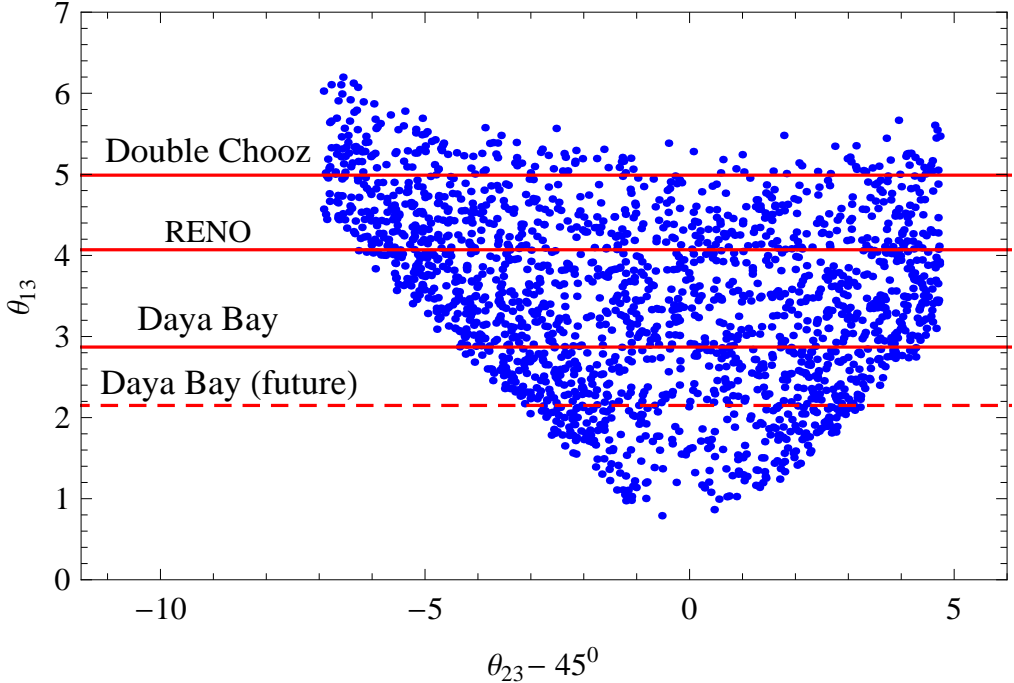}
\caption{The correlation between $\theta_{13}$ and
  $\theta_{23}-45^\deg$: all the inputs are the same as
  Fig.\,\ref{fig:dx-da},
  except requiring successful leptogenesis in the
  present figure, for 2000 samples.
  }
  \label{fig-deltax-deltaa-new}
%\vspace*{-1mm}
\end{figure}

Since the successful leptogenesis puts an additional nontrivial
constraint (\ref{eq:bound-deltaD})
on the parameter space, we inspect the correlation of
$\,\theta_{13}\,$ versus $\,\theta_{23}-45^\deg\,$ in Fig.\,\ref{fig:dx-da}.
We find that it is
now altered as in Fig.\,\ref{fig-deltax-deltaa-new}, after we
impose the condition (\ref{eq:bound-deltaD}) as well as $\,\,M_2/M_1\geqq 5$\,.\,
It is interesting to note that the successful leptogenesis results in
a general \,{\it lower bound}\, on the mixing angle $\,\theta_{13}$\,,\,
requiring
\beqa
\theta_{13} ~\gtrsim~ 1^\deg ~,
\eeqa
even for the region around $\,\theta_{23} = 45^\deg\,$.\,
This bound is still relatively weak due to the cubic power-dependence of
$\,\,\eta_B^{}\propto \zeta^3 \propto \theta_{13}^3\,$\, in (\ref{eq:etaB-M1})
for the present NH mass-spectrum of our model, which means $\,\eta_B^{}\,$
is not sensitive enough to the small mixing angle $\,\theta_{13}$\,.\,
Other extensions\footnote{S.-F. Ge, H.-J. He, F.-R. Yin, work in preparation.}
of our minimal model may lower the power-dependence
of $\,\eta_B^{}\,$ on $\,\theta_{13}\,$ and thus enhance this lower bound.

\begin{figure}[H]
  \hspace*{-4mm}
  \includegraphics[width=8.4cm,clip=true]{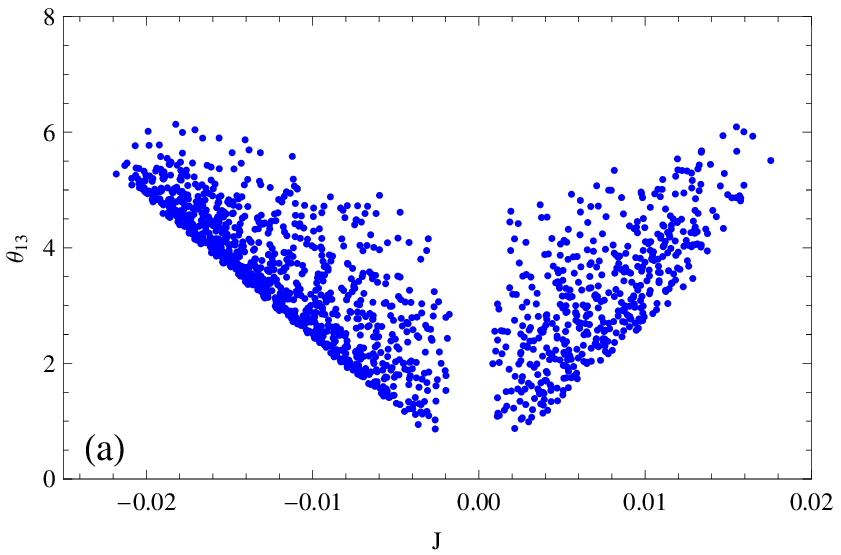}
  \includegraphics[width=8.3cm,clip=true]{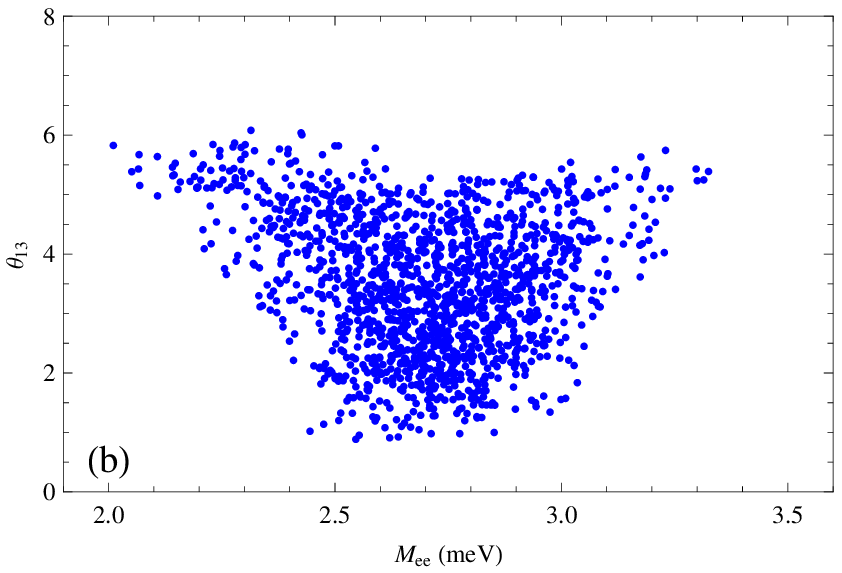}
  \caption{The correlations of $\theta_{13}$ with Jarlskog invariant
           $J$ [plot-(a)] and with neutrinoless double beta decay observable
           $M_{ee}$ [plot-(b)], where all the inputs are the same as
           Fig.\,\ref{fig:dx-J-Mee}, except requiring successful leptogenesis
           in the present figure, for 1500 samples.}
  \label{fig:dx-J-Mee-new}
\end{figure}

Then, in Fig.\,\ref{fig:dx-J-Mee-new} we replot the
the correlations of $\theta_{13}$ with the Jarlskog invariant $\,J\,$
and the neutrinoless double beta decay observable $\,M_{ee}$\,,\,
respectively.  This should be compared to Fig.\,\ref{fig:dx-J-Mee}
in Sec.\,\ref{sec:full-NLO}, where leptogenesis is not required.
We see that due to imposing the observed baryon asymmetry,
the parameter space around $\,J\sim 0\,$ in Fig.\,\ref{fig:dx-J-Mee-new}a
is significantly suppressed and more samples are distributed
along the upper left-wing (rather than the upper right-wing),
while Fig.\,\ref{fig:dx-J-Mee-new}b shows a clear lower bound on $\theta_{13}$.

\begin{figure}[H]
  \vspace*{2mm}
  \hspace*{-3mm}
  \includegraphics[width=8.2cm,clip=true]{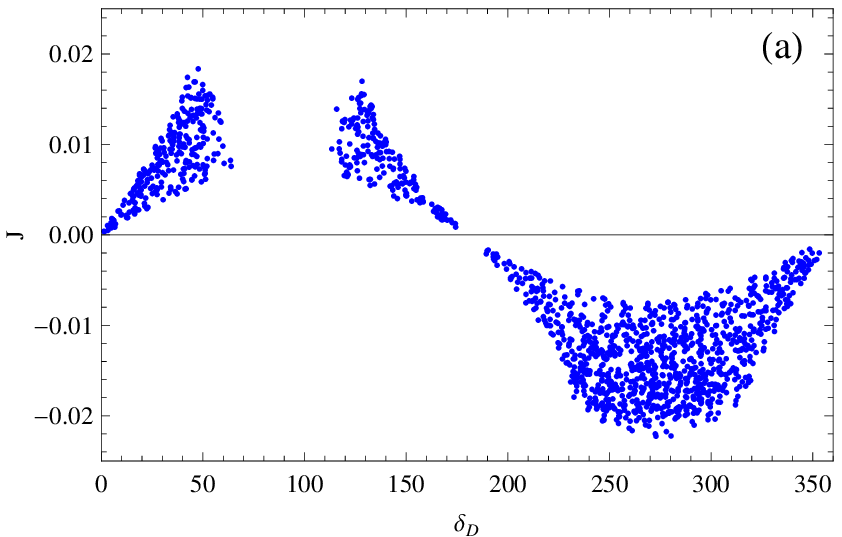}
  \hspace*{2mm}
  \includegraphics[width=8.0cm,clip=true]{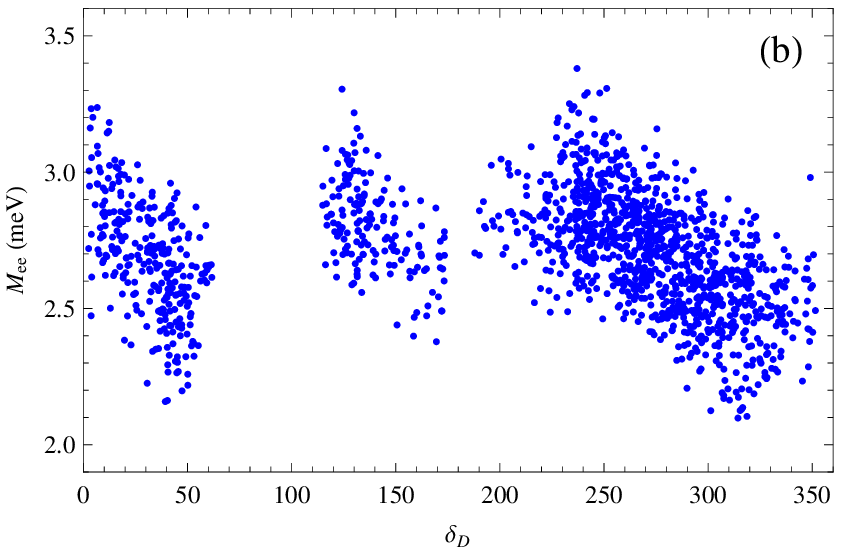}
  \caption{Under successful leptogenesis,
  the CP-violation Jarlskog invariant $J$ and the neutrinoless
  double beta decay observable $M_{ee}$ are shown as functions of the Dirac CP-phase
  \,$\delta_D$\,,\, where all experimental inputs
  are varied within their 90\%\,C.L. ranges, for 1500 samples.}
  \label{fig-J-Mee-dD-new}
\end{figure}

Under successful leptogenesis, the correlation between $\,J\,$ or
$\,M_{ee}\,$ with the Dirac CP-phase $\,\delta_D\,$  are plotted in
Fig.\,\ref{fig-J-Mee-dD-new}. As expected, we see that there are two
gaps around $\,\delta_D = \f{\pi}{2}\,$ and $\,\delta_D = \pi\,$,\,
respectively.  The first gap around $\,\delta_D = \f{\pi}{2}\,$ is
more significant.  This is consistent with
what we have observed from Eq.\,(\ref{eq:bound-deltaD})
and Fig.\,\ref{fig:rmp} earlier, as discussed
below Eq.\,(\ref{eq:bound-deltaD}).

\begin{figure}[h]
  \centering
  \includegraphics[width=10cm,clip=true]{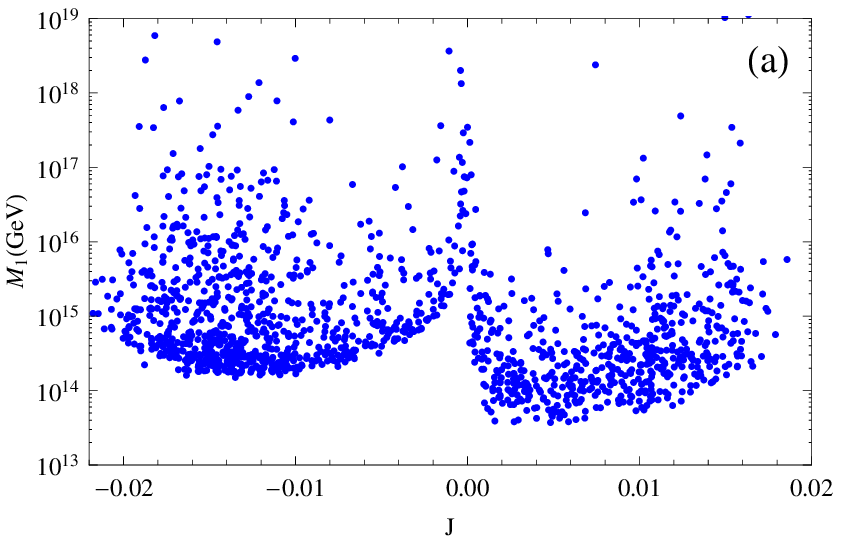}
  \\
  \includegraphics[width=10cm,clip=true]{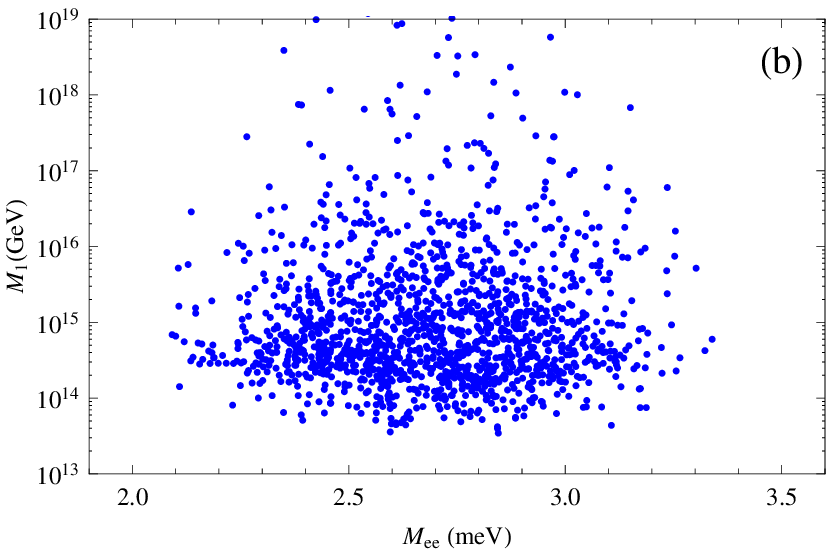}
  \caption{Correlation of the leptogenesis scale $M_1$ with
           the low energy Jarlskog invariant $J$ [plot-(a)]
           and the neutrinoless double beta decay observable $M_{ee}$ [plot-(b)]
           within 90\%\,C.L. ranges, for 3000 samples. }
  \label{fig-J-D-new}
\end{figure}

Finally, reversing Eq.\,(\ref{eq:etaB-M1}) we can express the leptogenesis scale $M_1$
in terms of baryon asymmetry $\eta_B^{}$ and other physical parameters,
\beqa
  M_1 ~=~
  \f{4\pi v^2\,}{\,d\, \kappa_f^{}\mh_3^{}\,}
  \f{128 (4 r^2 \!-\! 4 r \zeta \cos \delta_D \!+\! \zeta^2)}
    {\,3[ 4 y
           \!-\! (4 r^2 \!-\! 4 r \zeta \cos \delta_D \!+\! \zeta^2)^{\hf}
           ]^2\,}
  \f{\eta_B^{}}
    {\,( 4 r \cos \delta_D \!-\! \zeta )\sin\delta_D\,\zeta^2\,} \,.
  \label{eq:M1-etaB}
\eeqa
Since the low energy parameters $J$ and $M_{ee}$ in Eqs.\,(\ref{eq:J})-(\ref{eq:Mee})
are also predicted by our soft breaking model as functions of
$(\zeta,\,\d_D)$ and $\,\dx (=\theta_{13})$,\, we see that they must correlate with
the leptogenesis scale $\,M_1$.\,  This is plotted in Fig.\,\ref{fig-J-D-new}, and it
shows a robust lower bound on $M_1$, which is consistent with (\ref{eq:M1-LowerBound}),
as we inferred from Fig.\,\ref{fig-etaB/M1-omega}.


%\vspace*{4mm}
\newpage
\section{%\hspace*{-5.7mm}.\hspace*{-0.6mm}
Beyond $\boldsymbol{\mutau}$ Symmetry: Dictating Solar Mixing Angle}
\label{sec:solar}

In this section we present a general analysis about the derivation of solar mixing angle
and about how to dictate it from a new minimal hidden symmetry in the seesaw Lagrangian.
We have seen in Eq.\,(\ref{eq:t12}) that
the solar mixing angle $\theta_s\,(\equiv\theta_{12})$ is fully determined by the elements
of Dirac mass-matrix $\mD$ based on the expansion of $(r,\,\zeta)$ up to NLO.
In Sec.\,\ref{subsec:solar} we show how the formula (\ref{eq:t12})
can be derived in the $\mutau$ symmetric limit.
This feature can be proven without expanding the seesaw mass-matrix for light
neutrinos (Sec.\,\ref{subsec:solar-mixing-not-affected}).
Actually, we note that $\ts$ is controlled
by an extra new $\ZZ_2$ symmetry acting on the Dirac mass-matrix \,$\mD$\,,\,
as shown in Sec.\,\ref{sec:extra-z2}.

\vspace*{3mm}
\subsection{%\hspace*{-4.7mm}.\hspace*{-0.6mm}
Solar Mixing Angle from $\boldsymbol{\mutau}$ Symmetric Mass Matrix}
\label{subsec:solar}

Here we first derive the solar mixing angle $\ts$ as given by Eq.\,(\ref{eq:LO-Angle})
earlier in the $\mutau$ symmetric limit.
Let us rewrite Eq.\,(\ref{eq:Mnu-0}) as follows,
\beqa
  \ov{M}^{(0)}_\nu
&=&
  \frac{1}{(2 \!-\! r) r M_{22}}\!
  \begin{pmatrix}
    2 a^2 r & a (b + c) r & a (b + c) r \\
  & (b - c)^2 + 2 b c r & (b^2 + c^2) r - (b - c)^2 \\
  & & (b - c)^2 + 2 b c r
  \end{pmatrix}
\nn\\[3mm]
&\equiv&
  \begin{pmatrix}
    \overline A_0 & \overline B_0 & \overline B_0 \\
  & \overline C_0 & \overline D_0 \\
  & & \overline C_0
  \end{pmatrix} \!,~~~~~
  \label{eq:Mnu0-bar}
\eeqa
with an overline on $(A_0,\,B_0,\cdots)$ to distinguish from the LO elements in
Sec.\,\ref{sec:common-origin} (where further expansion in $r$ was made).
Thus we have,
%
% \begin{subequations}
% \begin{eqnarray}
 \beq
 \ba{ll}
  \dis\overline A_0
  ~=~
    \frac {2 a^2}{\,(2 - r) M_{22}\,} \,,
  ~~~&
    \dis\overline B_0
  ~=~
    \frac {a (b + c)}{\,(2 - r) M_{22}\,} \,,
  % \label{eq:A0-B0}
  \\[4mm]
    \dis\overline C_0  + \overline D_0
  ~=~
    \frac {(b + c)^2}{\,(2 - r) M_{22}\,} \,,
  ~~~&
    \dis\overline C_0  - \overline D_0
  ~=~
    \frac {\,(b - c)^2\,}{r M_{22}}  \,.
 % \label{eq:C0-D0}
 \ea
 \label{eq:A0B0C0D0}
 \eeq
 %\end{eqnarray}
 %\end{subequations}
 %

 As shown in Sec.\,\ref{sec:model-independent-reconstruction-formalism},
 any $\mutau$ symmetric mass matrix can be diagonalized by a two-step rotation,
 because of $\,\theta_{13} =0\,$.\,
 First, we make a $U_{23}$ rotation to partially diagonalize
 $\ov{M}^{(0)}_\nu$,
 with $\,\theta_{23}=45^\deg\,$ (as required by the $\mutau$ symmetry),
 \beqa
 \ov{M}^{(0)\prime}_\nu
 ~\equiv~
 U_{23}(45^\deg)^T \ov{M}^{(0)}_\nu\,U_{23}(45^\deg) ~=\,
  \(
 \ba{ccc}
   \overline A_0 & \sqrt{2} \, \overline B_0 & 0 \\[2mm]
   \sqrt{2} \, \overline B_0 & \overline C_0 + \overline D_0 & 0 \\[2mm]
 0 & 0 & \overline C_0 - \overline D_0
 \ea
 \)
  \label{eq:MnuS2}
 \eeqa
 where \,$U_{23}(\theta)$\, is introduced in Eq.\,(\ref{eq:U123}).
 Note that the above matrix $\,\ov{M}^{(0)\prime}_\nu$\, contains only
 a non-diagonal $2\times2$ sub-matrix involving $\{12\}$-mixing.
 Second, we further diagonalize $\,\ov{M}^{(0)\prime}_\nu$\,
 via the rotation $U_{12}$ with mixing angle $\,\ts\,(\equiv\theta_{12})$
 (which is independent of the $\mutau$ symmetry as noted at the end of
 Sec.\,\ref{sec:model-independent-reconstruction-formalism}),
 \beqa
 \label{eq:U12-diagonalization}
 U_{12}^T\, \ov{M}^{(0)\prime}_\nu\, U_{12}
 = \textrm{diag}(m_1^{},\,m_2^{},\,m_3^{}) \,,
 \eeqa
 where $U_{12}$ is introduced in (\ref{eq:U}).  Then, it is straightforward to
 derive\footnote{The expressions (\ref{eq:t12-general})-(\ref{eq:m12-general})
 agree with that given in Ref.\,\cite{Grimus1} earlier
 for general $\mutau$ symmetric mass matrix.} the mixing angle $\,\ts\,$,
 \beqa
 \label{eq:t12-general}
 \tan 2\ts ~=~ \f{2\sqrt{2}\,\ov{B}_0}{\,\ov{A}_0-(\ov{C}_0\!+\!\ov{D}_0)\,}\,,
 \eeqa
 as well as the mass-eigenvalues,
 \beqa
 \label{eq:m12-general}
 m_{1,2}^{} ~=~ \f{\,\ov{A}_0\!+\!\ov{C}_0\!+\!\ov{D}_0\,}{2}\!\[
 1\mp \sqrt{1-\f{\,4[\ov{A}_0(\ov{C}_0\!+\!\ov{D}_0)-2\ov{B}_0^2]\,}
           {(\ov{A}_0\!+\!\ov{C}_0\!+\!\ov{D}_0)^2}}\,\]\!, ~~~~~~
 m_3^{} ~=~ \overline{C}_0 - \overline{D}_0 \,.
 \eeqa
 In this section, all mass-eigenvalues are analyzed at the seesaw scale though
 we have suppressed the ``hat'' notation introduced around the end of
 Sec.\,\ref{sec:general}.

 Note that the minimal neutrino seesaw generally has vanishing determinant and thus,
 \begin{equation}
 0 ~=~ \det \ov{M}_{\nu}^{(0)} ~=~
 \left[
  \overline A_0 \left( \overline C_0 + \overline D_0 \right)
 - 2 \overline B^2_0
 \right]
 \left( \overline C_0 - \overline D_0 \right) \,,
 \end{equation}
 which, for $\,\ov{C}_0 \neq \ov{D}_0\,$ [as indicated in (\ref{eq:A0B0C0D0})],
 imposes a condition,
 \beqa
   \overline A_0 \left( \overline C_0 + \overline D_0 \right)
 ~=~ 2 \overline B^2_0 \,.
 \label{eq:det=0}
 \eeqa
 It is easy to verify this condition
 by explicitly inputting the LO predictions of (\ref{eq:A0B0C0D0}).
 With (\ref{eq:det=0}) we can readily deduce the mixing angle from
 (\ref{eq:U12-diagonalization}) or (\ref{eq:t12-general}),
 \begin{eqnarray}
 \tan \theta_s  ~=\,  -\f{\ov{A}_0}{\,\sqrt{2}\,\ov{B}_0\,} \,,
 \label{eq:tan2theta12}
 \end{eqnarray}
 as well as the mass-eigenvalues from (\ref{eq:t12-general}),
 \beqa
 \label{eq:m123-MSS}
 \ma ~=~ 0 \,, ~~~~~~
 \mb ~=~ \overline A_0 + \overline C_0 + \overline D_0 \,, ~~~~~~
 \mc ~=~ \overline C_0 - \overline D_0 \,.
 \eeqa
 We may also derive the above mass-spectrum by a different way. From the trace of
 $\,\ov{M}_{\nu}^{(0)}\,$ or $\,\ov{M}_{\nu}^{(0)\prime}\,$,\,
 we know the sum of three mass-eigenvalues,
 $\,\ma+\mb+\mc = \overline A_0 + 2 \overline C_0\,$,\, while Eq.\,(\ref{eq:MnuS2}) gives
 $\,\mc = \overline C_0 - \overline D_0\,$.\, So, we have
 $\,\ma+\mb = \overline A_0 + \overline C_0 + \overline D_0\,$.\,
 The minimal seesaw must have vanishing determinant of $\ov{M}_{\nu}^{(0)}$ and
 thus $\,\ma\mb\mc =0\,$.\,
 Hence, we have, $\,(\ma ,\,\mb)=(0,\,\overline A_0 + \overline C_0 + \overline D_0)$\,
 or $\,(\ma ,\,\mb)=(\overline A_0 + \overline C_0 + \overline D_0,\,0)$\,.\,
 But to determine whether $\ma$ or $\mb$ equals zero will require
 explicit derivation of the rotation matrix $U_{12}(\theta_s)$ and eigenvectors,
 as we did above, which predicts \,$\ma =0$\,,\, corresponding to the NH mass-spectrum for
 light neutrinos.

Using the explicit form of the $\mutau$ and CP symmetric
mass-matrix (\ref{eq:Mnu0-bar})-(\ref{eq:A0B0C0D0}),
we can readily deduce the solar mixing angle from (\ref{eq:tan2theta12}),
\beqa
  \tan \theta_s
~=~
  -\f{\sqrt{2}a}{\,b + c\,} \,,
  \label{eq:theta12-c}
\eeqa
where by a proper phase shift $\,\ab_1^{} \to \ab_1+n\pi\,$,
we can simply convert (\ref{eq:theta12-c}) to (\ref{eq:t12}) or (\ref{eq:LO-Angle})
which always ensures  $\,\tan\ts \geqq 0\,$.\,
The formula (\ref{eq:theta12-c}) or (\ref{eq:LO-Angle}) also shows that
solar angle $\ts$ derived from the $\mutau$ symmetric mass-matrix $\ov{M}_\nu^{(0)}$
is solely determined by the elements of the Dirac mass-matrix $\mD$ and is independent of
the Majorana mass-matrix $\,M_R$\,.\,
Besides, since $\ov{M}_\nu^{(0)}$ is real, all CP phases are irrelevant here.
Finally, substituting (\ref{eq:A0B0C0D0}) into (\ref{eq:m123-MSS}), we further derive
the mass-eigenvalues for light neutrinos in the $\mutau$ symmetric limit,
 \begin{eqnarray}
 m_{1}^{} \,=\, 0 \,,~~~~~~
 m_{2}^{} \,=\,
 \frac{\,2a^2\!+\!(b\!+\!c)^2\,}{M_{22}(2\!-\!r)} \,,~~~~~~
 m_{3}^{} \,=\, \frac{(b-c)^2}{\,M_{22}|r|\,} \,,
 \label{eq:m10m20m30}
 \end{eqnarray}
 where we can always ensure $\,m_3^{}>0\,$ by absorbing the sign of
 $r$ into the Majorana pahse $e^{i\phi_3^{}}$ in $U'$
 [cf. Eqs.\,(\ref{eq:Mnu-V-D}) and (\ref{eq:U'U''})].
 The mass formula (\ref{eq:m10m20m30}) proves what we gave earlier in Eq.\,(\ref{eq:LO-Mass})
 of Sec.\,\ref{sec:zeroth-seesaw} under the $\mutau$ symmetric limit (with the sign choice
 $\,\pb=+\,$).

\vspace*{3mm}
\subsection{%\hspace*{-4.7mm}.\hspace*{-0.6mm}
Solar Mixing Angle Not Affected by Soft Breaking}
\label{subsec:solar-mixing-not-affected}

 In this subsection, we present a more general proof that
 the solar mixing angle $\ts$ is determined by the zeroth order $\mutau$ symmetric
 mass-matrix $\,M_\nu^{(0)}\,$ only, independent of any $\mutau$ and CP breaking
 parameter in either $\,\d M_\nu^{s}\,$ or $\,\d M_\nu^{a}\,$
 [cf.\ (\ref{eq:Mnu-Split-0}) below]\footnote{A previous study\,\cite{Grimus2}
 analyzed a perturbative violation of a general low energy
 $\mutau$ symmetric mass-matrix of light neutrinos,
 it also noted that $\theta_{12}$ (and mass-eigenvalues)
 do not receive any $\mutau$ breaking correction.
 We thank an anonymous referee for kindly bringing \cite{Grimus2} to our attention.}.\,
 According to the decomposition method in
 Eqs.\,(\ref{eq:Mnu-decomposition})-(\ref{eq:Mnu-decomposition-all}),
 we can uniquely split any mass-matrix $M_\nu$ for light neutrinos into their
 $\mutau$ symmetric and anti-symmetric parts,
 \beqs
 \beqa
 \label{eq:Mnu-Split-0}
 && M_\nu ~=~ M_\nu^{s} + M_\nu^{a} \,,
 \\[2mm]
 \label{eq:Mnu-Split-s+a}
 && M_\nu^{s} ~=~ M_\nu^{(0)} + \d M_\nu^{s} \,,
    ~~~~~ M_\nu^{a} ~=~ \d M_\nu^{a} \,,
 \eeqa
 \eeqs
 where the superscripts $^s$ and $^a$ denote the $\mutau$ symmetric and anti-symmetric parts,
 respectively.
 The $\mutau$ symmetric $\,\d M_\nu^{s}\,$ is due to small $\O(y)$ correction
 (related to solar mass-squared difference $\Delta m_{21}^2$),
 and $\,\d M_\nu^{a}\,$ are induced by the $\mutau$ breaking
 effects characterized by the two small parameters $\,\d_x\,$ and $\,\d_a\,$.\,
 Since neutrino data require $(y,\,\d_x,\,\d_a)$ to be small, it is enough to expand
 its mass-eigenvalues (in $D_\nu$) and diagonalization matrix $V$ up to the linear order of
 $(y,\,\d_x,\,\d_a)$  for all practical purposes. Thus, we write,
 \beqs
 \beqa
 \label{eq:Dnu-s-a}
 D_\nu &\!\equiv\!&  D_\nu^{s} + \d D_\nu^{a}
 ~=~  (D_\nu^{(0)} + \d D_\nu^{s}) + \d D_\nu^{a} \,,
 \\[1mm]
 \label{eq:V-s-a}
 V &\!\equiv\!&
 V_s + \d V_a ~=~ (V_0 + \d V_s) + \d V_a \,,
 \eeqa
 \label{eq:Dnu-V}
 where $V_s$ corresponds to the $\mutau$ symmetric solution (\ref{eq:mutau-solution})
 at the end of Sec.\,\ref{sec:model-independent-reconstruction-formalism},
 \beqa
 \label{eq:Vs-def}
 V_s ~\equiv~ V\(\ts,\,\ta=45^\deg,\,\tx =0^\deg|\a_2^{}=\a_3^{}\) ,
 \eeqa
 \eeqs
and $\,\d V_a \propto (\d_x,\,\d_a,\,\a_2^{}-\a_3^{})\,$.\,
From (\ref{eq:Dnu-V})-(\ref{eq:Vs-def}), we derive the following,
\beqa
M_\nu
&=& (V_s^* + \d V_a^*) D_\nu (V_s^\dag + \d V_a^\dag)
\nn\\
&=& V_s^* D_\nu V_s^\dag
+ \(V_s^* D_\nu \d V_a^\dag + \d V_a^* D_\nu V_s^\dag + \d V_a^* D_\nu\d V_a^\dag\)\,.
\label{eq:Mnu-s0-ds1-da1}
\eeqa
Since the decomposition (\ref{eq:Mnu-Split-0}) is unique,
so we can deduce,
\beqs
\beqa
M_\nu^s
&=& V_s^* D_\nu V_s^\dag \,,
\label{eq:Recons-Mnu-s}
\\[1mm]
\d M_\nu^{a} &=&
V_s^* D_\nu \d V_a^\dag + \d V_a^* D_\nu V_s^\dag + \d V_a^* D_\nu\d V_a^\dag
\nn\\[1mm]
&=& V_0^* D_\nu^{(0)} \d V_a^\dag + \d V_a^* D_\nu^{(0)} V_s^\dag + \O (\d_j^2) \,,
\label{eq:dMnu-(a)}
\eeqa
\eeqs
where $\d_j$ denotes the generic NLO
parameters under consideration (such as $\d_x$, $\d_a$ and $y$, etc).

A key point here is to note that the $\mutau$ symmetric solution (\ref{eq:mutau-solution})
does not restrict $\ts$ and thus there is no need to expand $\ts$ in $V_s$ (despite
the $\mutau$ breaking parameters are small). Hence, we can determine $\ts$ by simply solving
$V_s$ from the diagonalization equation (\ref{eq:Recons-Mnu-s}).
In Sec.\,\ref{subsec:solar}, we have used $V_s$ to diagonalize the $\mutau$ symmetric mass-matrix
$\,\ov{M}_\nu^{(0)}\equiv M_\nu^s|_{\zeta =0}\,$ and found that the formula of $\ts$
is determined by $\mD$ only, but independent of $M_R$.\,
Now with full $\,M_\nu^s|_{\zeta \neq 0}\,$,
we will prove that the formula of $\ts$ remains unchanged.

Inspecting the general form of $\,M_\nu^s$\,,
\beqa
M_\nu^s ~=\,
  \begin{pmatrix}
    A & B_s & B_s \\
      & C_s & D \\
      &     & C_s
  \end{pmatrix} ,
\label{eq:Ms-full}
\eeqa
we can diagonalize it by the rotation $U_{23}(45^\deg)$,
 \beqa
 M^{s\prime}_\nu
 ~\equiv~
 U_{23}(45^\deg)^T M^s_\nu \,U_{23}(45^\deg)
 ~=\,
 \(
 \ba{ccc}
   A               & \sqrt{2} \, B_s &  0 \\[2mm]
   \sqrt{2} \, B_s & C_s \!+\! D           &  0 \\[2mm]
    0 & 0 & C_s \!-\! D
 \ea
 \) .
 \label{eq:Ms-diag1}
 \eeqa
Then the remaining $\{12\}$ sub-block can be readily diagonalized by the
rotation $U_{12}(\ts)$ with
\beqa
\tan \ts ~=\, -\f{A}{\,\sqrt{2}\,B_s\,} \,,
\label{eq:tantheta12-f}
\eeqa
which extends Eq.\,(\ref{eq:tan2theta12}) with the $\,\zeta \neq 0\,$ effects included.
The elements of (\ref{eq:Ms-full}) can be directly inferred from the complete
seesaw mass-matrix (\ref{eq:Mnu-all0}), {\it without any expansion,}
\beqs
\beqa
A &\!\!=\!\!& \f{a^2(2\!-\!X)}{~\,(2\!-\!r\!-\!X)M_{22}~}
      ~=~ \f{\,(2\!-\!r)(2\!-\!X)\,}{2(2\!-\!r\!-\!X)}\,\ov{A}_0 \,,
\label{eq:ABsCsD-1}
\\[2mm]
B_s &\!\!=\!\!& \f{a(b\!+\!c)(2\!-\!X)}{\,2(2\!-\!r\!-\!X)M_{22}\,}
      ~=~ \f{\,(2\!-\!r)(2\!-\!X)\,}{2(2\!-\!r\!-\!X)}\,\ov{B}_0 \,,
\label{eq:ABsCsD-2}
\\[2mm]
C_s+D &\!\!=\!\!& \f{(b\!+\!c)^2(2\!-\!X)}{\,2(2\!-\!r\!-\!X)M_{22}\,}
       ~=~ \f{\,(2\!-\!r)(2\!-\!X)\,}{2(2\!-\!r\!-\!X)}\,(\ov{C}_0+\ov{D}_0) \,,
\label{eq:ABsCsD-3}
\eeqa
\label{eq:ABsCsD}
\eeqs
where after the second equality of each equation we have used the $\mutau$ symmetric form
in (\ref{eq:Mnu0-bar})-(\ref{eq:A0B0C0D0}).
The relations (\ref{eq:ABsCsD-1})-(\ref{eq:ABsCsD-3}) prove that the elements
\,$(A,\,B_s,\,C_s+D)$\, differ from the LO results
\,$(\ov{A}_0,\,\ov{B}_0,\,\ov{C}_0+\ov{D}_0)$\,
only by a common overall factor which is irrelevant to the mixing angle $\,\theta_s\,$.\,
With these, we explicitly deduce from (\ref{eq:tantheta12-f}),
\beqa
\tan \ts ~=\, -\f{\ov{A}_0}{\,\sqrt{2}\,\ov{B}_0\,}
~=\, -\f{\sqrt{2}a}{\,b\!+\!c\,}
\,,
\label{eq:theta12-d}
\eeqa
in perfect agreement with Eqs.\,(\ref{eq:tan2theta12}) and (\ref{eq:theta12-c}),
which we derived earlier under the $\mutau$ symmetric limit in Sec.\,\ref{subsec:solar}.
As noted below (\ref{eq:theta12-c}), making a proper phase-shift
$\,\ab_1^{} \to \ab_1+n\pi\,$, we always ensure
$\,\tan\ts >0\,$ and thus $\,\ts\in [0,\,\f{\pi}{2}]$\,.
Thus we may rewrite (\ref{eq:theta12-d}) as
\beqa
\tan\ts ~=\, \f{\sqrt{2}|a|}{\,|b\!+\!c|\,}
        ~=\, \f{\sqrt{2}\,}{|k|} \,,
\label{eq:theta12-dd}
\eeqa
where we have used the definition of
$\,k\equiv (b+c)/a\,$ as in (\ref{eq:dBa-dCa-k}).

This completes our proof that the solar mixing angle $\theta_s$ is not affected
by the $\mutau$ breaking and depends on the Dirac mass-matrix $\mD$ only.
As a byproduct of the next subsection, we will further reveal
that this fact actually originates from the elegant feature of
a new hidden symmetry underlying the seesaw Lagrangian.

\vspace*{3mm}
\subsection{Beyond $\boldsymbol{\mutau}$ Symmetry -- A New Hidden Symmetry for
            $\boldsymbol{\theta_{12}^{}}$}
\label{sec:extra-z2}

In Sec.\,\ref{subsec:solar}-\ref{subsec:solar-mixing-not-affected},
we have revealed that the solar mixing angle $\theta_s\,(\equiv\theta_{12})$ is fully
determined by the structure of the Dirac mass-matrix $\mD$ only,
independent of the Majorana mass-matrix $\,M_R$\,.
In particular,  Eq.\,(\ref{eq:theta12-dd}) shows that $\,\ts$\, just depends on a ratio,
\beqa
\label{eq:k}
k ~\equiv\, \f{\,b+c\,}{a} \,.
\eeqa
Contemplating this striking feature, we wonder: {\it Is this unique ratio $k$ connected to
any new hidden symmetry underlying the structure of $\,\mD$\,?}

The simplest such hidden symmetry we could possibly imagine
is a new discrete $\ZZ_2$ symmetry which we will
denote as $\,\ZZ_2^s\,$ to indicate its connection to the solar angle \,$\ts$\,.\,
Under this new $\,\ZZ_2^s\,$ symmetry, we expect the three light neutrinos
$\nuL =(\nu_e^{},\,\nu_\mu^{},\,\nu_\tau^{})^T$ transform in its 3-dimensional representation
and the right-handed neutrinos are just singlets (as the above ratio $k$ is independent of
$M_R$),
\beqa
\label{eq:Z2-s}
  \nuL \,\to\,  T_s \,\nu_L \,,
&&~~~
  \mathcal{N} \,\to\,  \mathcal{N} \,.
\eeqa
If the hidden $\,\ZZ_2^s\,$ is a fundamental flavor symmetry of our theory,
it must keep the seesaw Lagrangian (\ref{eq:L-seesaw}) invariant under the transformation
(\ref{eq:Z2-s}).  This means,
\beqa
\label{eq:UL-mD}
  T_s^\dag \mD ~=~  \mD \,,
\eeqa
where the Dirac mass-matrix $\mD$ respects the $\mutau$ and CP symmetries
in our construction. Thus the $3\times 3$ unitary transformation matrix $\,T_s\,$
should be real and orthogonal,
$\,T_s^\dag T_s \,=\, T_s^T T_s ~=~ {\cal I}_3\,$,\,
with $\mathcal{I}_3$ the $3\times 3$ unit matrix.
Since $\,T_s \in \ZZ_2^s\,$,\, we must also have $\,T_s^2={\cal I}_3\,$.\,
Imposing these requirements we can solve out $\,T_s\,$ from the invariance
condition (\ref{eq:UL-mD}),
\beqa
\label{eq:UL-k}
  T_s(k) ~=~
  \f{1}{\,2 \!+\! k^2\,}\!
  \(\!\ba{crr}
    2 \!-\! k^2 & 2 k & 2 k \\[1.3mm]
    2 k & k^2 & - 2 \\[1.3mm]
    2 k & - 2 & k^2
  \ea\!\) ,
\eeqa
where $\,k=\f{\,b+c\,}{a}\,$ as defined in (\ref{eq:k}).
It is straightforward to explicitly verify that
$\,T_s(k)\,$ indeed obeys the invariance equation (\ref{eq:UL-mD})
as well as the desired conditions,
\beqa
T_s^\dag T_s \,=\, T_s^T T_s \,=\, {\cal I}_3 \,,
&& ~~~~
T_s^2 \,=\, {\cal I}_3 \,.
\eeqa
Hence, $T_s(k)$ does form a 3-dimensional real orthogonal representation of the hidden
symmetry $\ZZ_2^s$ in our seesaw Lagrangian (\ref{eq:L-seesaw}), and the
ratio $\,k=\f{\,b+c\,}{a}\,$ is just the group-parameter of $\,\ZZ_2^s\,$.

We further note that the new $\ZZ_2^s$ symmetry is also generally respected
by the low energy seesaw mass-matrix (\ref{eq:seesaw-formula}),
\beqa
  T_s^T M_\nu T_s^{}
~=~ T_s^T \mD M^{-1}_R (T_s^T\mD)^T
~=~ \mD M^{-1}_R m_D^T
~=~  M_\nu \,.
\eeqa
Revealing this hidden symmetry $\ZZ_2^s$, we can now fully understand why
solar mixing angle $\ts$ is not affected by the Majorana mass-matrix $M_R$ and the
soft $\mutau$ and CP breakings therein. This is because, as determined
in (\ref{eq:theta12-dd}), {\it the $\ts$ is generally protected
by the hidden symmetry $\,\ZZ_2^s\,$ via its group parameter $\,k\,$,\,
independent of the form of $\,M_R$\,.}

This is an intriguing and essential feature of our soft breaking seesaw model, which
poses two discrete flavor symmetries $\,\ZZ_2^{\mu\tau}\otimes \ZZ_2^s\,$ in addition
to the CP-invariance, among which $\,\ZZ_2^{\mu\tau}\,$ and CP receive small soft breakings
from a common origin in $\,M_R$\,.\,
The exact hidden symmetry $\,\ZZ_2^s\,$ dictates the solar angle
$\ts\,(\equiv\theta_{12})$, while the common soft breaking of $\mutau$ and
CP connects $\,\dx\,(\equiv\theta_{13})\,$
and $\,\da\,(\equiv\theta_{23}\!-\!45^\deg)\,$.\,
Under the symmetry group $\,\ZZ_2^{\mu\tau}\otimes \ZZ_2^s\,$
and using Eq.\,(\ref{eq:theta12-dd}),
we can reduce the neutrino mixing matrix $\,U(\ts,\,\ta,\,\tx)\,$ in (\ref{eq:U}) to
\beqa
\label{eq:Uk}
U(k) ~=~ U(\ts,45^\deg,0^\deg )
~=
\(\ba{rrr}
c_s^{} & -s_s^{} & 0 \\[2mm]
\f{s_s^{}}{\sqrt{2}} & \f{c_s^{}}{\sqrt{2}} & \f{-1}{\sqrt{2}} \\[2mm]
\f{s_s^{}}{\sqrt{2}} & \f{c_s^{}}{\sqrt{2}} & \f{1}{\sqrt{2}}
\ea\)
=
\(\ba{rrr}
\f{|k|}{\sqrt{2+k^2}} & \f{-\sqrt{2}}{\sqrt{2+k^2}} & 0 \\[2mm]
\f{1}{\sqrt{2+k^2}}   & \f{|k|}{\sqrt{2(2+k^2)}}    & \f{-1}{\sqrt{2}} \\[2mm]
\f{1}{\sqrt{2+k^2}}   & \f{|k|}{\sqrt{2(2+k^2)}}    & \f{1}{\sqrt{2}}
\ea\)  \!. ~~~~
\eeqa

We note that this $\,\ZZ_2^s\,$ symmetry has a simple geometric interpretation.
In the invariance equation (\ref{eq:UL-mD}) the Dirac mass-matrix $\mD$ consists of two
columns which can be viewed as two vectors, i.e., $\,\mD = (u_1^{},\,u_2^{})\,$
with $\,u_1^{}=(a,\,b,\,c)^T\,$ and $\,u_2^{}=(a,\,c,\,b)^T\,$
in a 3-dimensional coordinate frame. Then, the invariance equation (\ref{eq:UL-mD}) just
means the eigenvalue equations,
\beqa
T_s u_1^{} ~=~ u_1^{} \,,
&~~~&
T_s u_2^{} ~=~ u_2^{} \,.
\eeqa
The vectors $u_1^{}$ and $u_2^{}$ must be linearly independent for any realistic $\,\mD$\,,\,
and thus determine a plane $S$.  The equation for the plane $S$ reads,
\beqa
\label{eq:S-equation}
x - \f{1}{k}(y+z) ~=~ 0\,,
\eeqa
which is characterized by a single parameter
$\,k\,$ as defined in (\ref{eq:k}).
For any other given vector $\,u'\,$, so long as it lives in the plane $S$,
it can be expressed as a linear combination of $(u_1^{},\,u_2^{})$, and thus is also
an eigenvector of $\,T_s\,$,\, leading to $\, T_su' = u' \,$.
If $\,u'\,$ lives outside the plane $S$, i.e.,
disobeys Eq.\,(\ref{eq:S-equation}), then the transformation $\,T_s\,$ reflects
the vector $\,u'\,$ respect to the plane $S$, as $\,T_su'\,$.\,
Since the plane $S$ has the normal vector
$~n_{\bot}^{}=(k,-1,-1)/\sqrt{2+k^2}~$ perpendicular to itself,
it is clear that $\,T_s n_{\bot}^{} = -n_{\bot}^{}\,$ holds.
Then, we can project the vector $\,u'\,$ along the directions
parallel and perpendicular to the plane $S$,
$\,u' = u'_{\|}+u'_{\bot}\,$,\, so we have
\beqa
\label{eq:reflection}
T_s u' ~=~ T_su'_{\|}+T_su'_{\bot} ~=~ u'_{\|}-u'_{\bot}\,,
&&~~~
T_s^2 u' ~=~ u'\,,
\eeqa
where $\,u'_{\|}\,$ lies in the plane $S$ and is thus a linear
combination of vectors $\,(u_1^{},\,u_2^{})$\,,\, while
$\,u'_{\bot}\propto n_{\bot}^{}\,$.\,
Hence, Eq.\,(\ref{eq:reflection}) proves that the operation
$\,T_s\,$ does reflect the arbitrary vector
$\,u'\,$ respect to the plane $S$, and $\,u'\,$
goes back to itself after applying $\,T_s\,$ twice.
It is clear that such transformations form the 3-dimensional representation
of a discrete $\ZZ_2$ group.

For the case of general seesaw with three right-handed neutrinos
$\,{\cal N}=(N_1,\,N_2,\,N_3)^T\,$,\, the Dirac mass-matrix $\mD$ is extended to
a $3\!\times\! 3$ matrix which, under the $\mutau$ symmetry, takes the form,
\beqa
\label{eq:mD-3x3}
m_D' ~=~
\(\ba{lll}
a' & a & a
\\[1mm]
b' & b & c
\\[1mm]
b' & c & b
\ea\)
\eeqa
where all elements are real due to CP-conservation.
Now, to hold the invariance equation (\ref{eq:UL-mD}) we just need the first column
$\,u_3^{}=(a',\,b',\,b')^T\,$ to live in the $S$ plane. Thus the plane equation
(\ref{eq:S-equation}) requires,
\beqa
\label{eq:a'b'-cond}
\f{a'}{\,2b'\,} ~=~ \f{a}{\,b+c\,} ~=~ \f{1}{k} \,,
\eeqa
under which the symmetry $\,\ZZ_2^s\,$ holds for general three-neutrino-seesaw,
and the solar mixing angle is still determined by the group parameter $k$
via Eq.\,(\ref{eq:theta12-dd}).
We note that the $m_D'$ in (\ref{eq:mD-3x3}) contains
two more parameters $(a',\,b')$ than the minimal seesaw form of $\,\mD$\,,
but the $\ZZ_2^s$ symmetry requires only one of them be independent due to
the condition (\ref{eq:a'b'-cond}).
Furthermore, from Eq.\,(\ref{eq:a'b'-cond}) we can express $\,u_3^{}\,$
as a combination of $(u_1^{},\,u_2^{})$,
\beqa
u_3 ~=~ \f{b'}{\,b+c\,}(u_1^{} + u_2^{}) \,,
\eeqa
which forces  $m_D'$ to be rank-2 and thus $\,\det m_D'=0\,$.
So, the corresponding seesaw mass-matrix for light neutrinos,
\beqa
M_\nu' ~=~ m_D' M_R^{\prime -1} m_D^{\prime \,T} \,,
\eeqa
must have vanishing determinant because
$\,\det M_\nu' = (\det m_D')^2(\det M_R')^{-1} = 0\,$
for any $3\times 3$ heavy Majorana mass-matrix $\,M_R'\,$.\,
It is clear that this feature depends only on $\,m_D'\,$.\,

Hence, we can state a general theorem:
{\it for the $\,\ZZ_2^s\,$ symmetric seesaw Lagrangian,
the Dirac mass-matrix $\,m_D'\,$ is rank-2 at most, enforcing the seesaw
mass-matrix $\,M_\nu'\,$ to be rank-2, having zero determinant and thus
one zero-mass-eigenvalue\footnote{The possibility of $\,m_D'\,$ or $\,M_\nu'\,$ being rank-1
is already ruled out by the oscillation data where both $\Delta_a$ and $\Delta_s$
are nonzero (Table-\ref{tab:1}),
requiring at least two non-vanishing mass-eigenvalues for light neutrinos.}.\,}
This means that $\,M_\nu'\,$ shares a similar feature with the minimal seesaw
(including two right-handed heavy neutrinos),
and they both have, $\,m_1^{}m_2^{}m_3^{}=0\,$.

Finally, we note that our $\,\ZZ_2^s\,$ symmetry allows its group-parameter $k$
to take different values,
and thus gives different values of the solar mixing angle $\,\ts$\,.
For instance, to predict the conventional tri-bimaximal
mixing\,\cite{TBM} only needs to assign $\,|k|=2\,$ in our construction, leading to
$\,\tan\ts = \f{1}{\sqrt{2}\,}$ ($\,\ts \simeq 35.3^\deg\,$);\,
while we can make another equally simple assignment of $\,|k|=\f{3}{\sqrt{2}\,}\,$ to generate
\,$\tan\ts =\f{2}{3}\,$ ($\,\ts\simeq 33.7^\deg$)\,,\,
which is well within the $1\sigma$ range of $\ts$ and
agrees even better to its central value in Table-\ref{tab:1}.

The hidden symmetry $\,\ZZ_2^s\,$ points to a very encouraging direction although it is
not yet powerful enough to fix the group-parameter $k$ on its own.
We expect that at a higher scale the two elegant discrete symmetries
$\,\ZZ_2^{\mu\tau}\otimes \ZZ_2^s\,$
can be unified into a larger flavor symmetry which is more restrictive
and thus fully fixes \,$k$\,.\,
For instance, let us consider a larger group $\,S_4$\,\cite{S4},
the permutation group of four objects, which has five irreducible representations
$\,\{\bf{1},\,\bf{1}',\,\bf{2},\,\bf{3},\,\bf{3}'\}\,$,\, and assign the left-handed neutrinos
to $\,\bf{3}$\,.\,  With boldface superscript denoting a given irreducible representation,
the 3-dimensional representations $G_j$ ($j=2,3$) of \,$S_4$\, are,
\beqa
G_2^{\bf 3} ~=~ G_2^{\bf 3'} \,=
\(\!\!\ba{rrr}
-\f{1}{3} & \f{2}{3} & \f{2}{3}
\\[1mm]
\f{2}{3} & -\f{1}{3} & \f{2}{3}
\\[1mm]
\f{2}{3} & \f{2}{3} & -\f{1}{3}
\ea\) \!,
&&
G_3^{\bf 3} ~=\, -G_3^{\bf 3'} \,=
\(\ba{rrr}
1 & 0 & 0
\\[1mm]
0 & 0 & 1
\\[1mm]
0 & 1 & 0
\ea\) \!,
\eeqa
and the $\,G_1^{\bf 3} = G_2^{\bf 3}G_3^{\bf 3}\,$ is non-independent.
With these, we can thus identify,
\beqs
\beqa
&& G_3^{\bf 3} ~=\, -G_3^{\bf 3'} ~=~ T_L^{} \,,
\\[1mm]
&& G_2^{\bf 3} ~=~  G_2^{\bf 3'} ~=~ T_s(k=2) \,,
\eeqa
\eeqs
where $T_L^{}$ is the $\ZZ_2^{\mu\tau}$ transformation matrix we defined in
(\ref{eq:T3new}).  We note that $\,T_s(k=2)\,$ is just a special case of
our 3-dimensional representation $\,T_s(k)\,$ of the hidden
$\ZZ_2^s$ symmetry, fixing $\,k=2\,$ and corresponding to the
conventional tri-bimaximal ansatz\,\cite{TBM}, $\,\tan\ts =\f{1}{\sqrt{2}\,}\,$,\,
via our Eq.\,(\ref{eq:theta12-dd}).
This also supports the previous studies of tri-bimaximal ansatz that leads to
the $S_4$ group\,\cite{S4}.

Strikingly, our general construction of the hidden $\,\ZZ_2^s\,$ symmetry
contains the tri-bimaximal mixing as {\it a special case ($k=2$) and allows deviations from it,}
e.g., as noted above, another simple choice,
\,$k=\f{3}{\sqrt{2}\,}\,$,\, leads to $\,\tan\ts =\f{2}{3}\,$ with a better agreement to
the measured value of $\,\ts\,$ (Table-\ref{tab:1}).
This direction will be further explored in a forthcoming publication\,\cite{dgh}.

 \vspace*{4mm}
 \section{Conclusions}
 \label{sec:conclusions}

 The oscillation data have provided compelling evidence for the
 $\mutau$ symmetry as a good approximate symmetry in the neutrino sector.
 The $\mutau$ symmetry requires vanishing mixing angle $\,\theta_{13}\,$
 and thus the Dirac CP-conservation. So, the $\mutau$ breaking term must be small
 and also serve as the source of Dirac CP-violation.
 On the theory ground, it is natural and tempting to expect all
 CP violations arising from a common origin, implying that both
 Dirac and Majorana CP violations vanish in the $\mutau$ symmetric limit.
 In this work, based upon these
 we have conjectured that both discrete $\mutau$ and CP symmetries are fundamental
 symmetries of the seesaw Lagrangian (respected by interaction terms) and
 are {\it only softly broken from a common origin.} Such soft breaking has to
 arise from a {\it unique dimension-3 Majorana mass-term} of the
 heavy right-handed singlet neutrinos, as shown in Sec.\,\ref{sec:common-origin}.
 This conjecture can hold for general seesaw with three right-handed neutrinos;
 we have chosen the minimal neutrino seesaw\,\cite{MSS} in Sec.\,2 and shown
 that the soft $\mutau$ and CP breakings
 are further characterized by a single complex parameter in the mass-matrix $M_R$
 of right-handed neutrinos [cf.\ (\ref{eq:mD-MR-FF})].
 (In Sec.\,6.3 we have proven that the general three-neutrino-seesaw under the
 $\mutau$ symmetry $\,\mathbb{Z}_2^{\mu\tau}\,$ and hidden symmetry $\,\mathbb{Z}_2^s\,$
 also predicts a massless light neutrino, sharing the same feature as the minimal seesaw.)

 From this conceptually attractive and simple construction,
 we have predicted the soft $\mutau$ and CP breaking effects at low energies.
 This gives the quantitative {\it correlations}
 between the {\it two apparently small deviations}
 $\,\tbc \!-45^\deg\,(\equiv\da )$ and $\,\tac\!-0^\deg\,(\equiv\dx )$.\,
 For any nonzero $\,\tbc \!-45^\deg\,$,\,
 we can place a generic {\it lower limit} on the mixing angle $\,\theta_{13}$\,
 in Eq.\,(\ref{eq:Lbound-dx}) of Sec.\,\ref{sec:prediction-low-energy-parameter}.
 This feature is demonstrated in Fig.\,\ref{fig:dx-da} (Sec.\,\ref{sec:full-NLO})
 without requiring leptogenesis, where $\,\theta_{13}$\, is shown as a function of
 the deviation $\,\tbc \!-45^\deg\,$ in its \,90\%\,C.L.\, range.
 Fig.\,\ref{fig:dx-da} also predicts a new {\it upper bound,} $\,\theta_{13}\lesssim 6^\deg\,$,\,
 for full range of $\,\theta_{23}\,$.\,
 Adding the successful leptogenesis in Fig.\,\ref{fig-deltax-deltaa-new}
 (Sec.\,\ref{sec:model-prediction-leptogenesis})
 further predicts a nontrivial {\it lower bound,} $\,\theta_{13}\gtrsim 1^\deg\,$, even for
 $\,\theta_{23}\sim 45^\deg\,$.\,
 Fig.\,\ref{fig:dx-da} and Fig.\,\ref{fig-deltax-deltaa-new}
 quantitatively connects the on-going measurements of
 mixing angle $\,\theta_{23}\,$ with the upcoming probe of $\,\theta_{13}\,$
 at the Double-Chooz, Daya Bay, T2K and NO$\nu$A experiments, etc.
 We note that the current measurements of
 $\,\theta_{23}\,$ already show an interesting deviation from
 the maximal value $45^\deg$ in its central value (due to the subleading effects
 in the oscillation\,\cite{D23-Smirnov,Lisi-09}),  but still has sizable errors which are larger
 than that of $\,\theta_{12}\,$ by a factor of $\,3\sim 4\,$ and are also comparable to that of
 $\,\theta_{13}\,$ [cf.\ Table-\ref{tab:1} and Eq.\,(\ref{eq:da-dx-exp})].\,
 Fig.\,\ref{fig:dx-da} and Fig.\,\ref{fig-deltax-deltaa-new} reveal that a more precisely measured
 deviation $\,\theta_{23}-45^\deg\,$  will definitely put stronger lower limit on $\,\theta_{13}\,$.\,
 Hence, {\it our findings strongly encourage experimental efforts to improve the precision
 of $\,\theta_{23}\,$ as much as possible.}

 We have further derived the low energy Dirac and Majorana CP-violations from a common soft-breaking
 phase associated with $\mutau$ breaking in the neutrino seesaw. In particular, the low energy Dirac
 phase angle $\,\d_D\,$ and Majorana phase angle $\,\phi_{23}^{}\,$ are predicted in terms of the original
 soft CP-breaking phase angle $\,\om\,$ in $\,M_R\,$,\,
 by Eqs.(\ref{eq:sol-deltaD'}) and (\ref{eq:phi23}), respectively.
 The correlations of the low energy Jarlskog invariant $J$
 and neutrinoless double decay observable $M_{ee}$ with the mixing angle $\theta_{13}$ are analyzed,
 as depicted in Fig.\,\ref{fig:dx-J-Mee} (without requiring leptogenesis)
 and Fig.\,\ref{fig:dx-J-Mee-new} (with successful leptogenesis).
 Then, we studied the origin of matter (cosmological baryon asymmetry)
 via leptogenesis, which leads to a robust lower bound
 on the leptogenesis scale, $\,M_1\gtrsim 3\!\times\! 10^{13}\,$GeV,
 as in (\ref{eq:M1-LowerBound}) of Sec.\,\ref{sec:solution}.
 We further analyzed the interplay between the leptogenesis scale $M_1$ and the low energy
 Jarlskog invariant $\,J\,$ as well as
 the neutrinoless double-beta decay observable $\,M_{ee}\,$,\, shown in
 Fig.\,\ref{fig-J-D-new}a-b.

 Finally, we proved in Sec.\,\ref{subsec:solar}-\ref{subsec:solar-mixing-not-affected}
 that the solar mixing angle $\,\theta_{12}\,$ is independent of the
 soft $\mutau$ breaking and only depends on the structure of the Dirac mass-matrix $\,\mD\,$
 as shown in Eq.\,(\ref{eq:theta12-c}) or (\ref{eq:theta12-dd}).  In Sec.\,\ref{sec:extra-z2},
 we further revealed a new hidden symmetry $\,\ZZ_2^s\,$ that dictates the solar mixing angle
 $\,\theta_{12}\,$ in terms of its group parameter $\,k\,$
 [cf.\ (\ref{eq:theta12-dd}) and (\ref{eq:UL-k})].
 The new $\,\ZZ_2^s\,$ can further hold for general three-neutrino-seesaw
 under the condition (\ref{eq:a'b'-cond}).
 This restricts the $3\!\times\! 3$ Dirac mass-matrix \,$m_D'$\,
 to have only one more independent parameter than the corresponding $\,\mD\,$ in the
 minimal seesaw. We further proved that the $3\!\times\!3$ Dirac mass-matrix $\,m_D'\,$ and
 the seesaw mass-matrix $\,M_\nu'\,$ must be rank-2,  predicting a zero mass-eigenvalue
 for $\,M_\nu'\,$;\,  this same feature is also shared in the minimal two-neutrino-seesaw (Sec.\,2).
 In addition, we revealed that the conventional tri-bimaximal mixing (TBM) ansatz\,\cite{TBM}
 is realized as a special case of the $\,\ZZ_2^s\,$ group with $\,k=2\,$,\,
 under which $\,\ZZ_2^s\otimes\ZZ_2^{\mutau}$\, [together with the $\,G_{\ell L}\,$
 symmetry for leptons in (\ref{eq:GL-lep})] is naturally unified into a larger group
 $\,S_4$\, (supporting previous $S_4$ studies\,\cite{S4}).
 We note that the $\,\ZZ_2^s\,$ symmetry allows deviations
 from the TBM and a different choice of our group parameter $k$
 (such as $\,k=\f{3}{\sqrt{2}\,}\,$)
 can give a better agreement to the current data of $\,\theta_{12}$\,.\,
 Further explorations along this direction will be given elsewhere\,\cite{dgh}.

 \vspace*{5mm}
 \noindent
 {\Large\bf Note Added in Proof:}\\
 After the submission of this paper, a newly updated global
 analysis of solar, atmospheric, reactor and accelerator neutrino data for
 three-neutrino oscillations appeared\,\cite{fit2009}, which gives the following
 fitted ranges of the low energy neutrino parameters at $1\sigma$ ($3\sigma$) level,
 \beqa
 && \Delta m_{21}^2 \,=\, \[7.59\pm 0.20\,(^{+0.61}_{-0.69})\]\times 10^{-5}\,\textrm{eV}^2\,,
 \non\\[2mm]
 && \Delta m_{31}^2 \,=\, \left\{\ba{l}
 \[+2.47\pm 0.12\,(\pm 0.37)\]\times 10^{-3}\,\textrm{eV}^2~~(\textrm{NH}),
 \\[2mm]
 \[-2.36\pm 0.07\,(\pm 0.36)\]\times 10^{-3}\,\textrm{eV}^2~~(\textrm{IH}),
 \ea\right.
 \\[2mm]
 &&
 \theta_{12}\,=\, 34.5\pm 1.0^\deg\,\(^{+3.2^\deg}_{-2.8^\deg}\) ,~~~~
 \theta_{23}\,=\, 42.9^{+4.1^\deg}_{-2.8^\deg}\,\(^{+11.1^\deg}_{-7.2^\deg}\) ,~~~~
 \theta_{13}\,=\,  6.8^{+2.2^\deg}_{-2.8^\deg}\,\(\leqq 12.8^\deg \)\, ,
 \non
 \eeqa
 where the abbreviations NH and IH stand for the ``normal hierarchy"
 ($m_1<m_2<m_3$) and ``inverted hierarchy" ($m_1 \sim m_2 > m_3$)
 of light neutrino mass spectra, respectively. Since these updated values change very little
 from the previous global fit summarized in our Table-1\,\cite{Fogli-08}, we find no visible
 correction to all the numerical analyses presented in Figs.\,1-11.

 \vspace*{8mm}
 \noindent
 \addcontentsline{toc}{section}{Acknowledgements}
 {\bf{\Large Acknowledgements}}
 \\[2mm]
 We thank Stefan Antusch, Duane Dicus,  Ferruccio Feruglio, Alexei Yu.\ Smirnov and
 Thomas Weiler for valuable discussions, and especially,
 to Alexei Yu.\ Smirnov for explaining the analysis of
 the deviation \,$(\theta_{23} - 45^\deg ) < 0$\, in \cite{D23-Smirnov},
 and to Gianluigi Fogli and Eligio Lisi for discussing their global analysis
 and the subleading effects related to
 nonzero $\theta_{13}$ and $(\theta_{23} - 45^\deg)$ \cite{Lisi-09}.
 We are grateful to Yi-Fang Wang for discussing the Daya Bay experiment\,\cite{DayaBay},
 Sacha Kopp and Karol Lang for discussing the MINOS experiments\,\cite{MINOS}
 and Josh Klein for discussing the SNO experiment\,\cite{SNO}.
 We also thank Werner Rodejohann and Rabi Mohapatra for discussion and bringing our attention
 to \cite{M-R} which discussed the potential link of $\mutau$ and CP breakings in a different
 context.  This work was supported by the NSF of China (under grants 10625522 and 10635030),
 the National Basic Research Program of China (under grant 2010CB833000),
 and Tsinghua University.
 We also thank Kavli Institute for Theoretical Physics China (KITPC) for
 partial support and for organizing the stimulating workshop
 ``Neutrino Physics and New Physics Beyond the Standard Model'' during the finalization of
 this manuscript. SFG is supported in part by the China Scholarship Council.

\vspace*{8mm}
%\newpage
\appendix

\noindent
{\Large\bf Appendices}
\vspace*{2mm}

\section{\hspace*{0.05mm} Mass Diagonalization for Right-handed Majorana Neutrinos}
\label{Appendix:diagonalization-MR}

In this Appendix, we perform the mass-diagonalization for heavy right-handed Majorana neutrinos.
The Majorana mass matrix \,$M_R$\, and its diagonal form $\,D_R\,$
are connected by the $2\times 2$ unitary rotation $\,V_R$\,,
\beqa
  V^T_R M_R V_R ~=~  D_R
 ~\equiv~
  \begin{pmatrix}
    M_1 & 0 \\[1.5mm]
    0   & M_2
  \end{pmatrix} \!,
\qquad \textrm{or,} %\Longrightarrow
\qquad
  M_R ~=~
  V^*_R D_R V^\dag_R \,.
  \label{eq:diagonalization-MR}
\eeqa
The transformation matrix $V_R$ can be parametrized as
\beqs
\begin{eqnarray}
 && V_R ~\equiv~ U_R'' U_R U_R' ~=\,
  \begin{pmatrix}
    \cR e^{i\gamma_1^{}} & -\sR e^{i\gamma_2^{}} \\[1.5mm]
    \sR e^{i(\beta+\gamma_1^{})} & \cR e^{i(\beta+\gamma_2^{})}
  \end{pmatrix} \!,
  \label{eq:reconstruction-VR}
  \\[3mm]
&&  U_R ~=\,
  \begin{pmatrix}
    \cR & -\sR \\[1.5mm]
    \sR &  \cR
  \end{pmatrix} \!,~~~~
  U_R' ~=\,
  \begin{pmatrix}
   e^{i\gamma_1^{}} & 0 \\[1.5mm]
   0                & e^{i\gamma_2^{}}
  \end{pmatrix} \!,~~~~~
    U_R'' ~=\,
  \begin{pmatrix}
    1 & 0\\[1.5mm]
    0 & e^{i\beta}
  \end{pmatrix} \!,~~~~~~
\end{eqnarray}
\eeqs
where $\,(\sR,\,\cR) \equiv (\sin\theta_R^{},\,\cos\theta_R^{})\,$.\,
Note that there is no independent Dirac CP phase in the $2\!\times\! 2$ unitary matrix $\,U_R\,$
and all possible CP-phases are included in the diagonal phase matrices
$\,U_R'\,$ and $\,U_R''\,$.\,
For notational convenience, we define,
\begin{subequations}
\begin{eqnarray}
  \widetilde{M}_R  & \!\!\!\equiv\!\!\! &
  {U_R''}^T M_R U_R''
~=\,
  \begin{pmatrix}
    \widetilde M_{22} & \widetilde M_{23} \\[1.5mm]
    \widetilde M_{23} & \widetilde M_{33}
  \end{pmatrix}
\,=~
  M_{22}
  \begin{pmatrix}
    1 & (1 \!-\! r) e^{i \beta} \\[1.5mm]
    (1 \!-\! r) e^{i \beta} & (1 \!-\! \zeta e^{i \omega}) e^{i2\beta}
  \end{pmatrix} \!,~~~~~~
  \label{eq:tilde-MR}
\\[3mm]
  \widetilde{D}_R  & \!\!\!\equiv\!\!\! &
  {U_R'}^* D_R {U_R'}^\dag
~\equiv\,
  \begin{pmatrix}
    \widetilde M_1 & 0 \\[1.5mm]
    0 & \widetilde M_2
  \end{pmatrix}
\,=\,
  \begin{pmatrix}
    M_1 e^{-i2\gamma_1^{}} & 0 \\[1.5mm]
    0 & M_2 e^{-i2\gamma_2^{}}
  \end{pmatrix} \!.
  \label{eq:tilde-DR}
\end{eqnarray}
\label{eq:tilde-R}
\end{subequations}
Then the diagonalization equation (\ref{eq:diagonalization-MR}) takes the
following form,
\begin{equation}
  U^T_R \widetilde{M}_R U_R
~=~
  \widetilde D_R  \,,
\qquad  ~\textrm{or,}~  \qquad  %\Longrightarrow
  \widetilde{M}_R
~=~
  U_R^* \widetilde{D}_R U_R^\dag \,.
\label{eq:diagonalization-tilde-MR}
\end{equation}
Using Eq.\,(\ref{eq:diagonalization-tilde-MR})
we can reconstruct $\,\widetilde{M}_R$\,,
\begin{equation}
  \widetilde M_R
~=~
  \begin{pmatrix}
    c^2_R \widetilde M_1 + s^2_R \widetilde M_2
  & c_R s_R \left( \widetilde M_1 - \widetilde M_2 \right) \\[1.5mm]
    c_R s_R \left( \widetilde M_1 - \widetilde M_2 \right)
  & s^2_R \widetilde M_1 + c^2_R \widetilde M_2
  \end{pmatrix}
  \label{eq:reconstruction-tilde-MR}
\end{equation}
Thus, comparing the right-hand-sides (RHS's)
of Eqs.\,(\ref{eq:tilde-MR}) and (\ref{eq:reconstruction-tilde-MR})
we can generally solve the masses $\,\widetilde M_1\,$ and $\,\widetilde M_2$\,,
\begin{subequations}
  \begin{eqnarray}
    \widetilde M_1
  & \!\!=\!\! &
  \hf \left( \widetilde M_{22} + \widetilde M_{33}
  \right)+
    \frac {\widetilde M_{23}}{\,2\cR\sR\,} \,,
    \label{eq:expression-DR1}
  \\[1mm]
    \widetilde M_2
  & \!\!=\!\! &
  \hf \left( \widetilde M_{22} + \widetilde M_{33}
  \right)-
    \frac {\widetilde M_{23}}{\,2\cR\sR\,} \,.
    \label{eq:expression-DR2}
  \end{eqnarray}
  \label{eq:expression-DR}
\end{subequations}
as well as the rotation angle $\,\theta_R^{}\,$,
\begin{equation}
  \tan 2 \theta_R^{}
~=~
  \f{2 \widetilde M_{23}}
    {\,\widetilde M_{22} - \widetilde M_{33}\,}
~=~
  \f{2(1 - r)}
    {\,\( e^{- i \beta} - e^{i \beta}\) + \zeta e^{i (\omega + \beta)}\,} \,.
  \label{eq:tan-2thetaR}
\end{equation}
As $\,\tan 2\theta_R^{}\,$ is real, the imaginary part on the RHS
of Eq.\,(\ref{eq:tan-2thetaR}) must vanish. This allows us to resolve
$\,\beta\,$ and $\,\theta_R^{}\,$ from (\ref{eq:tan-2thetaR}),
\beqs
  \begin{eqnarray}
    \beta  & = &
    \hf\, \zeta\sin\omega + \O(\zeta^2) \,,
    \label{eq:solution-beta}
  \\
    \theta_R^{} & = &
    \pm \f{\pi}{4}
  - \frac{1}{4}\,\zeta \cos\omega \,.
    \label{eq:solution-thetaR}
  \end{eqnarray}
\eeqs
The sign of $\,\pm\f{\pi}{4}\,$ will be fixed from $\,M_1 < M_2\,$,\, as required by (\ref{eq:q2+}).
Substituting (\ref{eq:solution-thetaR}) back into (\ref{eq:expression-DR}), we deduce,
\begin{subequations}
  \begin{eqnarray}
    M_1   & \!\!=\!\! &
    M_{22} e^{i (2\gamma_1^{}+\beta)}
   \[ 1\pm 1\mp r-\f{\zeta}{2}e^{i\om} +\O(\zeta^2)\]
    \!,
    \label{eq:expression-DR1-b}
  \\[2mm]
    M_2 & \!\!=\!\! &
    M_{22} e^{i (2\gamma_2^{}+\beta)}
     \[ 1\mp 1\pm r-\f{\zeta}{2}e^{i\om} +\O(\zeta^2) \]
    \!.
    \label{eq:expression-DR2-b}
  \end{eqnarray}
  \label{eq:expression-DR-b}
\end{subequations}
We see that requiring $\,M_1\ll M_2\,$ as in (\ref{eq:q2+}) will peak up the minus sign
for $\f{\pi}{4}$ on the RHS of Eq.\,(\ref{eq:solution-thetaR}).
Thus we further simplify (\ref{eq:expression-DR1-b})-(\ref{eq:expression-DR2-b}) as
\beqs
\beqa
  M_1 & \!\!\simeq\!\! &
   M_{22} e^{i (2\gamma_1^{}+\beta)}
   \[ r-\hf{\zeta}e^{i\om} \]
    \!,
\label{eq:MR-M1-f}
  \\[1mm]
      M_2 & \!\!\simeq\!\! &
    M_{22} e^{i (2\gamma_2^{}+\beta)}
    \[ 2 - r-\hf{\zeta}e^{i\om} \] \!.
\label{eq:MR-M2-f}
\eeqa
\label{eq:MR-M1M2-f}
\eeqs
The mass eigenvalues $M_1$ and $M_2$ are real and positive, so taking the absolute value
on both sides of (\ref{eq:expression-DR-b}) we can derive $M_1$ and $M_2$ under the expansion
of $r$ and $\zeta$. The formulas turn out to fully coincide with that in Eq.\,(\ref{eq:MR-M1M2})
which we derived earlier. Finally, requiring the RHS of Eq.\,(\ref{eq:MR-M1M2-f}) be real and
noting $\,\beta=\O(\zeta)\,$,\,
we deduce the phase angles,
%
% \beqs
% \begin{eqnarray}
% && \hspace*{-20mm}
% \sin(2\gamma_1^{}\!+\!\beta) \,=\,
% \f{\f{1}{2}\zeta\sin\omega}
%   {\,[r^2 \!-\! r\zeta\cos\omega \!+\!\f{1}{4}\zeta^2]^{\frac{1}{2}}\,} \,, ~~~~
% \cos(2\gamma_1^{}\!+\!\beta) \,=\,
% \f{r-\f{1}{2}\zeta\cos\omega}
%   {\,[r^2 \!-\! r\zeta\cos\omega \!+\! \f{1}{4}\zeta^2]^{\f{1}{2}}\,} \,,
%      \label{eq:solution-gamma1}
% \\[3mm]
% &&  \hspace*{-20mm}
%   \tan(2\gamma_2^{} \!+\! \beta) \,=\,
%    \f{\zeta\sin\omega}
%      {\,4 \!-\! 2 r \!-\! \zeta \cos\omega\,}
%      ~\simeq~ \f{1}{4}\zeta\sin\om  \,.
%    \label{eq:solution-gamma2}
% \end{eqnarray}
% \label{eq:solution-gamma12}
% \eeqs
%
% Noticing $\,\beta=\O(\zeta)\,$,\, we can further resolve (\ref{eq:solution-gamma12}) as,
%
\beqs
\beqa
&& \hspace*{-20mm}
\sin 2\gamma_1^{} \,\simeq\,
 \f{\f{1}{2}\zeta\sin\omega}
   {\,[r^2 \!-\! r\zeta\cos\omega \!+\!\f{1}{4}\zeta^2]^{\frac{1}{2}}\,} \,, ~~~~
\cos 2\gamma_1^{} \,\simeq\,
 \f{r-\f{1}{2}\zeta\cos\omega}
   {\,[r^2 \!-\! r\zeta\cos\omega \!+\! \f{1}{4}\zeta^2]^{\f{1}{2}}\,} \,,
      \label{eq:solution-gamma11}
\\[2mm]
&&  \hspace*{-20mm}
   \gamma_2^{}
      \,\simeq\, -\f{1}{8}\zeta\sin\om \,=\, \O(\zeta) \,.
    \label{eq:solution-gamma22}
\eeqa
\eeqs
As expected, here we see that the three phase angles
\,$(\gamma_1^{},\,\gamma_2^{},\,\beta)$\,
would vanish as $\,\zeta\sin\om\to 0\,$.

\section{\hspace*{0.05mm} Positivity Constraints from Baryon
                          Asymmetry $\bd{\eta_B^{}}$ }
\label{Appendix:Positivity}

In this Appendix we systematically derive the positivity constraints
from the baryon asymmetry $\,(\eta_B^{}>0)\,$, on the CP-phase
$\,\d_D$\, of our model, as given in
Eqs.\,(\ref{eq:bound-deltaD}) of
Sec.\,\ref{sec:model-prediction-leptogenesis}.
For convenience, we rewrite Eq.\,(\ref{eq:rr}) as
\begin{eqnarray}
\label{eq:r+-}
 r_\pm
~=~
  \frac \zeta 2
  \[
    \cos \delta_D
  \pm
    \sqrt{\frac {s^4_s}{16} \frac {y^2 \zeta^2}{\delta^4_x} -
    \sin^2\delta_D}~
  \] \!, \label{eq:a-185}
\end{eqnarray}
where we have used $\,r_{\pm}\,$ to denote the two sign-combinations
in the brackets. With the above formula we further derive the
positivity condition (\ref{eq:Cond-etaB>0}) of $\,\eta_B^{}\,$ as,
\beqa \label{eq:Cond-etaB-1} \zeta\sin\d_D\(\cos2\delta_D \pm
2\cos\delta_D \sqrt{\f{s^4_s}{16} \frac {y^2 \zeta^2}{\delta^4_x} -
\sin^2\delta_D}~\) \,>~ 0 \,, \eeqa
which reflects the requirement of the baryon asymmetry $\,\eta_B^{}
> 0\,$. Below we will solve this inequality and derive the physical
regions for the Dirac CP-phase $\d_D$ as allowed by the successful
leptogenesis.

\vspace*{2mm}
\subsection{Solution of Eq.\,(\ref{eq:Cond-etaB-1}) for
$\,\boldsymbol{r=r_+^{}=r_-^{}}$ Branch} \label{sec:r+-}

The case of \,$r=r_+^{}=r_-^{}$\, corresponds
$~\zeta\sin\d_D\,\cos2\delta_D \,>\,0\,$.\,
Since our convention always holds
$\,\zeta>0\,$, the condition (\ref{eq:Cond-etaB-1}) reduces to
\beqa
\sin\d_D\cos2\delta_D ~>~ 0 \,,
\eeqa
which requires
\beqa
   \d_D
  ~\in\,
    \( 0,\,\f{\pi}{4} \)
  \bd{\cup}
    \( \f{3\pi}{4},\, \pi\)
  \bd{\cup}
    \( \f{5\pi}{4},\,\f{7\pi}{4} \) .
\eeqa

\vspace*{2mm}
\subsection{Solution of Eq.\,(\ref{eq:Cond-etaB-1}) for
$\,\boldsymbol{r=r_+^{}\neq r_-^{}}\,$ Branch} \label{sec:r+}

The case of  $\,r=r_+^{}\,$ corresponds to $+$ sign in
(\ref{eq:Cond-etaB-1}). With$\,\zeta>0\,$, the condition
(\ref{eq:Cond-etaB-1}) reduces to
\beqa \label{eq:Cond-etaB-r+} \sin\d_D\(\cos 2\d_D + 2\cos\d_D
\sqrt{\frac {s^4_s}{16} \f{y^2 \zeta^2}{\delta^4_x} -
\sin^2\delta_D}~\) ~>~ 0 \,. \eeqa
The defined range of $\d_D$ is $\,[0,\,2\pi)\,$, but the inequality
(\ref{eq:Cond-etaB-1}) requires $\,\sin\d_D\neq 0\,$ and thus
excludes $\,\d_D= 0,\,\pi\,$.\, So, we will analyze the ranges
$\,\d_D\in (0,\,\pi)\,$ and $\,\d_D\in (\pi,\,2\pi)\,$,
respectively.

%\subsubsection{%\hspace*{-5mm}.\hspace*{-0.6mm}
%For the Range $\bd{\delta_D \in (0,\,\pi)}$}
\vspace{5mm}
\noindent {\bfseries B.2.1. For the Range $\bd{\delta_D \in (0,\,\pi)}$ }

\vspace{3mm}
 For $\,\d_D \in (0,\,\pi)\,$, the condition
(\ref{eq:Cond-etaB-r+}) becomes,
\beqa \label{eq:Cond-etaB-r+A} \cos2\delta_D \,>\,
-2\cos\delta_D\sqrt{\frac {s^4_s}{16}
 \f{y^2\zeta^2}{\delta^4_x} - \sin^2\delta_D} \,~,
\eeqa
which can be resolved as below.

\begin{itemize}

\item  For the sub-range $\,\d_D \in \(0,\,\frac{\pi}{4}\)\,$,\,
we have $\,\cos2\d_D > 0\,$ and $\,-\cos\d_D <0\,$,\, which always
hold the condition (\ref{eq:Cond-etaB-r+A}).

\item  For the sub-range $\,\d_D \in \(\f{\pi}{4},\,\f{\pi}{2}\)\,$,\,
we have $\,\cos2\delta_D <0\,$ and $\,-\cos\delta_D <0\,$, leading
to the solution from (\ref{eq:Cond-etaB-r+A}),
\begin{eqnarray}
\label{eq:etaB-sol-1} \cos^2\delta_D ~>~ \f{4}{s^4_s}
\f{\delta^4_x}{y^2\zeta^2} \,.
 \end{eqnarray}

\item
For the sub-range $\,\delta_D \in \(\f{\pi}{2},\,\f{3\pi}{4}\)\,$,\,
we have $\,\cos 2\d_D < 0\,$ and $\,-\cos\d_D >0\,$,\, showing that
the condition (\ref{eq:Cond-etaB-r+A}) has no solution.

\item
For the sub-range $\,\delta_D  \in \(\f{3\pi}{4},\,\pi\)\,$,\, we
have $\,\cos 2\d_D >0\,$ and $\,-\cos\delta_D >0\,$,\, leading to
the solution from (\ref{eq:Cond-etaB-r+A}),
\begin{eqnarray}
\label{eq:etaB-sol-2} \cos^2\delta_D ~<~ \f{4}{s^4_s}
\f{\delta^4_x}{y^2\zeta^2} \,.
\end{eqnarray}
\end{itemize}

%\subsubsection{%\hspace*{-4.7mm}.\hspace*{-0.6mm}
%For the Range $\bd{\delta_D \in (\pi,\,2\pi)}$}
\vspace{5mm}
 \noindent
 {\bfseries B.2.2. For the Range $\bd{\delta_D \in (\pi,\,2\pi)}$ }

\vspace{3mm}

For $\,\d_D \in (\pi,\,2\pi)\,$, the condition
(\ref{eq:Cond-etaB-r+}) becomes,
\beqa \label{eq:Cond-etaB-r+B} \cos2\delta_D \,<\,
-2\cos\delta_D\sqrt{\frac {s^4_s}{16}
 \f{y^2\zeta^2}{\delta^4_x} - \sin^2\delta_D} \,~,
\eeqa
which can be resolved as below.

\begin{itemize}

\item
For the sub-range $\,\d_D \in \(\pi,\frac{5\pi}{4}\)$,\, we have
$\,\cos2\d_D >0\,$ and $\,-\cos\d_D >0\,$.\, From the condition
(\ref{eq:Cond-etaB-r+B}), we deduce the solution just as in
(\ref{eq:etaB-sol-1}).

\item
For the sub-range  $\,\d_D \in \(\f{5\pi}{4},\,\f{3\pi}{2}\,\)$,\,
we have  $\,\cos2\delta_D <0\,$ and $\,-\cos\delta_D >0\,$, which
always hold the condition (\ref{eq:Cond-etaB-r+B}).

\item
For the sub-range $\,\d_D \in \(\f{3\pi}{2},\,\f{7\pi}{4}\)\,$,\, we
have $\,\cos 2\d_D <0\,$ and $\,-\cos\d_D<0\,$.\, From the condition
(\ref{eq:Cond-etaB-r+B}), we deduce the solution just as in
(\ref{eq:etaB-sol-2}).

\item
For the sub-range  $\,\d_D  \in \(\f{7\pi}{4},\,2\pi\)$,\, we have
$\,\cos 2\d_D >0\,$ and $\,-\cos\d_D <0\,$,\, showing that the
condition (\ref{eq:Cond-etaB-r+B}) has no solution.
\end{itemize}

\vspace*{2mm}
\subsection{Solution of Eq.\,(\ref{eq:Cond-etaB-1}) for
            $\,\boldsymbol{r=r_-^{}}\,\neq r_+$ Branch}
\label{sec:r-}

The case of  $\,r=r_-^{}\,$ corresponds to $-$ sign in
(\ref{eq:Cond-etaB-1}). With $\,\zeta>0\,$ in our convention, we
reduce the condition (\ref{eq:Cond-etaB-1}) to
\beqa \label{eq:Cond-etaB-r-} \sin\d_D\(\cos 2\d_D - 2\cos\d_D
\sqrt{\frac {s^4_s}{16} \f{y^2 \zeta^2}{\delta^4_x} -
\sin^2\delta_D}~\) ~>~ 0 \,, \eeqa
which excludes $\,\d_D= 0,\,\pi\,$.\, So, we will analyze the ranges
$\,\d_D\in (0,\,\pi)\,$ and $\,\d_D\in (\pi,\,2\pi)\,$,
respectively.

%\subsubsection{%\hspace*{-4.7mm}.\hspace*{-0.6mm}
%For the Range $\bd{\delta_D \in (0,\,\pi)}$}

\vspace{5mm}
\noindent
{\bfseries B.3.1. For the Range $\bd{\delta_D \in (0,\,\pi)}$ }

\vspace{3mm}

 For $\,\d_D \in (0,\,\pi)\,$, the condition
(\ref{eq:Cond-etaB-r-}) becomes,
\beqa \label{eq:Cond-etaB-r-A} \cos2\delta_D \,>\,
2\cos\delta_D\sqrt{\frac {s^4_s}{16}
 \f{y^2\zeta^2}{\delta^4_x} - \sin^2\delta_D} \,~,
\eeqa
which can be resolved as below.

\begin{itemize}

\item
For the sub-range  $\,\d_D \in \(0,\,\f{\pi}{4}\)$,\, we have
$\,\cos 2\d_D > 0\,$ and $\,\cos\d_D >0\,$.\, From the condition
(\ref{eq:Cond-etaB-r-A}), we deduce the solution just as in
(\ref{eq:etaB-sol-2}).

\item
For the sub-range  $\,\d_D \in \(\f{\pi}{4},\,\f{\pi}{2}\)$,\, we
have $\,\cos 2\d_D < 0\,$ and $\,\cos\d_D >0\,$,\, showing the
condition (\ref{eq:Cond-etaB-r-A}) has no solution.

\item
For the sub-range $\,\d_D \in \(\f{\pi}{2},\,\f{3\pi}{4}\)$,\, we
have  $\,\cos 2\d_D <0\,$ and $\,\cos\d_D <0\,$.\, From the
condition (\ref{eq:Cond-etaB-r-A}), we deduce the solution just as
in (\ref{eq:etaB-sol-1}).

\item
For the sub-range $\,\d_D \in \(\f{3\pi}{4},\,\pi\)$,\, we have
$\,\cos 2\d_D >0\,$\, and $\,\cos\d_D<0\,$,\, which always hold the
condition (\ref{eq:Cond-etaB-r-A}).

\end{itemize}

%\subsubsection{%\hspace*{-4.7mm}.\hspace*{-0.6mm}
%For the Range $\bd{\delta_D \in (\pi,\,2\pi)}$}

\vspace{5mm}
 \noindent
 {\bfseries B.3.2. For the Range $\bd{\delta_D \in (\pi,\,2\pi)}$ }
\vspace{3mm}

For $\,\d_D \in (\pi,\,2\pi)\,$, the condition
(\ref{eq:Cond-etaB-r-}) becomes,
\beqa \label{eq:Cond-etaB-r-B} \cos2\delta_D \,<\,
2\cos\delta_D\sqrt{\frac {s^4_s}{16}
 \f{y^2\zeta^2}{\delta^4_x} - \sin^2\delta_D} \,~,
\eeqa
which can be resolved as below.

\begin{itemize}

\item
For the sub-range $\,\d_D \in \(\pi,\,\f{5\pi}{4}\)$,\, we have
$\,\cos 2\d_D >0$\, and $\,\cos\d_D <0$\,,\, showing that the
condition (\ref{eq:Cond-etaB-r-B}) has no solution.

\item
For the sub-range $\,\d_D \in \(\f{5\pi}{4},\,\f{3\pi}{2}\)$,\, we
have $\,\cos 2\d_D <0$\, and \,$\cos\d_D <0$\,.\, From the condition
(\ref{eq:Cond-etaB-r-B}), we deduce the solution just as in
(\ref{eq:etaB-sol-2}).

\item
For the sub-range $\,\d_D \in \(\f{3\pi}{2},\,\f{7\pi}{4}\)$,\, we
have $\,\cos 2\d_D <0$\, and $\,\cos\d_D>0$\,,\, which always hold
the condition (\ref{eq:Cond-etaB-r-B}).

\item
For the sub-range $\,\d_D \in \(\f{7\pi}{4},\,2\pi\)$,\, we have
$\,\cos2\d_D >0\,$ and $\,\cos\d_D >0\,$.\, From the condition
(\ref{eq:Cond-etaB-r-B}), we deduce the solution just as in
(\ref{eq:etaB-sol-1}).

\end{itemize}

%\subsection{%\hspace*{-5mm}.\hspace*{-0.6mm}
%Summary of All Solutions to Positivity Condition}
%\label{sec:summary-solutions}

\vspace*{3mm}
Finally, we can summarize all solutions to the positivity condition (\ref{eq:a-185})
into a compact form,
for $\,r=r_+^{}\,=r_-^{}\,$, $\,r=r_+^{}\neq\,r_-^{}\,$ and
$\,r=r_-^{}\,\neq \,r_+^{}\,$,
\begin{subequations}
\begin{eqnarray}
\label{eq:bound-deltaD-00}
&& \hspace*{-28mm}
\textrm{for} ~~r\,=\,r_+^{}\,=\,r_-\,,
\nonumber\\[2mm]
&& \hspace*{-19mm}
  \d_D \in
    \( 0,\,\f{\pi}{4} \)
  \bd{\cup}
    \( \f{3\pi}{4},\, \pi\)
  \bd{\cup}
    \( \f{5\pi}{4},\,\f{7\pi}{4} \)  ;
\\[3mm]
&& \hspace*{-28mm}
\textrm{for} ~~r \,=\, r_+^{} \,\neq\, r_- \,,
\nonumber\\[2mm]
 && \hspace*{-19mm}
\d_D \in
    \( 0,\,\f{\pi}{4} \]
  \bd{\cup}
    \( \f{\pi}{4},\, \f{\pi}2\)_>
  \bd{\cup}
    \( \f{3\pi}{4},\,\pi \)_<  \cup
    \( \pi,\,\f{5\pi}{4} \)_>  \cup
    \[\frac{5\pi}{4},\,\frac{3\pi}{2}\] \cup
    \( \frac{3\pi}{2},\,\f{7\pi}{4}
    \)_< \,;
\\
\label{eq:bound-deltaD-aa}
&& \hspace*{-28mm}
\textrm{for} ~~r=r_-^{}\neq r_+
\nonumber \\[2mm]
&& \hspace*{-19mm}
\d_D \in
    \( 0,\,\f{\pi}{4} \)_<
  \bd{\cup}    \( \f{\pi}{2},\, \f{3\pi}4\)_>
  \bd{\cup}    \[ \f{3\pi}{4},\,\pi \)
  \bd{\cup}    \( \frac{5\pi}{4},\,\f{3\pi}{2} \)_<
  \bd{\cup}    \[\frac{3\pi}{2},\,\frac{7\pi}{4}\]
  \bd{\cup}    \( \frac{7\pi}{4},\,2\pi \)_> \,;
 \label{eq:bound-deltaD-bb}
  \end{eqnarray}
\label{eq:bound-deltaD-2}
\end{subequations}
where the regions with subscript ``$>$"  require imposing the condition
(\ref{eq:etaB-sol-1})
and those with subscript ``$<$" are constrained by the condition
(\ref{eq:etaB-sol-2}). The above constraints just confirm what we have presented in
Eq.\,(\ref{eq:bound-deltaD}) of Sec.\,\ref{sec:model-prediction-leptogenesis}.

 %\vspace*{5mm}

 \end{document}